%% file: ms.tex
\pdfoutput=1

\documentclass[12pt,preprint]{aastex}

\usepackage[colorinlistoftodos]{todonotes}
\usepackage[normalem]{ulem}
\usepackage[nolist]{acronym}  

\slugcomment{}

\shorttitle{The EBEX Instrument}
\shortauthors{EBEX Collaboration}

\include{document_commands}

\usepackage{ulem}  

\begin{document}


\title{The EBEX Balloon Borne Experiment -- Detectors and Readout}


\author{The EBEX Collaboration:
Maximilian~Abitbol\altaffilmark{1},
Asad~M.~Aboobaker\altaffilmark{2, 3*},
Peter~Ade\altaffilmark{4}, 
Derek~Araujo\altaffilmark{1},
Fran\c{c}ois~Aubin\altaffilmark{5, 2*, \dag},
Carlo~Baccigalupi\altaffilmark{6,7},
Chaoyun~Bao\altaffilmark{2}, 
Daniel~Chapman\altaffilmark{1},
Joy~Didier\altaffilmark{1, 8*},
Matt~Dobbs\altaffilmark{5,9},
Stephen~M.~Feeney\altaffilmark{10, 11*},
Christopher~Geach\altaffilmark{2},
Will~Grainger\altaffilmark{4},
Shaul~Hanany\altaffilmark{2}, 
Kyle~Helson\altaffilmark{12, 13*},
Seth~Hillbrand\altaffilmark{1},
Gene~Hilton\altaffilmark{14},
Johannes~Hubmayr\altaffilmark{14}, 
Kent~Irwin\altaffilmark{14, 15*},
Andrew~Jaffe\altaffilmark{10},
Bradley~Johnson\altaffilmark{1},
Terry~Jones\altaffilmark{2},
Jeff~Klein\altaffilmark{2},
Andrei~Korotkov\altaffilmark{12},
Adrian~Lee\altaffilmark{16},
Lorne~Levinson\altaffilmark{17},
Michele~Limon\altaffilmark{1}, 
Kevin~MacDermid\altaffilmark{5},
Amber~D.~Miller\altaffilmark{1,8*},
Michael~Milligan\altaffilmark{2},
Kate~Raach\altaffilmark{2},
Britt~Reichborn-Kjennerud\altaffilmark{1},
Carl~Reintsema\altaffilmark{14},
Ilan~Sagiv\altaffilmark{17},
Graeme~Smecher\altaffilmark{5}, 
Gregory~S.~Tucker\altaffilmark{12},
Benjamin~Westbrook\altaffilmark{16},
Karl~Young\altaffilmark{2},
Kyle~Zilic\altaffilmark{2}}


\altaffiltext{1}{Physics Department, Columbia University, New York, NY 10027} 
\altaffiltext{2}{School of Physics and Astronomy, University of Minnesota-Twin Cities, Minneapolis, MN 55455} 
\altaffiltext{3}{Jet Propulsion Laboratory, California Institute of Technology, Pasadena, CA 91109} 
\altaffiltext{4}{Rutherford Appleton Lab, Harwell Oxford, OX11 0QX, United Kingdom} 
\altaffiltext{5}{Department of Physics, McGill University, Montreal, H3A 2T8, Canada} 
\altaffiltext{6}{Astrophysics Sector, SISSA, Trieste, 34014, Italy} 
\altaffiltext{7}{INFN, Sezione di Trieste, Via Valerio 2, I-34127 Trieste, Italy} 
\altaffiltext{8}{Department of Physics and Astronomy, University of Southern California, Los Angeles, CA  90089} 
\altaffiltext{9}{Canadian Institute for Advanced Research, Toronto, M5G 1Z8, Canada} 
\altaffiltext{10}{Department of Physics, Imperial College, London, SW7 2AZ, United Kingdom} 
\altaffiltext{11}{Center for Computational Astrophysics, Flatiron Institute, New York, NY 10010} 
\altaffiltext{12}{Department of Physics, Brown University, Providence, RI 02912} 
\altaffiltext{13}{NASA Goddard Space Flight Center, Greenbelt, MD 20771} 
\altaffiltext{14}{National Institute of Standards and Technology, Boulder, CO 80305} 
\altaffiltext{15}{Department of Physics, Stanford University, Stanford, CA 94305} 
\altaffiltext{16}{Department of Physics, University of California, Berkeley, CA 94720} 
\altaffiltext{17}{Faculty of Physics, Weizmann Institute of Science, Rehovot, 76100, Israel} 
\altaffiltext{*}{Current affiliation} 
\altaffiltext{\dag}{Corresponding Author: Fran\c{c}ois~Aubin (faubin@umn.edu)}


\begin{abstract}

EBEX was a long-duration balloon-borne experiment to measure the polarization of the cosmic 
microwave background. The experiment had three frequency bands centered at 150, 250, 
and 410~GHz and was the first to use a kilo-pixel array of transition edge sensor (TES) bolometers aboard 
a balloon platform; shortly after reaching float we operated 504, 342, and 109 TESs at each of 
the bands, respectively. 
 The array was read out with a frequency domain multiplexing (FDM) system.  We describe 
the design and characterization of the array and the readout system.  We give 
the distributions of measured thermal conductances, normal resistances, and transition 
temperatures. From the lowest to highest frequency, 
the median measured average thermal conductances were 39, 53, and 63~pW/K, the medians of 
transition temperatures were 0.45, 0.48, and 0.47~K, and the medians of 
normal resistances were 1.9, 1.5, and 1.4~$\Omega$. With the exception 
of the thermal conductance at 150~GHz, all measured values are within 30\% of their 
design. We measured median low-loop-gain time constants $\tau_{0}$ = 88, 46, and 57~ms
and compare them to predictions. Two measurements of bolometer absorption 
efficiency gave results consistent within 10\% and showing high ($\sim$0.9)
efficiency at 150~GHz and medium ($\sim$0.35, and $\sim$0.25) at the two 
higher bands, respectively. We measure a median total optical load 
of 3.6, 5.3 and 5.0~pW absorbed at the three bands, respectively. EBEX pioneered
the use of the digital version of the FDM system. The system multiplexed the bias and 
readout of 16 bolometers onto two wires. We present accounting of the measured noise equivalent
power and  compare it to predictions. The median per-detector noise equivalent temperatures 
referred to a black body with a temperature of 2.725~K  
are 400,  920, and 14500 $\mu$K$\sqrt{s}$ for the three bands, respectively.  We 
compare these values to our pre-flight predictions and to a previous balloon payload, 
discuss the sources of excess noise, and the path for a future payload to make full 
use of the balloon environment.
\end{abstract}



\maketitle


\input{introduction2}  
\input{focal_plane_layout}
\input{time_domain_data}
\input{readout}
\input{detectors}
\input{detector_optimization}
\input{detector_characterization}
\input{time_constants}
\input{optical_load}
\input{optical_efficiency}
\input{noise}
\input{net_and_maps}
\input{summary2}


\input{acronyms2}

\acknowledgments
Support for the development and flight of the EBEX instrument was provided by 
NASA grants NNX12AD50G, NNX13AE49G, 
NNX08AG40G, and NNG05GE62G, and by NSF grants AST-0705134 and ANT-0944513.   
We acknowledge support from the Italian INFN INDARK Initiative.  Ade and Tucker acknowledge the 
Science \& Technology Facilities Council for its continued support of the underpinning 
technology for filter and waveplate development.  
We also acknowledge support by the Canada Space Agency, the Canada Research 
Chairs Program, the Natural Sciences and 
Engineering Research Council of Canada, the Canadian Institute for Advanced Research, 
the Minnesota Supercomputing Institute, the National Energy Research Scientific Computing 
Center, the Minnesota and Rhode Island 
Space Grant Consortia, our collaborating institutions, and Sigma Xi the Scientific Research Society. 
Research described in this paper used facilities of the Midwest 
Nano Infrastructure Corridor (MINIC), a part of the National 
Nanotechnology Coordinated Infrastructure (NNCI) program of the National 
Science Foundation.
Baccigalupi acknowledges support from the RADIOFOREGROUNDS grant of the European Union's Horizon 2020 
research and innovation program (COMPET-05-2015, grant agreement number 687312). 
Didier acknowledges a NASA NESSF fellowship NNX11AL15H.  Reichborn-Kjennerud acknowledges an NSF 
Post-Doctoral Fellowship AST-1102774, 
and a NASA Graduate Student Research Fellowship. Raach and Zilic acknowledge 
support by the Minnesota Space Grant Consortium.
The Flatiron Institute is supported by the Simons Foundation. Feeney was partially supported 
by the UK Science and Technology Facilities Council (STFC).
We very much thank Danny Ball and his colleagues at the Columbia Scientific Balloon Facility for their dedicated 
support of the EBEX program.  We thank Darcy Baron and Kaori Hattori for inputs on the stray inductance of the microstrips 
and an anonymous referee for his/her careful and thoughtful review. 


\newpage
\appendix
\input{reference_maps}
\clearpage

\bibliographystyle{aasjournal}
\bibliography{EBEXPaper2}

\end{document}

%% file: document_commands.tex
\usepackage[nolist]{acronym}
\usepackage[margin=0.75in]{geometry}                
\usepackage{graphicx}
\usepackage{subfigure}
\usepackage{amssymb}
\usepackage{xcolor}
\usepackage{fixltx2e}
\usepackage{amsmath}
\usepackage{float}
\usepackage{tikz}
\usepackage{subfiles}
\usepackage{multirow}
\usepackage{multicol}
\usepackage{calrsfs}
\usepackage{morefloats}

\usetikzlibrary{calc,fit,shadows}
\newcommand{\planck}{{\it Planck}}
\newcommand{\ebexLDB}{EBEX2013}

\newcommand{\PinBRC}{593}
\newcommand{\PoutBRC}{723}
\newcommand{\PperDfMUX}{21}
\newcommand{\PperPcrate}{296}
\newcommand{\PperCh}{0.33}

%% file: introduction2.tex
\section{Introduction}

\label{sec:introduction}

Measurements of the \ac{CMB} have provided a wealth of information about the physical 
mechanisms responsible for the evolution of the Universe.
In recent years, experimental efforts have focused on measuring the \ac{CMB}'s polarization 
patterns: {\it E}-modes and {\it B}-modes~\citep{zaldarriaga97_physrev}. 
The level and specific shape of the angular power spectrum of CMB {\it E}-mode
polarization can be predicted given the measured intensity anisotropy. 
Lensing of {\it E}-modes by the large scale structure of the Universe produces
cosmological {\it B}-modes at small angular scales. An inflationary phase at sufficiently 
high energy scales near the big bang is predicted to leave another perhaps detectable {\it B}-mode signature at large 
and intermediate angular scales~\citep{Baumann_2009}.  

The {\it E}-mode polarization of the \ac{CMB}
was first detected by the DASI experiment~\citep{DASI_Emodes}, and other
experiments soon followed suit~\citep{Scott_2010}.  
The combination of all measurements is in excellent agreement with
predictions. {\it B}-mode polarization from gravitational lensing 
of {\it E}-modes and from Galactic dust emission has also recently been 
detected~\citep{SPT_Bmodes, PolarBear_Bmodes, ACT_Bmodes, bicep2detection, bicep+planck}. 
Intense efforts are ongoing by ground- and balloon-based instruments to
improve the measurements, separate the Galactic from the cosmological signals, and 
identify the inflationary {\it B}-mode signature.


The \ac{EBEX} was a balloon-borne CMB polarimeter designed to 
detect or constrain the levels of the inflationary gravitational wave and lensing {\it B}-mode power spectra.  
\ac{EBEX} was also designed to be a technology pathfinder for future \ac{CMB} space missions.
To achieve instrument sensitivity, we implemented a kilo-pixel array of \ac{TES} bolometers and planned
for a long duration balloon flight. We included three spectral bands centered on 150, 250, and 410~GHz to 
give sensitivity to both the CMB and the Galactic dust foreground. 
The combination of the 400~deg$^2$ intended survey size and an optical system with 0.1$\degr$ resolution gave 
sensitivity to the range 30~$<$~$\ell$~$<$~1500 of the angular power spectrum. 
Polarimetry was achieved with a continuously rotating achromatic \ac{HWP} and a wire grid analyzer.


Design and construction of the experiment began in 2005.
A ten-hour engineering flight was launched from Ft.\ Sumner, NM on June 11, 
2009, and the long-duration science flight was launched from McMurdo
Station, Antarctica on December 29, 2012. Data collection ceased when cryogens expired after 11 days, as planned. 
Because the majority of the flight occurred during January 2013, we refer to this flight as EBEX2013.


This paper is one of a series of papers describing the experiment and its in-flight performance.
This paper, called \ac{EP2}, discusses the detectors and readout; \ac{EP1}~\citep{EBEXPaper1} describes the 
optical system, the receiver, and the polarimetric approach; \ac{EP3}~\citep{EBEXPaper3}, gives information
about the gondola, the attitude control system, and other support systems. A fourth paper, \ac{EP4}~\citep{EBEXPaper4}
discusses effects associated with detector non-linearity and their influence on polarimetric measurements.  
Several other publications give additional details about \ac{EBEX}. Some are 
from earlier stages of the program~\citep{Oxley_EBEX2004, hubmayr_ebex2008, Grainger_EBEX2008, Aubin_TESReadout2010, Milligan_Software, ReichbornKjennerud_EBEX2010, klein_HWP, Sagiv_MGrossman2012, Westbrook_2012}, and others discuss some
subsystems in more detail~\citep{Dan_thesis, Britt_thesis, Sagiv_thesis, aubin_thesis, MacDermid_thesis, MacDermid_SPIE2014, Westbrook_thesis, Zilic_thesis, chappy_spie, chappy_thesis, chappy_ieee_paper, joy_ieee_paper, joy_thesis, Aubin_MGrossman2015}. 

Several new technologies have been implemented and tested for the first time in the \ac{EBEX} instrument. 
It was the first astrophysical instrument to implement and operate 
a \ac{SMB}, which was used to levitate the \ac{HWP}. This system was discussed in \ac{EP1}. 
\ac{EBEX} was also the first balloon-borne experiment to implement a kilo-pixel array of \ac{TES} bolometric detectors  
and the first experiment to implement a digital frequency domain multiplexing system to read out the \ac{TES} arrays. 
The implementation of these new technologies aboard a balloon flight presented several new challenges. 
These included the development of detectors that were optimized for the smaller optical 
loading at the stratosphere and that were tuned to operate at higher frequency bands compared to 
those typically used on ground-based instruments,
and the development of a readout system with as low power dissipation per channel as realistically 
implementable while still giving low noise and low 1/$f$ knee. 

This paper describes our implementation of the readout system, and the design and characterization 
of the bolometers. In Section~\ref{sec:focal_plane_layout} we give a brief overview of the instrument, its optical system, and the design 
of the focal plane. Section~\ref{sec:timedomaindata} gives the characteristics of the time domain data and describes the processing 
that was necessary to extract time-domain noise levels. The implementation of the readout system and an assessment of its noise
and gain stability are the subject of Section~\ref{sec:detector_readout}. In Section~\ref{sec:detectors} we discuss the design 
of the detectors and give results from laboratory characterization measurements. We also describe the measured in-flight radiative 
load and use this measurement as one of three approaches to extract the bolometer absorption efficiencies. We end this 
Section with a discussion of the measured \ac{NEP}. In Section~\ref{sec:temperaturenoise} we report 
our measurements of the \ac{NET} and the final map depth at each of the three frequency bands. We 
discuss and summarize our findings in Section~\ref{sec:summary}.


%% file: focal_plane_layout.tex

\section{The EBEX Instrument and Focal Planes}
\label{sec:focal_plane_layout}

\begin{figure}[ht!]
  \includegraphics[width=\textwidth]{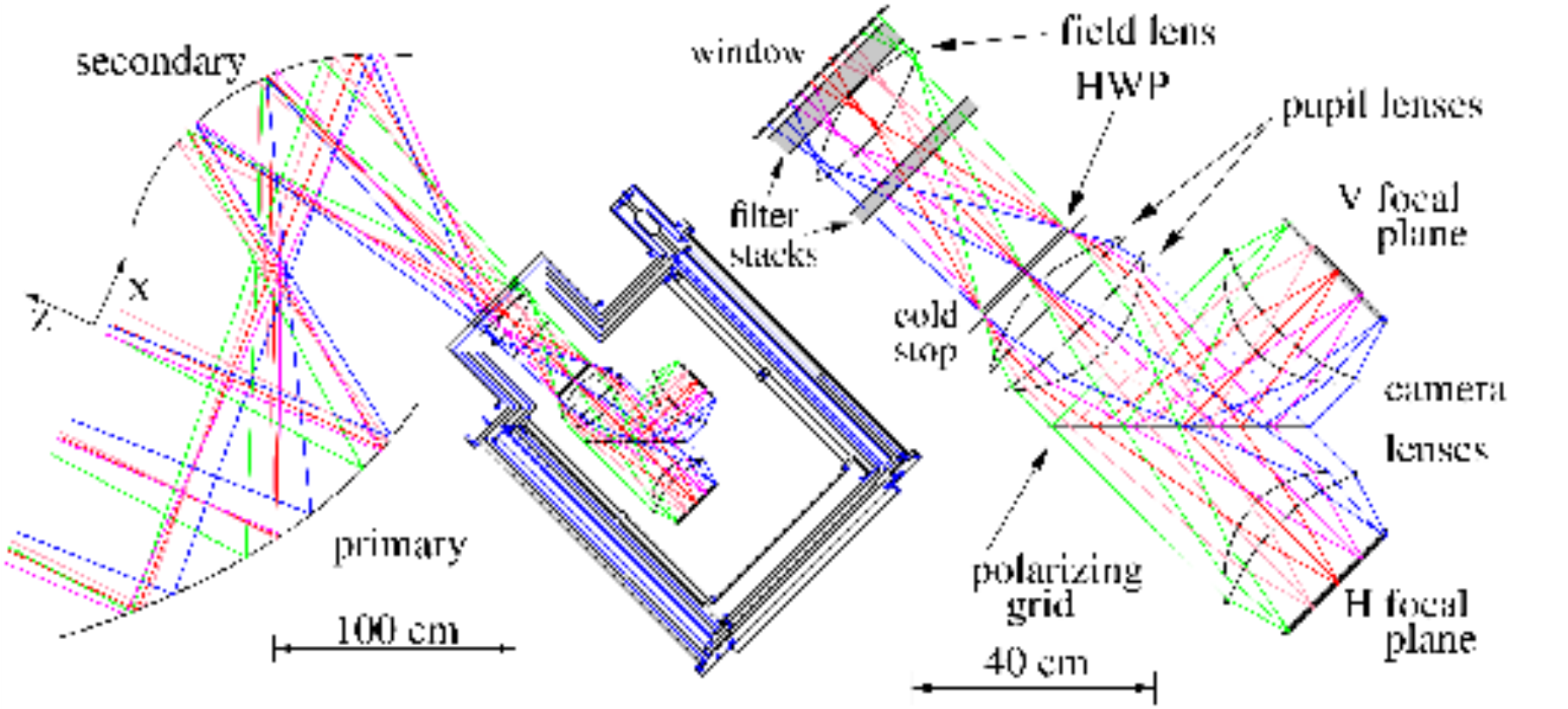}
  \caption[\ac{EBEX} optics ray trace]{Left: ray tracing of the \ac{EBEX}
    optical design consisting of two ambient temperature reflectors in 
    a Gregorian configuration and a cryogenic receiver. Right: inside the 
    receiver, cryogenically cooled polyethylene lenses formed a cold stop
    and provided diffraction limited performance over a flat, telecentric, $6.6\degr$ field 
    of view. Two identical focal planes (H and V) terminated the optical path.}
  \label{fig:raydiagram}
\end{figure}

The \ac{EBEX} optical system consisted of a 105~cm aperture off-axis Gregorian telescope
coupled to a cryogenic receiver containing refractive optics, a rotating achromatic \ac{HWP} at a
cold aperture stop, and a polarizing grid that directed independent polarizations to each 
of two focal planes cooled to a bath temperature of $\sim$0.25~K; see Figure~\ref{fig:raydiagram}. 
The focal planes were referred to as the H (horizontal) and V (vertical) focal planes. The optical and 
cryogenic systems are described in detail in \ac{EP1}. 

\begin{figure}[ht]
\centering
\includegraphics[width=0.9\textwidth]{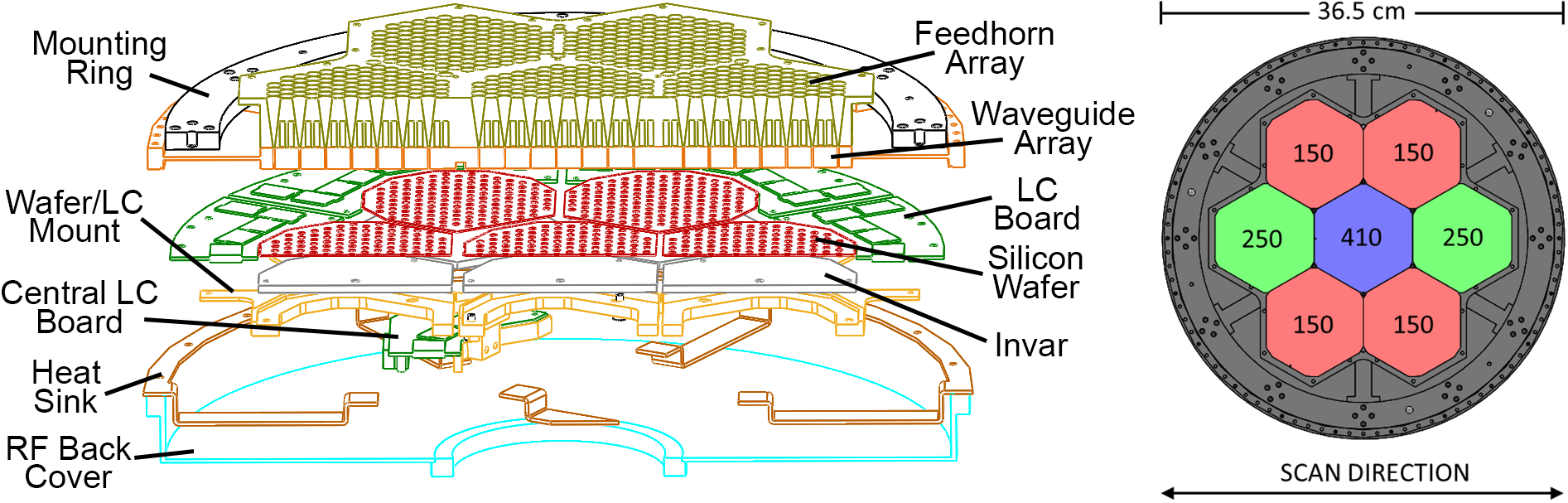} 
\caption{Left: a solid model cross section through one of the EBEX focal planes. Right: sketch of a photon view.  The instrument had 
two identical focal planes. The colors of the hexagons (and numbers) encode frequency bands (in GHz). 
\label{fig:EBEX_Focal_Plane} }
\end{figure}

Figure~\ref{fig:EBEX_Focal_Plane} shows the elements of a focal plane. It consisted of a layer of 
electromagnetic filters, a monolithic
array of feedhorns attached to a monolithic array of waveguides, 7 detector wafers coupled to wafer holders and 
\ac{LC} boards, and a back cover.
The back cover, together with the array of horns, completed a Faraday cage around 
the focal plane. The electromagnetic filters and waveguides 
defined frequency bands centered on 150, 250, and 410~GHz. Each focal plane was arranged such that 
4 wafers operated at 150~GHz, 2 at 250~GHz, and 1 at 410~GHz; see Figure~\ref{fig:EBEX_Focal_Plane}. 
The \ac{LC} boards were part of the 
multiplexed frequency domain bias and readout of the detectors, see Section~\ref{sec:detector_readout}. 
Each detector wafer had 140 lithographically fabricated bolometers of which up to 127 
could be biased and read out with 8 pairs of wires, see Section~\ref{sec:detector_characterization}.  

Each \ac{TES} bolometer had a pair of electrical lines lithographed with niobium connected to bond pads at the edge 
of its wafer. Wire bonds connected the wafer to an \ac{LC} board, which contained inductors and 
capacitors that were part of the frequency 
domain multiplexing readout; see Section \ref{sec:detector_readout}.  We placed each silicon wafer 
onto a wafer-shaped piece of Invar thinly 
coated with warmed Apiezon N grease. For the edge wafers, the Invar was screwed into an aluminum mount holding 
the \ac{LC} board in the same plane as the wafers. For the central wafer, 
we mounted the \ac{LC} board below this plane with standoffs. 
To align the wafers with the feedhorns, we doweled the wafer's aluminum mount to a jig that simulated the 
waveguide array but had fewer, larger holes. 
We warmed the Invar and nudged the wafers until the bolometers were aligned with the alignment holes. The same 
alignment dowels that were used 
with the jig were later used with the waveguide and feedhorn array. 
We soldered to the \ac{LC} boards a commercially made microstrip with copper traces that was terminated with a micro-D connector. 
An additional set of self-made microstrips with niobium wires, which are 
described in Section~\ref{sec:microstrips}, transmitted the 
signals to the \ac{SQUID} boards. Commercial wire harnesses with manganin twisted pairs embedded in a Nomex weave
connected the \ac{SQUID} boards to the room temperature readout electronics.




%% file: time_domain_data.tex

\section{The Time Domain Data}
\label{sec:timedomaindata}

\begin{figure}[ht]
\centering
\includegraphics[width=0.97\textwidth]{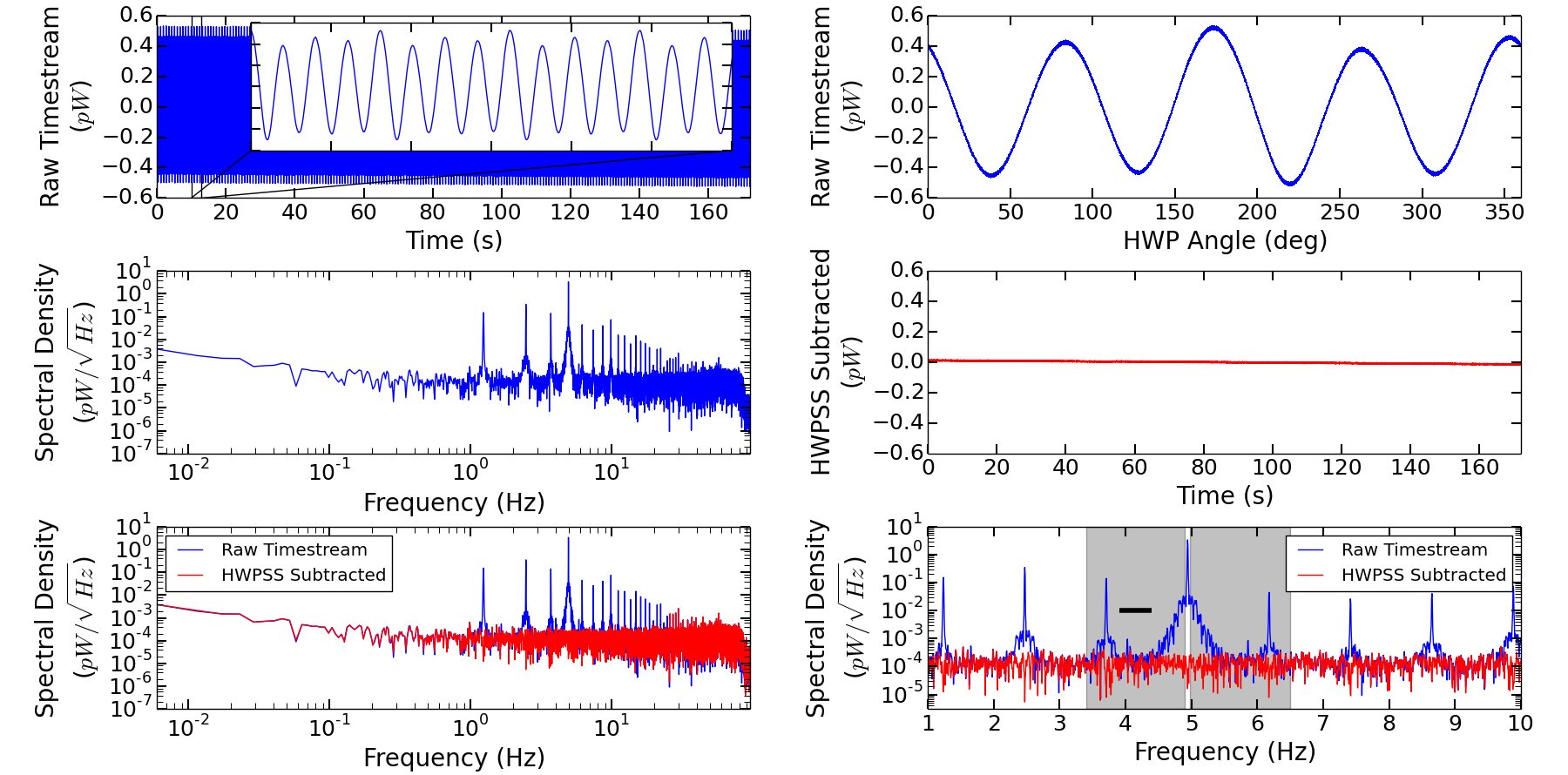} 
\caption{ Time ordered and spectral density data for 172~s section of a 150~GHz detector before (blue) 
and after (red) HWPSS removal.  The \ac{TOD} are given in units of power incident on the telescope. 
For the 150~GHz band, 0.31~pW power fluctuation is equivalent to 1~K$_{\mathrm{CMB}}$.
Upper left: calibrated time domain data and a magnified view of 3~s (inset) showing a sinusoidal behavior in time. 
Upper right: the data in HWP angle demonstrating a stable signal 
that is synchronous with the rotation of the \ac{HWP}.  
Middle left: spectral density of the data.  
Middle right: the data (upper left) after subtraction of the \ac{HWPSS}.  
Lower left: spectral densities before and after \ac{HWPSS} removal. 
Lower right: spectral densities before and after \ac{HWPSS} removal on a linear frequency scale, 
showing the first eight harmonics of the \ac{HWP} rotation frequency. The shaded region indicates the sidebands surrounding 
the fourth harmonic, in which the $Q$ and $U$ signal reside.  The black bar indicates the frequency range (from 3.9 to 4.4~Hz) 
used to quantify noise levels. 
\label{fig:template1} }
\end{figure}

During polarimetric observations we rotated the \ac{HWP} at $f_{HWP}$~=~1.235~Hz. 
This frequency was chosen considering the intended sky scan speed and the pre-flight measured detector 
time constants.
The combination of the \ac{HWP} rotation 
and the fixed polarizing grid modulated the polarized radiation incident on the \ac{HWP} at frequency 4$f_{HWP}$ in the bolometer \ac{TOD}. 
Sources of incident polarized radiation included polarized sky sources, such as the Galaxy and the \ac{CMB}, polarized emission by the 
instrument, primarily arising from the two reflectors, and sources of instrumental polarization.
As described in~\ac{EP1},
the primary source of instrumental polarization was the field lens but there were also smaller 
contributions from the reflectors and the vacuum window. The instrument polarized emissions and instrumental polarization were 
much more intense than both polarized sky emissions and instrument noise. 
Figure~\ref{fig:template1} shows the typical characteristics of the calibrated \ac{TOD} during a period in which 
the \ac{HWP} was rotating.  A large amplitude signal that is synchronous with the rotation of the \ac{HWP}, 
and which we call \ac{HWPSS}, is apparent at harmonics of the \ac{HWP} rotation frequency and is due to the mostly 
time stable instrumental polarization and instrumental polarized emission. The spectral density of the \ac{TOD}, also 
given in Figure~\ref{fig:template1}, shows
that the amplitude of the \ac{HWPSS} is orders of magnitude larger than the white noise level. The maximum 
Galactic signal at 150~GHz is expected to be 0.3~fW, also much smaller than the \ac{HWPSS}. 
\citet{joy_thesis}, \citet{derek_thesis}, and \citet{EBEXPaper4} give more information about the \ac{HWPSS}.

In this paper we give an assessment of the noise properties of the instrument including during astrophysical observations. 
To assess these noise properties we first removed the \ac{HWPSS}. 
The \ac{HWPSS} is modeled as the sum of the first 20 harmonics of the \ac{HWP} rotation frequency.  The \ac{HWPSS} for a given 
detector is estimated by performing a maximum likelihood fit of the sine and cosine amplitudes of the harmonics on 60~s 
chunks of data. In addition to a constant term, the amplitudes are also allowed to vary linearly with time. This procedure 
follows the one used for MAXIPOL~\citep{johnson_apj_2007} and more details are given in \citet{joy_thesis} and in~\citet{derek_thesis}.
The \ac{HWPSS} is subtracted from the raw detector timestream to yield the ``\ac{HWPSS}-subtracted'' timestream.  
When we quote noise properties we give special attention to a narrow band of frequencies between 3.9 and 4.4 Hz.
This band is within the side-band of the 4th harmonic of the \ac{HWP} rotation which contains half of the Stokes 
$Q$ and $U$ sky synchronous signals. (The other half is contained in a symmetric side-band placed above the 4th harmonic 
and we find the noise level there to be essentially identical to the lower side-band.) 

Analysis of data collected throughout flight showed that the amplitude of the \ac{HWPSS} was sufficiently large
to introduce a non-linear detector response. (A similar effect was observed by the Polarbear 
team~\citep{takakura17}.) \ac{EP4} discusses this non-linear response and a technique to mitigate its 
effects.

%% file: readout.tex
\section{Readout}
\label{sec:detector_readout}

The \ac{TES} bolometers were voltage biased~\citep{lee_appliedoptics_1998}.
Incident optical power fluctuations modulated the current across the bolometer and across a series inductor and capacitor.
A \ac{SQUID} transimpedance pre-amplifier converted the modulated current to a voltage signal that was subsequently 
further amplified, digitized, and filtered~\citep{yoon_apl_2001}.

We used \ac{FDM} to couple several detectors to a bias and a readout line~\citep{yoon_apl_2001}.
When the \ac{EBEX} program began, power dissipation of state-of-the-art analog \ac{FDM} readout boards consumed 
approximately 5~W per readout channel, giving a total consumption of nearly 10~kW, a prohibitive level for a balloon-borne payload.
We have therefore embarked on implementing a new \ac{DfMUX} scheme that had approximately one tenth the power 
consumption~\citep{dobbs_ieee_2008}.
\ac{EBEX} was the first experiment to implement this scheme, multiplexing 8 detectors with two wires in its 2009 engineering 
flight, and the first to multiplex 16 detectors in its \ebexLDB\ flight. 

The \ac{EBEX} readout system required investment of effort along three dimensions:  development and testing of 
the new \ac{DfMUX} electronic boards, the development of automated on-board software to bias and read the \ac{TES} 
bolometers and the \ac{SQUID} pre-amplifiers, and the development and implementation of a cooling system that 
dissipated the \PinBRC~W consumed by the \ac{DfMUX} boards.
We discuss only the first element in this paper.
The software and cooling system are discussed in \ac{EP3}.   
\citet{Aubin_TESReadout2010}, \citet{aubin_thesis}, \citet{MacDermid_SPIE2014}, and \citet{MacDermid_thesis} 
give more details about the \ac{EBEX} readout system.

\subsection{Superconducting Broad-side Coupled Microstrips} 
\label{sec:microstrips}

During operation the EBEX \ac{TES}s had a resistance of approximately 1~$\Omega$. Stray impedance in series with this 
resistance reduces the detector's voltage bias and causes crosstalk. Therefore 
we kept the bias and readout wires between the \ac{SQUID}s and the \ac{LC} boards superconducting. Some sections 
of the wiring were commercially made, but for the longest sections we 
designed and constructed superconducting broad-side coupled microstrips~\citep{Dan_thesis,dobbs_revSciInst_2012}.

\begin{figure}[t]
  \centering
  \includegraphics[width=0.55\textwidth]{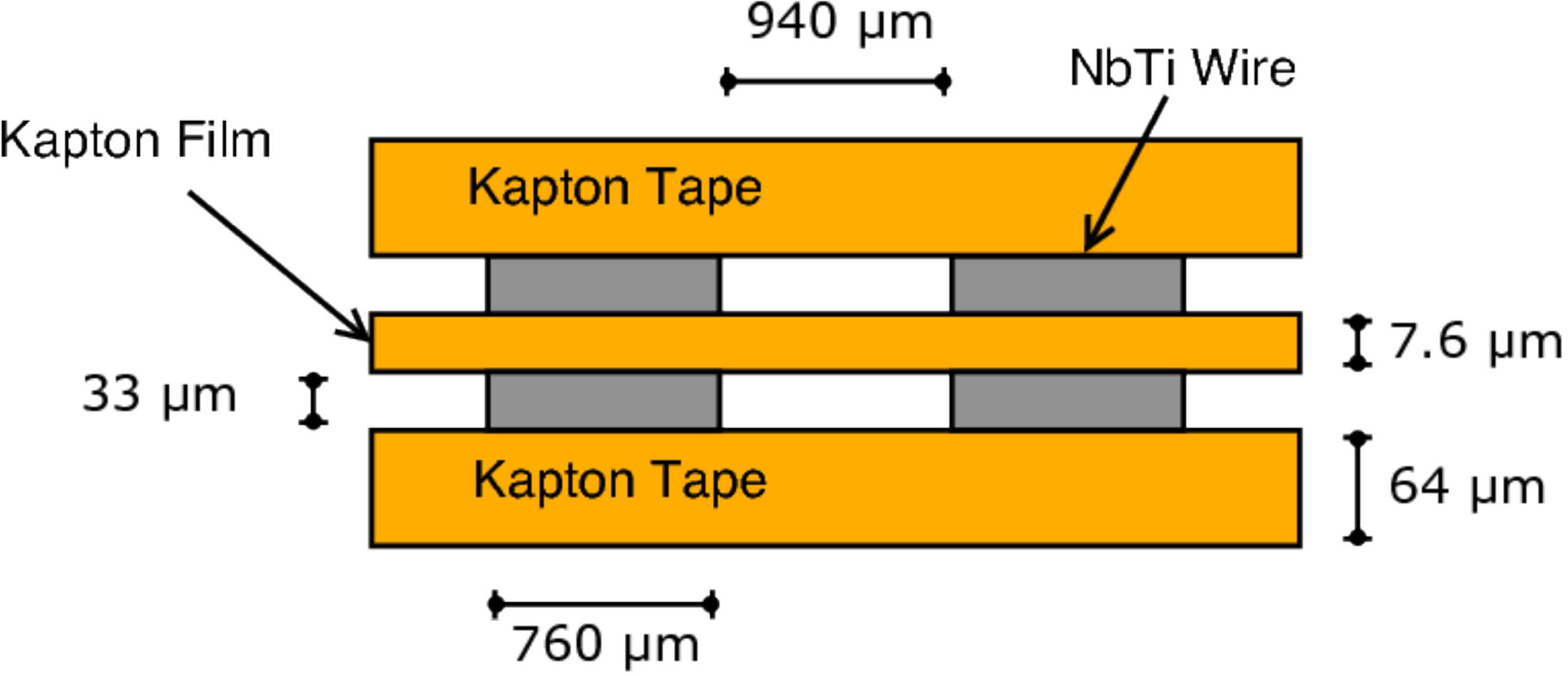}
  \includegraphics[width=0.3\textwidth]{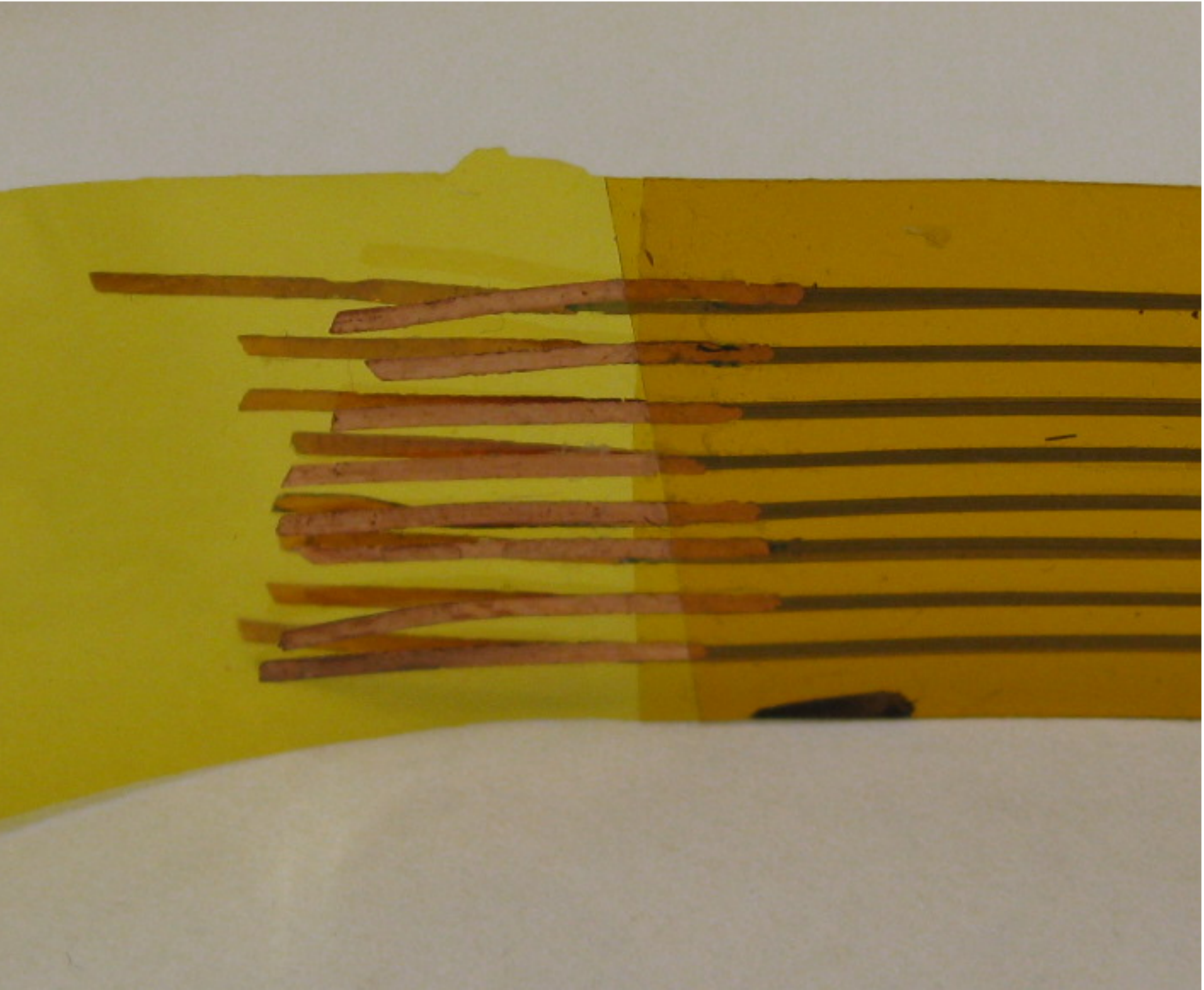}
  \includegraphics[width=0.35\textwidth]{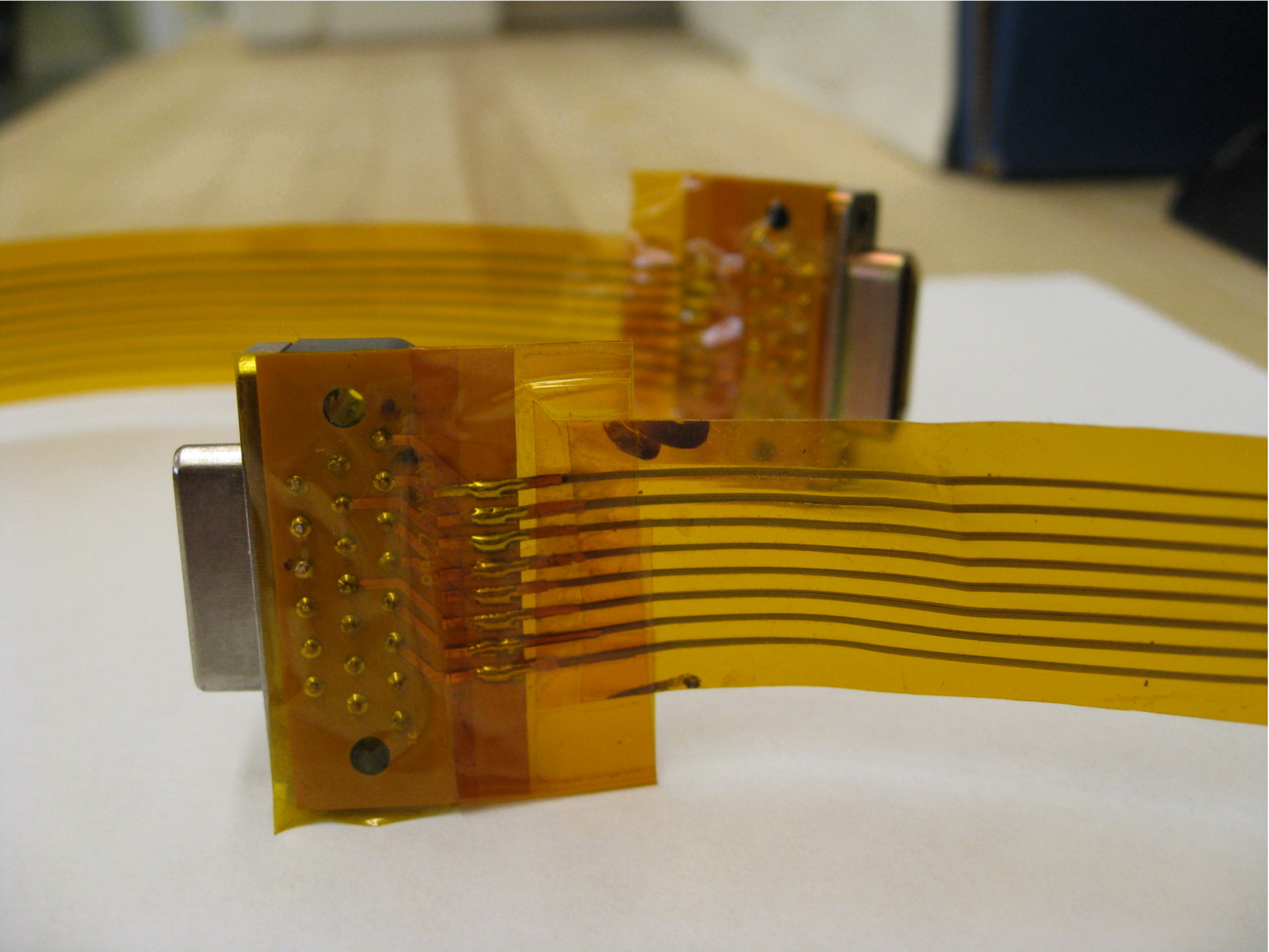}
  \caption{Upper left: diagram of the cross-section of the microstrip geometry. Two of the 8 pairs of wires are shown.
    Upper right: image of copper-clad niobium-titanium wire ends.
    Lower panel: image of the microstrip soldered to micro-D connectors.}
  \label{fig:ustrip}
\end{figure}

The microstrips consisted of interleaved layers of kapton and niobium-titanium wires. We selected niobium-titanium because 
it has a critical temperature of 9.2~K
and low thermal conductance in the superconducting state. We chose a parallel-plate waveguide geometry to minimize 
stray inductance; see Figure \ref{fig:ustrip}.
We fabricated the flat niobium-titanium wire from 191~$\mu$m diameter niobium-titanium wire which was copper-clad\footnote{Supercon, Inc.} 
to a total diameter of 305~$\mu$m and rolled flat\footnote{H. Cross Company} to thickness of about 33~$\mu$m.
We dissolved the copper-cladding from the entire length of the flattened wire except for the ends using a nitric acid wash. 
The copper-cladding remaining at the ends was used to facilitate soldering the wires.
The final cross-sectional dimensions of the wires are 760 by 33~$\mu$m.
To electrically isolate the two wires in each pair, a film of 7.6~$\mu$m thick Kapton HN was used as a spacing layer.
Adjacent pairs were spaced apart by 940~$\mu$m and were secured in their position by a layer of 64 $\mu$m thick Kapton 
HN tape with silicone adhesive.
Each microstrip had 8 pairs of wires.
We measured inductances of 24.5 and 33.3~nH for the 70 and 95~cm \ac{EBEX} microstrips, respectively.

For each microstrip the ends of the wires were soldered to small flexible circuit boards that had a micro-D connector. The 
micro-D connectors mated the microstrips to the \ac{SQUID} boards on one end, and to the \ac{LC} boards on the other. 
The microstrips were coupled to the SQUID and \ac{LC} boards with 4 connectors. We calculated an inductance 
of 9~nH per connector giving an estimated total stray inductance of 60.5 and 69~nH for the two lengths.
In \ac{EP1} we describe the `RF tower' assembly through which the microstrips passed between these two sets of boards. 
The RF tower gave heat sinking for the microstrips and was part of the RF protection of the receiver.

\subsection{Digital Frequency Domain Multiplexing}
\label{sec:dfmux}

\begin{figure}[htbp]
\begin{center}
\includegraphics[width=0.9\textwidth]{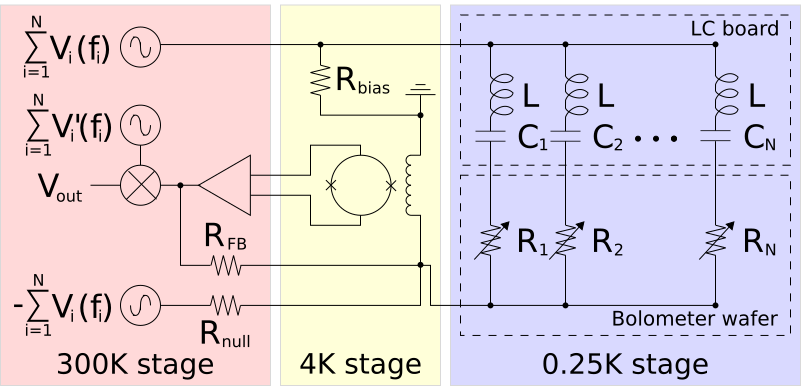}
\caption{Schematic of the frequency domain readout system (see also~\citet{aubin_thesis} and~\citet{dobbs_ieee_2008}).
Colors encode different temperature stages. For the engineering flight $N$ was 8; for the \ebexLDB\ flight $N$ was 16.
\label{fig:dfmux_schematic} }
\end{center}
\end{figure}

The \ac{EBEX} detectors were read out with the \ac{DfMUX} readout system described in~\citet{dobbs_ieee_2008}.  
A schematic diagram is shown in Figure~\ref{fig:dfmux_schematic}. For the \ebexLDB\ flight we implemented a 
multiplexing factor of 16~\citep{aubin_thesis}, the highest used at the time with this readout approach.
Ceramic capacitors $C_i$  in series with 24~$\mu$H inductors $L$ and the detectors, which had resistance $R_i$, 
defined resonant frequencies $f_i$ between 200 and 1200~kHz.
The detectors were voltage biased with a 30~m$\Omega$ resistor $R_{bias}$ and their output current was nulled 
to reduce dynamic range requirements.
The \ac{SQUID}s ~\citep{Huber2001} were operated in a shunt-feedback with a low-noise transistor op-amp and a 
feedback resistor $R_{FB}$ of 5~k$\Omega$ located on a custom \ac{SQUID} controller circuit board~\citep{lanting_thesis}.
The wires between the \ac{SQUID}s and the room temperature \ac{SQUID} controller boards were constrained to be shorter than 
33~cm~\citep{dobbs_ieee_2008}. For longer wire lengths, the phase of the signal shifted by more than 45$^o$ 
giving positive instead of negative feedback. For EBEX the length of these wires\footnote{Tekdata Interconnections Limited} was 19.5~cm. 
The amplified output bolometer currents were digitized at 25~MHz, demodulated, filtered, and decimated to 
190.73~Hz by a \ac{FPGA}\footnote{Xilinx Virtex-4 \ac{FPGA}} on each \ac{DfMUX} board.
The data were packetized by a Microblaze virtual processor which was part of the \ac{FPGA} and were streamed to the 
flight control computer over Ethernet for storage, as described in \ac{EP3}.

We operated 28 \ac{DfMUX} readout boards and 112 \ac{SQUID}s during the \ebexLDB\ flight.
Each board provided biases to and read out 4 \ac{SQUID}s and therefore 64 detectors. 
The boards were separated into four \ac{BRC} each containing either 6 or 8 boards.
In addition to the readout boards, each \ac{BRC} had a VME backplane, a clock distribution board, and an ethernet communication ring switch.
One of the \ac{BRC}s also held boards that were used to cycle the sub-kelvin refrigerators and read out receiver temperatures. 
The clock distribution board distributed the 25~MHz clock signal, as well as the commands to turn on/off 
the boards, and the commands that triggered the re-programming of the boards' firmware.
Each \ac{BRC} shell acted as a Faraday cage, shielding the readout electronics from RF signals.
Two other electronic crates each populated with DC/DCs\footnote{Interpoint Series from Crane Aerospace and Electronics}
converted unregulated 28~V to regulated 6 and 10~V~\citep{Sagiv_thesis, Sagiv_MGrossman2012}.
Each crate supplied \PperPcrate~W to two BRCs; this is \PperDfMUX~W per \ac{DfMUX} board or \PperCh~W per readout channel.
Factoring in the 82\% efficiency of the DC/DC the total power consumption of the readout system was \PoutBRC~W.

The \ac{DfMUX} on-board `algorithm manager' software listened for \ac{JSON} formatted requests over TCP/IP issued by the flight 
control computer via Ethernet communication and executed the requested Python-coded task.
The virtual processor of each \ac{DfMUX} board was programmed to perform tasks in parallel on up to 
two \ac{SQUID}s or the 32~detectors wired to them.
These tasks were stored on the \ac{DfMUX} board flash memory and included tuning the \ac{SQUID}s, configuring the 
voltage biases for the detectors, and ensuring the \ac{SQUID}s were operated within their dynamic range (see 
more details in~\citet{macdermid_aip2009}).
With this architecture, we parallelized the array tuning process, saving observing time and reducing the load on the flight control computer.
Since each \ac{DfMUX} board operated four \ac{SQUID}s and could perform up to two tasks in parallel, 
tuning time of the detector array was equivalent to the time to tune two \ac{SQUID}s and their associated 32 detectors, 
and was independent of the total number of detectors populating the focal plane.
Once the task was completed, the result was sent back to the flight control computer through Ethernet for storage. 
The interaction between the \ac{DfMUX} boards and the flight control computer is further described in \ac{EP3}.

\subsection{Readout System Noise and Stability}
\label{sec:readout_performance}

\subsubsection{Noise}
\label{sec:readoutnoise} 

The term `readout noise' refers to noise contributions from the \ac{SQUID} pre-amplifier and the electronic 
components of the readout system.
It excludes the contributions from the detectors that are described in Section~\ref{sec:noise_performance}.
The performance of `dark \ac{SQUID}s' and resistor channels
inform the readout noise level independent of detector noise contributions.
Unless otherwise noted, we refer to data that was collected during flight.
The data was converted from raw digital units (counts) to current through the \ac{SQUID} by 
applying the measured transfer function of the readout system $dI_{b}^{A}/dI_{b}^{c}$~\citep{aubin_thesis}
\begin{equation}
X~[ \mbox{A} / \sqrt{ \mbox{Hz} } ]  = 
        X~[ \mathrm{counts} / \sqrt{\mathrm{Hz} } ] \cdot dI_{b}^{A}/dI_{b}^{c} \left[ \mathrm{A} / \mathrm{count} \right] .
\label{eqn:counttocurrent}
\end{equation}
The process to generate noise predictions, which are compared to measurements, is described 
in~\citet{Aubin_TESReadout2010} and in~\citet{aubin_thesis}.
We also include a `current sharing' noise term that is not described in these publications. This noise term arises
because current driven through the \ac{SQUID} feedback 
wire is shared between the \ac{SQUID} input coil and 
the bolometer. Noise sources originating downstream of the \ac{SQUID} but inside the feedback loop will result in 
a feedback current that is enhanced by this sharing. The negative feedback will increase the current, enhancing the 
noise from these sources~\citep{silvafever2017}. Current sharing depends on the relative size of the \ac{SQUID} 
input reactance and the bolometer impedance. To calculate the effect for noise expectations, we
use a measurement  of the \ac{SQUID} input inductance for the same SQUID model implemented in a different instrument, $\sim$350~nH, and include it in our noise expectations \citep{silvafever2017}.

\begin{figure}[htbp]
\begin{center}
\includegraphics[width=0.49\textwidth]{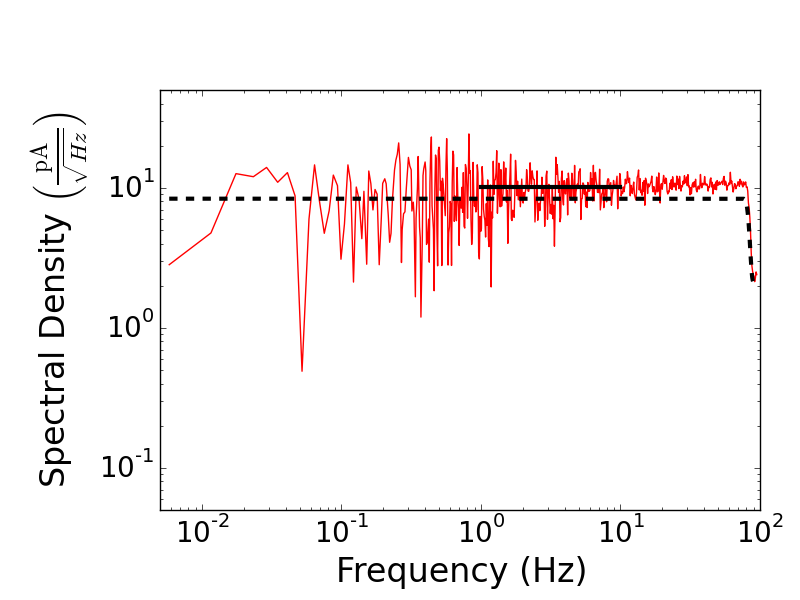}
\includegraphics[width=0.49\textwidth]{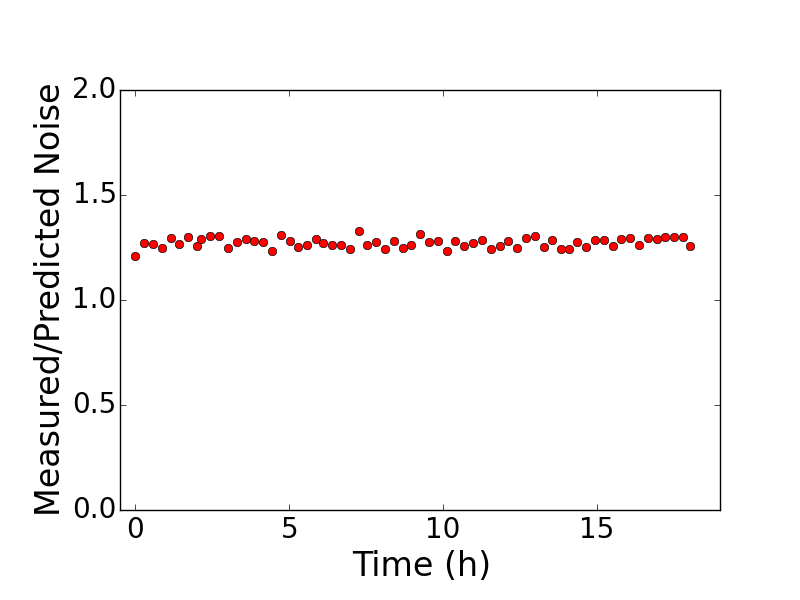}
\caption{Left: the current spectral density of a dark \ac{SQUID} channel demodulated at 495~kHz.
The average level between 1 and 10~Hz (solid red) is 10~pA/$\sqrt{\mathrm{Hz}}$. The predicted 
level (dashed line) is 8.4~pA/$\sqrt{\mathrm{Hz}}$.
Right: the ratio of measured-to-predicted noise as a function of time for the same channel.
\label{fig:squidNoiseSpectrum} }
\end{center}
\end{figure}

A dark \ac{SQUID} is a \ac{SQUID} whose input coil was not connected to detectors or to the sub-kelvin \ac{LC} boards. 
Generally, the input wires for a dark SQUID are left open (unterminated) at the 4 K stage; 
see Figure~\ref{fig:dfmux_schematic}. 
We demodulated each of the two dark \ac{SQUID}s at 16 frequencies within the 200 -- 1200~kHz range 
typically used to bias detectors. We similarly demodulated two other \ac{SQUID}s which became 
inadvertently dark when they developed an electrical open wire once cooled to 4~K.
The current spectral density through the \ac{SQUID} input coil of a typical dark \ac{SQUID} for 172~s of 
data is shown in Figure~\ref{fig:squidNoiseSpectrum}.
We select 172~s of data to compute spectral densities in this paper since we high-pass filter signals above 33~mHz for data analysis.
Median predicted noise contributions for the 64 channels are given in Table~\ref{tab:readoutNoise}. 
The ratio of average measured to predicted noise between 1 and 10~Hz 
for the SQUID channel in Figure~\ref{fig:squidNoiseSpectrum} is 1.2, which we consider to be 
broadly consistent with expectations from Table~\ref{tab:readoutNoise}, given uncertainties in the transfer functions, 
component values, and SQUID noise. We use a relatively broad bandwidth to average the noise because we 
find no evidence for the \ac{HWPSS} signal with the dark \ac{SQUID}s. 
This noise level is reasonably stable over the entire flight. We show data spanning 18 hours in 
Figure~\ref{fig:squidNoiseSpectrum}. 


\begin{table}[ht!]
\begin{center}
\begin{tabular}{| l | c |}
\hline
Noise source &  Value \\ 
&  pA/$\sqrt{\mathrm{Hz}}$  \\ \hline \hline 
\ac{SQUID} & $\sqrt{2}$~$\cdot$~3.5 \\ \hline
Demodulation chain & $\sqrt{2}$~$\cdot$~3.8 \\ \hline
Nuller chain & $\sqrt{2}$~$\cdot$~2.4 \\ \hline \hline
Total & $\sqrt{2}$~$\cdot$~5.7 \\ \hline
\end{tabular}
\end{center}
\caption{ Components of the predicted post-demodulation noise for dark \ac{SQUID}s.
SQUID noise of 3.5~pA/$\sqrt{\mathrm{Hz}}$ is typical for our devices~\citep{KentPrivate}.
The explicit factors of $\sqrt{2}$ are due to uncorrelated power on both sides of the demodulation frequency.
The nuller chain noise is dominated by resistor Johnson noise when not providing nulling currents.
\label{tab:readoutNoise} }
\end{table}%

\begin{figure}[htbp]
\begin{center}
\includegraphics[width=0.49\textwidth]{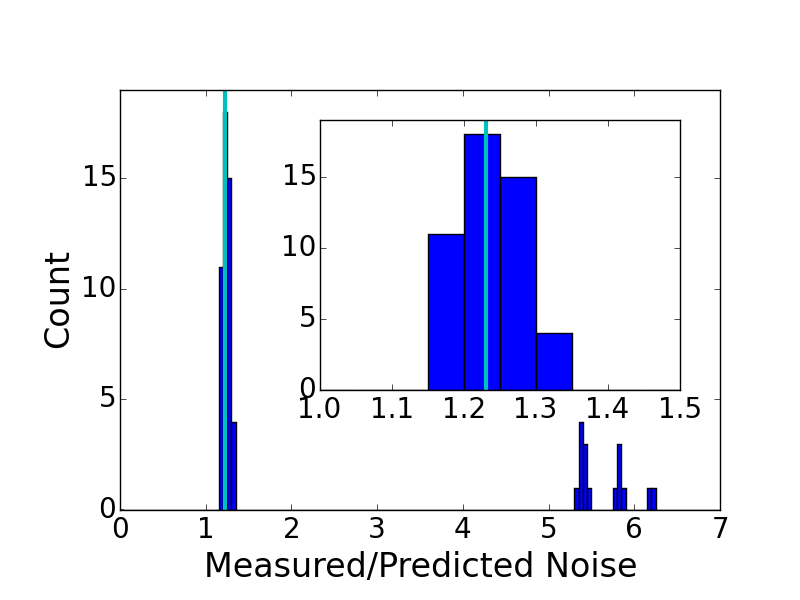}
\includegraphics[width=0.49\textwidth]{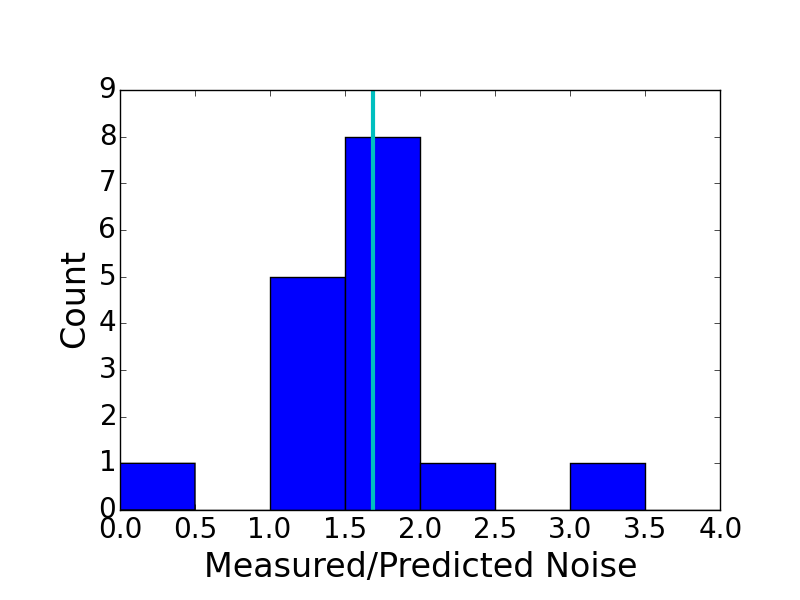}
\caption{Left: the distribution of measured-to-predicted noise ratio for the 64 dark \ac{SQUID} channels.
The median for the 48 well-behaved channels (vertical cyan) is 1.2. 
Right: the distribution of measured-to-predicted noise ratio for the 16 resistor channels.
The median (vertical cyan) is 1.7. 
\label{fig:readoutNoiseSpectrumDist} }
\end{center}
\end{figure}

To quantify the readout noise over the entire flight and for all dark \ac{SQUID}s we 
calculated power spectral densities for 172~s sections of data every 20~min throughout flight. 
We found the median measured-to-predicted ratios throughout flight for a given dark SQUID channel. 
We histogrammed all the channel medians and extracted a single median value representing the 
performance of the 64 dark \ac{SQUID} channels; see Figure~\ref{fig:readoutNoiseSpectrumDist}.
The 48 well behaved channels had a median value of 1.2 
when the noise is averaged between 1 and 10 Hz.
One dark \ac{SQUID} showed ratios larger than 4 for all of its channels and is considered an aberration.
This high-noise SQUID was connected to detector modules but exhibited an open connection in one of its two input wires. 
We speculate the other wire acts as a high impedance antenna, coupling unwanted EMI into the SQUID input resulting in high noise.

We also found a consistent median ratio of 1.2 when using measurements on the launch pad, just before flight.
Similar measurements performed in two different test cryostats with different \ac{SQUID}s gave nominal noise 
levels
%
suggesting that the 20\% higher median ratio could have a contribution from environmental pickup, differences in the 
SQUIDs, or signals coupled from elsewhere in the instrument.
A $\sim$5\% variation is measured between the lowest and highest frequency channels which is likely caused by the 
omission of stray inductance in the noise prediction. 

Resistor channels had the entire readout chain coupled to 1~$\Omega$ resistors that were mounted on the \ac{LC} boards. 
The noise for the resistor channels consists of both Johnson and readout noise terms. The total noise 
with these channels represents a cross-check on measurements of the detectors' noise when they were biased above their
superconducting transition, as discussed in Section~\ref{sec:noise_performance}. The analysis 
of the data parallels that of the dark \ac{SQUID}s. 
Figure~\ref{fig:readoutNoiseSpectrumDist} shows the distribution of the median
measured-to-predicted noise ratio for 16 well-behaved resistor channels throughout flight; 5 additional 
resistor channels had a noise ratio more than 3.5 times the median and were excluded from the analysis.
The resistor channels have a median measured-to-predicted noise ratio of 1.7. 
The resistors are mounted on the 0.25~K stage. We hypothesize that 
wires of 70 -- 95~cm long from the \ac{SQUID}s act as antennas allowing electro-magnetic pickup 
to couple to the SQUID input, producing excess noise.
Coupling such as this can produce narrow-band features in noise spectra if the EMI lies in the carrier bandwidth, 
resulting in detector-by-detector variations, or broadband white noise if the EMI is away from the carrier 
bandwidth but coupled to the SQUID. Another potential increase 
in noise is another detector, or detectors, sharing the same SQUID and thus the bias lines for 
same comb of frequencies. For example, a `latched' 
detector -- a detector operating below the superconducting temperature -- increases the noise level for all other 
readout channels on the same comb. 

After the EBEX North American engineering flight noise was measured in the laboratory
with detectors above the superconducting transition.  As we discuss in more detail in Section~\ref{sec:JRN}, 
in terms of readout-chain noise expectations, this configuration is similar to the resistor channel setup for 
the science flight. The ratio of measured to predicted 
noise was 1.0$\pm0.1$~\citep{Aubin_TESReadout2010}, which suggests that the excess resistor noise 
measured in the science flight is an environmental effect, rather than inherent to the readout chain. We revisit and discuss
this again in Section~\ref{sec:JRN}.  

We find no evidence for correlated noise between the data of any of the dark channels.
Using 18 contiguous time sections, 86~s each, we calculate the coherency~\citep{PriestlySpectralAnalysis} of 
all pairs of channels\footnote{The coherency of two time series is the ratio between their cross-spectra and the product of their auto-spectra.}.
We compare the average coherency between all pairs to the expectation for white noise, which with our data is 0.029. 
We find a coherency of 0.0253 with a dispersion of 0.0002 for dark \ac{SQUID}s and 0.035 with a dispersion of 0.009 for resistor channels.

\subsubsection{Gain Variations} 
\label{sec:readoutgain}

\begin{figure}[htbp]
\begin{center}
\includegraphics[width=0.32\textwidth]{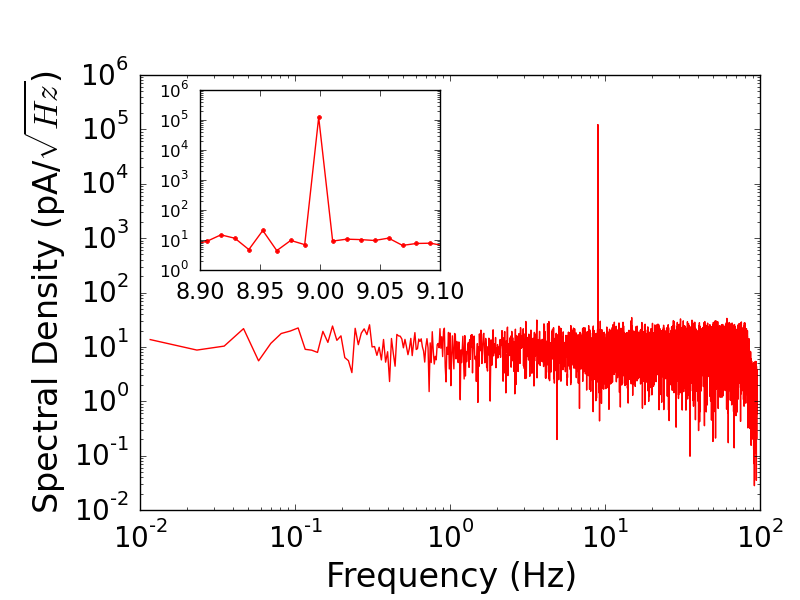}
\includegraphics[width=0.32\textwidth]{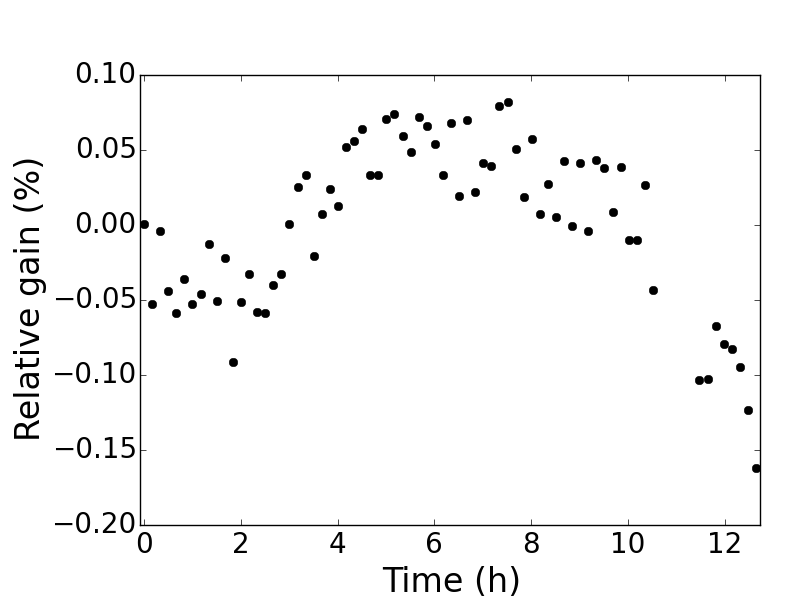}
\includegraphics[width=0.32\textwidth]{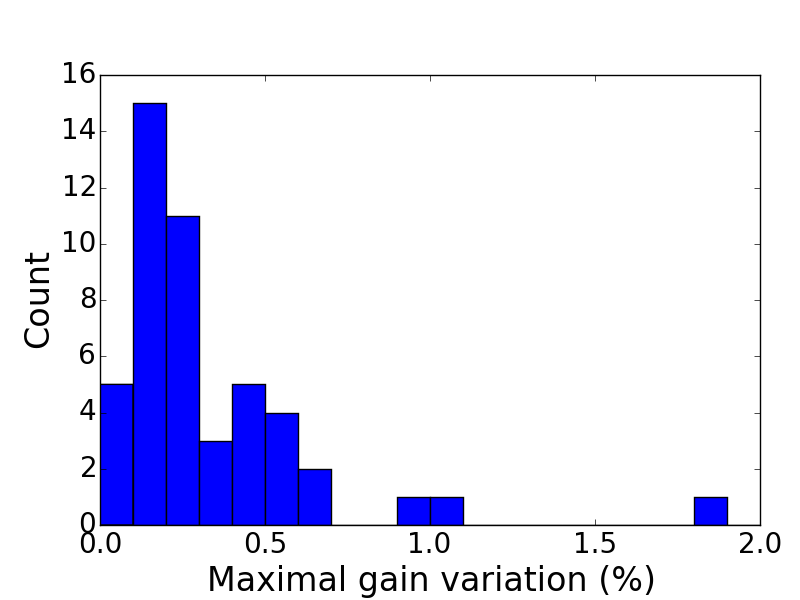}
\caption{Left: current spectral density for one gain-monitor channel. The sharp line at 9~Hz is the demodulated 
nuller current provided to the \ac{SQUID} coil at 110~kHz and demodulated at 110~kHz~+~9~Hz.
Middle: the time variation of the readout gain for one channel during 12.7 hours.
The maximal peak-to-peak gain variation is 0.25\%, 
which is negligible compared to the overall calibration uncertainty.
Right: the distribution of the maximal gain variation measured at 2 frequencies for 12 \ac{SQUID}s for multiple segments of the \ebexLDB\ flight.
\label{fig:squidGainStability} }
\end{center}
\end{figure}

We monitored gain fluctuations of the readout system by using 24 unused readout channels distributed over 
12~\ac{SQUID}s, which also read out live detector channels, a pair per \ac{SQUID}. 
For these channels, no biases were supplied through the carrier wire 
since they are operated far from any LC resonance in the cold multiplexer; see Figure~\ref{fig:dfmux_schematic}.
Instead, we injected two small constant sinusoidal currents, one at 110~kHz and the other at 1,260~kHz, 
into the input coil of the \ac{SQUID} using the nuller wire. These frequencies were 60~kHz above 
and below our standard lowest and highest bias frequencies. The amplitudes were either 30 or 100~nA.
The signals were demodulated at a frequency 9~Hz higher than the nulling current resulting in a 9~Hz 
demodulated sinusoidal current; 
see Figure~\ref{fig:squidGainStability}. These data were then processed like all other bolometer data. 

Since the injected current was constant, amplitude modulation of the demodulated 
signal is an indication of readout gain variation. 
The variation of gain for one channel over 12.7~hours of flight is shown in Figure~\ref{fig:squidGainStability}.
This particular channel experiences a gain change of 0.25\%.
The data points are from 85.9~s sections of data separated by 10~minutes. A histogram for all gain monitoring channels 
gives a stability that is mostly better than 1\%. 
Over the same period of time temperature sensors on the readout boards indicate board temperature changes 
of up to 20$^\circ$C. 
We measure in the laboratory that a similar temperature change of one readout board 
causes its gain to change by up to 0.3\%. We conclude that gain variations of the readout system are negligible 
compared to the overall calibration uncertainty, which is $\gtrsim$10\%.

%% file: detectors.tex
\section{Detectors}
\label{sec:detectors}

\ac{EBEX} uses spider-web \ac{TES} bolometers.
We optimized the parameters of the bolometers for cosmological observation from a balloon platform. 
In this section we describe the design characteristics of the \ac{EBEX} bolometers 
and report on measurements of their normal resistance, thermal conductance, critical temperature, time constant, and optical efficiency.
We use these measurements in combination with the measured absorbed power by the detectors to predict the noise performance of the detectors.
We finally compare these predictions to data to assess their in-flight performance.


%% file: detector_optimization.tex

\subsection{Detector Design}
\label{sec:detector_optimization}

\begin{figure}[ht!]
  \centering
  \includegraphics[width=0.95\textwidth]{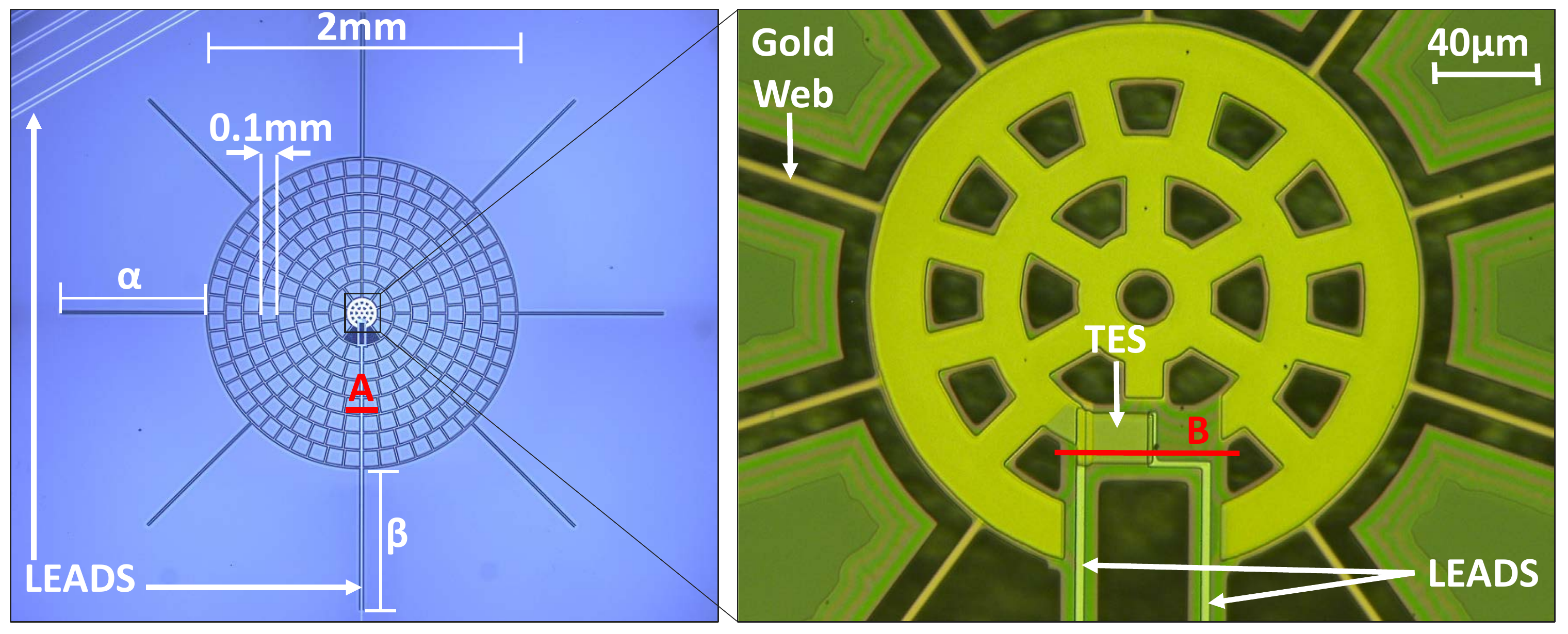}
  \caption{Left: photograph of an \ac{EBEX} bolometer.  Right: an enlargement of its central region.  
    The leg with length $\beta$ was 17~$\mu$m wide to accommodate the niobium leads.
    The other 7 legs (with length $\alpha$) were 6~$\mu$m wide.
    Table~\ref{tab:Design_Params} gives the dimensions of the parameters $\alpha$ and $\beta$.  
    Red lines labeled ``A"  and ``B" indicate cross-sections which are shown in Figure~\ref{fig:EBEX_Fab_Stack}. 
    \label{fig:Bolometer_Overview} }
\end{figure}

The detectors had spider-web absorber architecture similar to the \ac{TES}s used for the 
APEX-SZ and SPT-SZ experiments~\citep{Lee_1996, Chang_2009, Schwan_2010, Westbrook_2012, Westbrook_thesis}. 
The design of a single \ac{TES}, shown in Figure~\ref{fig:Bolometer_Overview}, consisted of a low-stress silicon nitride, 
gold-metalized web that served as the absorber for millimeter-wave radiation. The web was suspended above the silicon 
wafer by 8 legs, which provided thermal isolation, and were used to tune the thermal conductance of the bolometers. 
A \ac{TES} made of aluminum-titanium superconducting proximity bilayer in the middle of the web was 
operated at its transition temperature, typically between 0.4 and 0.5~K. Two niobium leads delivered 
constant voltage bias and monitored current fluctuations.

Table~\ref{tab:Design_Params} gives the bolometer design parameters for the 
three \ac{EBEX} frequency bands, including the \ac{TES} electrical resistance in the normal state ($R_{n}$), \ac{TES} transition 
temperature ($T_{c}$), average thermal conductance ($\overline{G}$), bolometer geometry parameters $\alpha$ and $\beta$ 
described in Figure~\ref{fig:Bolometer_Overview},  the expected heat capacity ($C$) of the bolometers, and 
the intrinsic optical time constant ($\tau_0$). 
The choice of some of the design parameters, most notably the thermal conductance, reflects the unique demands of a 
balloon flight; the in-flight optical loading is not a-priori known, and once the instrument is conducting observations there is 
no opportunity for further detector optimization. Therefore the thermal conductance must have sufficient margin, that is, must
be large enough, to ensure the detectors
can operate over a relatively broad range of optical loadings.  A consequence is that with nominal optical load a 
larger detector voltage bias is required, which reduces the responsivity and amplifies readout noise equivalent power.

\begin{figure}[ht!]
  \centering
  \includegraphics[width=0.95\textwidth]{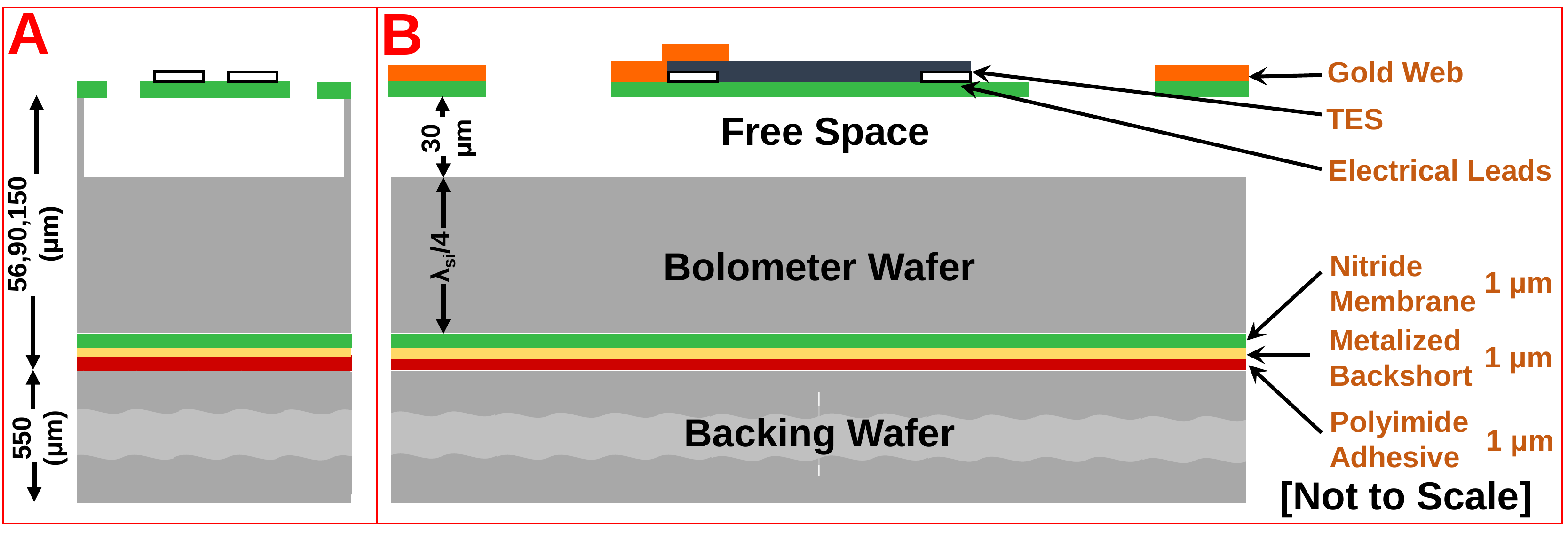}
  \caption{Cross-sections through the lines ``A" and ``B" from Figure \ref{fig:Bolometer_Overview} showing the 
    layers of the \ac{EBEX} detectors 
    through the thermal isolation leg carrying the electrical leads (A) and the central portion of the bolometer which 
    carries the \ac{TES} (B).  The thickness of bolometer wafer depended on frequency band; see text. The bolometer 
    wafer was bonded to a thicker backing wafer to provide mechanical stability during fabrication.  
    A metallic layer deposited on the backside of the detector wafer provided a backshort for increased absorption.  An aluminum-titanium
    bilayer forms the \ac{TES} of the bolometer at the center of the pixel. 
    \label{fig:EBEX_Fab_Stack}
    \label{subfig:Fab_Stack}}
\end{figure}

\begin{table}[ht!]
\centering
\begin{tabular}{| c | c c | c c | c c |}\hline
\multicolumn{1}{|c}{Band (GHz)}   &  \multicolumn{2}{|c}{150}   & \multicolumn{2}{|c}{250}   & \multicolumn{2}{|c|}{410}  \\
                                     & Design & Measured & Design & Measured & Design & Measured  \\ \hline
$R_{n}$ ($\Omega$)            & 1.5  & 1.9  & 1.5  & 1.5  & 1.5  & 1.4  \\
$T_{c}$ (K) $^\ddagger$       & 0.44        & 0.45 & 0.44  & 0.48 & 0.44 & 0.47  \\ 
$\overline{G}$ (pW/K)       &  19   & 39 & 45   & 54 & 63   & 63  \\
$\tau_{0}$ (ms)                 & 17 & 88$^\dagger$  & 13  &  46$^\dagger$  &  10 &  57$^\dagger$  \\
$C$ (pJ/K)*                         &  0.5  & 3.8$^\dagger$ &  0.9  & 3.3$^\dagger$  & 1.0   & 8.4$^\dagger$  \\ \hline 
Wafer thickness ($\mu$m)   & \multicolumn{2}{|c|}{150}  & \multicolumn{2}{|c|}{90}  & \multicolumn{2}{|c|}{56}  \\
$\alpha$ (mm)                &  \multicolumn{2}{|c|}{1.05}   & \multicolumn{2}{|c|}{1.0}   &  \multicolumn{2}{|c|}{0.5}  \\
$\beta$ (mm)                & \multicolumn{2}{|c|}{1.45} & \multicolumn{2}{|c|}{1.0}  & \multicolumn{2}{|c|}{0.5}  \\ \hline
\multicolumn{7}{l}{\footnotesize$^\ddagger$ Design values calculated early in the program using thermal conductivity power index $n=3$ and}\\
\multicolumn{7}{l}{\footnotesize a bath temperature of 260~mK.}\\
\multicolumn{7}{l}{\footnotesize$^\dagger$ Median of measurements on a single wafer at each frequency; see Section~\ref{sec:time_constants}.}\\
\multicolumn{7}{l}{\footnotesize* Calculated from time constant and thermal conductance.}\\
\end{tabular}
\caption{Designed and measured detector parameters for each of the frequency
bands.  The values in the `measured' columns are median values for 
all detectors on wafers used for flight.  Description of the measurements, histograms, and further discussion 
of the measured values are given in Section \ref{sec:detector_characterization}.
For the parameters $\alpha$ and $\beta$ shown in Figure~\ref{fig:Bolometer_Overview} we give the design values. 
The lithography was generally accurate to within 0.5~$\mu$m.
\label{tab:Design_Params} }
\end{table}

\subsubsection{TES, Absorber, and Backshort}
\label{sec:tes}

The \ac{TES}s were constructed from a 110~nm titanium layer deposited on top of a 40~nm 
aluminum layer in a single vacuum step.
We chose a transition temperature of 0.44~K to minimize phonon noise 
given the expected 0.25~K focal plane temperature~\citep{Suzuki_thesis}.  The normal resistance for all bands was
1.5~$\Omega$ to ensure that the detector operated in the stable regime~\citep{dobbs_revSciInst_2012}. 

A metalized spider-web structure appears as a continuous sheet of metal with resistivity $\rho_{web}$ to electromagnetic radiation 
of wavelength $\lambda$ incident upon it as long as the grid spacing $g$ satisfies $g \ll \lambda$~\citep{BockNovelSpiderWebs}.  
In addition, the return loss remains below -10~dB 
when $\rho_{web}$ is between 150 and 700~$\Omega$/sq~\citep{glenn_appliedoptics_2002}.  The \ac{EBEX} web grid spacing was 
100~$\mu$m, 7 times smaller than the shortest wavelength admitted, and we used 200~$\mathrm{\AA}$ thick layer of gold to 
achieve $\rho_{web}$~=~$350~\Omega$/sq. 

The absorption efficiency of the detector depends on the distance of the backshort from the absorber. 
Within a single uniform silicon medium the optimal distance is $\lambda_{c}$/4, 
where $\lambda_{c}$ is the wavelength in solid silicon behind the web absorber;  
$\lambda_{c}$~=~$\lambda_{0}/n_{Si}$, where $\lambda_{0}$ is the wavelength in 
empty space and $n_{Si}$~=~3.4 is the index of refraction of silicon~\citep{LambMaterials, SpiderWebFeedHorns}. 
To achieve these backshort thicknesses we 
etched the detector wafers to thicknesses of 150, 90, and 56~$\mu$m for the 150, 250, and 410~GHz bands,  
respectively. However, these wafer thicknesses were too thin to survive standard micro-fabrication techniques and we 
thus bonded the bolometer wafer to a thicker backing wafer to provide additional mechanical 
stability~\citep{Westbrook_2012, Westbrook_thesis}; see Figure~\ref{fig:EBEX_Fab_Stack}.

When we include in the absorption simulation the 30~$\mu$m cavity behind the web absorber, which makes
the backshort a combination of two media, 
we find that the expected absorption efficiency is 35\% at 410 GHz, and 75\% for the other two bands.

\subsubsection{Thermal Conductance}
\label{sec:thermal_isolation}

Table~\ref{tab:Design_Params} gives our specifications for the average thermal conductance for 
each of the frequency bands. We arrived at these specifications by conservatively requiring that 
the bolometer handle total absorbed powers of up to 4, 9 and 12~pW for each of the 150, 250, 
and 410~GHz bands, respectively, before the TES exits the superconducting transition. 
We set these requirements early in the development of instrument by calculating 
the expected optical loads from both sky and instrument sources and by applying a safety 
factor to account for unexpected loads. 
We find that our conservative specifications are met when comparing to the measurement of the in-flight total optical load described in Section~\ref{sec:optical_load}.

Of the two contributions to the thermal conductance of the bolometer, (i) the silicon nitride legs that connected
the web to the wafer, and (ii) the two niobium electrical leads, the thermal conductance of the niobium leads was 
negligible~\citep{ThermalConductivityNb,  Westbrook_thesis}. Therefore, we achieved different thermal 
conductances for each frequency band through changes 
in the geometry of the silicon nitride legs. The geometry of the legs and the calculated average thermal 
conductances are given in Figure~\ref{fig:Bolometer_Overview} and in Table~\ref{tab:Design_Params}.

\subsubsection{Intrinsic Time Constant}
\label{sec:optical_time_constant}

The intrinsic optical time constant of the bolometer $\tau_{0}$ is set by the ratio of the heat capacity $C$ at the 
operating temperature to the differential thermal conductance $G = dP/dT$: $\tau_{0} = C/G$. 
The predicted heat capacity of the 
bolometer was dominated by the contributions of the TES and the gold absorber layer, which gave 80\% and 20\%
of the total, respectively. The calculated 
contributions of the silicon nitride web and niobium leads were calculated to be 
negligible~\citep{Westbrook_thesis, VanSciverHeliumCryogenics}. Since the TES and gold layer were identical 
among the frequency bands, changes in the time constant with frequency band were a result of differences in 
thermal conductance.  Table~\ref{tab:Design_Params} gives the design values of $\tau_{0}$ for each of the 
frequency bands. The table quotes the average thermal conductance $\overline{G}$, which is related to the 
differential value through
\begin{equation}
G = (n+1) \left( \frac{1 - \left(\frac{T_0}{T_{c}}\right)} {1 - \left(\frac{T_0}{T_{c}}\right)^{n+1} } \right) \overline{G},
\label{eqn:G_dyn}
\end{equation}
where $T_0$ is the bolometer bath temperature and following 
conventions we assumed power law dependence for the differential thermal 
conductance $G = g T_{c}^{n}$. 
We used $n$~=~2 since thermal conductivity measurements of engineering flight wafers gave 
values of 2.2~$\pm$~0.3, 1.9~$\pm$~0.2, and 2.1~$\pm$~0.2 for the 150, 250, and 410 GHz bands, 
respectively~\citep{hubmayr_thesis, Aubin_TESReadout2010}.


%% file: detector_characterization.tex
\subsection{Detector Characterization}
\label{sec:detector_characterization}

We characterized the properties of all 14 detector wafers that were included in the \ebexLDB\ flight. 
Characterization included testing for electrical continuity of each detector
before a wafer was wire-bonded with an \ac{LC} board and a standard set of measurements at cryogenic temperatures 
in dark conditions. In dark conditions, the wafer was mounted inside a sealed box cooled to 0.32~K and thus was 
subjected to an optical power load of $\sim$9~fW, which is two orders of magnitude smaller than expected in-flight conditions.  
	
There were 140 bolometers fabricated on a single \ac{EBEX} wafer, of which we could read out 124 
with the edge, and 127 with the central, \ac{LC} boards. This gave a total of 1960 bolometers from which 1742 could be read out, 
as listed in the first two lines of Table~\ref{yield_table}. We now enumerate further reductions in detector yield. 
At room temperature, each bolometer was inspected visually under a microscope and also probed across its bond pads for 
electrical continuity (line 3). 
To monitor noise, some electronic channels read out resistors located on the \ac{LC} boards (two per wafer),
four combs were attached to dark \ac{SQUID}s that did not have bolometers, and one \ac{SQUID} had an electronic malfunction (line 4). 
Some bolometers which had electrical continuity at room temperature 
did not have continuity during the cold (0.8~K) network analysis (line 5). 
Of the 112 \ac{SQUID}s 5 failed to tune (line 6). Those bolometers whose operation drastically worsened the 
noise performance of their entire comb were disconnected (line 7).  
The number of bolometers that had successful I-V curves during first tuning at float is given in line 8. 

\begin{table}[ht!]
\begin{center}
\begin{tabular}{| c | c | c | c | c | c |}
\hline
 &   & \multicolumn{3}{c |}{ {\bf Frequency (GHz)} } & \\
 &  {\bf Count of} & {\bf 150 } & {\bf 250 } & {\bf 410 } & {\bf Total}  \\
\hline 
1 & detectors on wafers & 1120 & 560 & 280 & 1960 \\
\hline 
2 & maximum detectors to read out & 992 & 496 & 254 & 1742 \\
\hline 
3 & detectors which passed warm electrical \& visual inspections & 908 & 455 & 232 & 1595 \\
\hline 
4 & channels wired to detectors which &  &  &  &  \\
 & passed warm electrical \& visual inspections & 861 & 447 & 213 & 1521 \\
\hline 
5 & detectors which passed the 0.8~K network analysis test & 805 & 430 & 187 & 1422 \\
\hline 
6 & detectors after SQUID failures removed & 773 & 414 & 155 & 1342 \\
\hline 
7 & detectors after noise polluters removed & 676 & 371 & 133 & 1180 \\
\hline 
8 & detectors with successful flight IV curves & 504 & 342 & 109 & 955 \\
\hline
\end{tabular}
\end{center}
\caption{Detector yield tally by frequency band. 
\label{yield_table} }
\end{table}%

\subsubsection{Resonant Frequencies}

\begin{figure}[htbp]
\begin{center}
\includegraphics[width=0.49\textwidth]{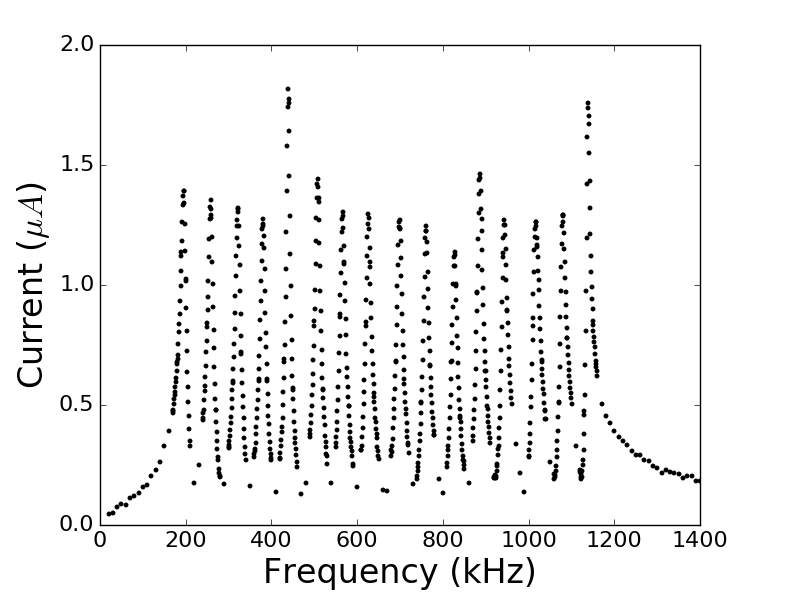}
\includegraphics[width=0.49\textwidth]{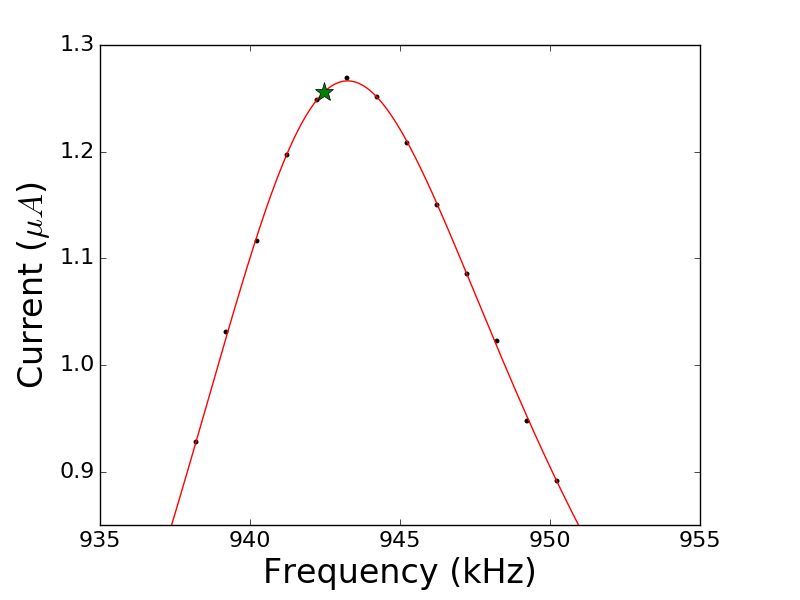}
\caption{Left: current response of a multiplexed detector circuit as a function of bias frequency.
Right: zoom on one peak (black dots) with the fitted response (red line) and the optimal bias frequency (green star) minimizing crosstalk~\citep{aubin_thesis}.
\label{fig:network_analysis} }
\end{center}
\end{figure}

We determined the resonant frequencies for each comb of frequencies by maintaining the detectors at 0.8~K, that is, above their transition temperature, and sweeping the bias frequencies 
between 0.1 and 1.3 MHz. The RLC circuit resonated at its characteristic 
frequency such that the entire sweep gave a response as shown in Figure~\ref{fig:network_analysis}. 
A fit of the peak of the resonance with a model of the circuit that accounts for current leakage to all resonators 
gave the resonant frequencies and the normal resistances of the 
TESs in the comb~\citep{aubin_thesis}.
Missing peaks in this measurement corresponded to opens in electrical lines.

\begin{figure}[ht!]
\centering
\includegraphics[width=0.31\textwidth]{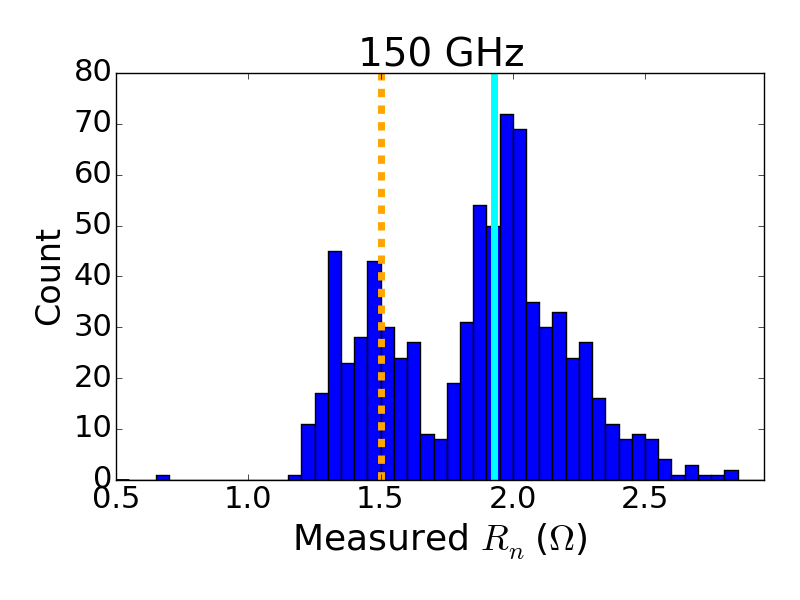}
\includegraphics[width=0.31\textwidth]{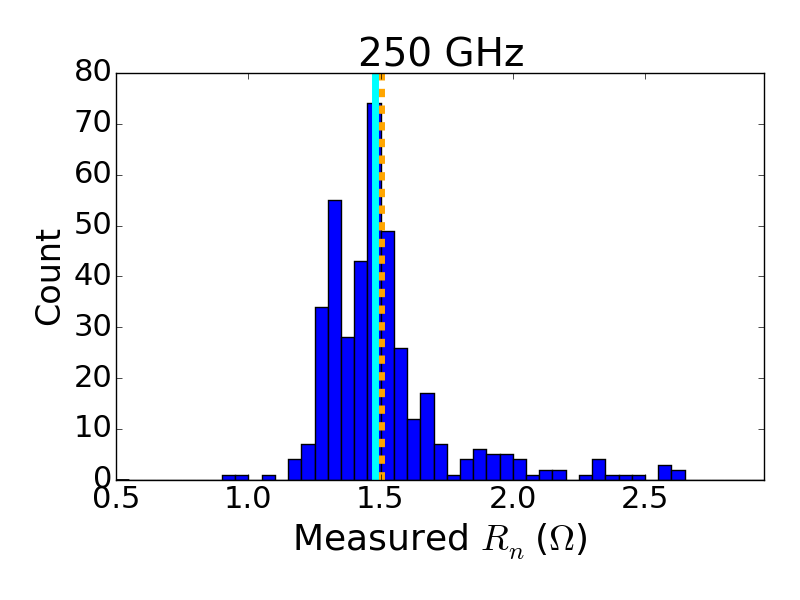}
\includegraphics[width=0.31\textwidth]{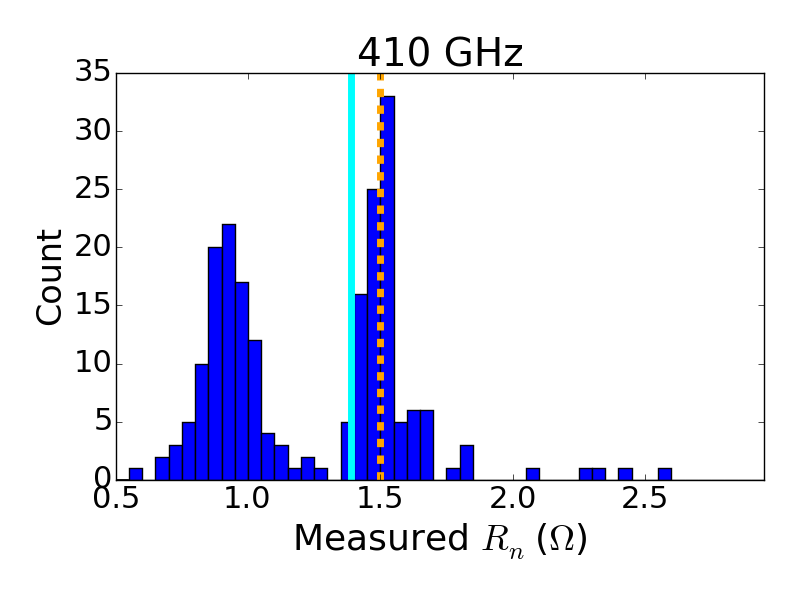}
\caption{Histogram of measured normal resistances $R_{n}$ for each of the frequency bands, including the median (vertical cyan)  
and design (vertical gold dashed) values.
\label{fig:rn_histograms} }
\end{figure}

Figure~\ref{fig:rn_histograms} gives histograms of the measured normal resistance $R_{n}$ values for each frequency band. 
The 150 and 410~GHz bimodal distributions were due to detector parameters being closely grouped within a single 
fabrication run, but varying between fabrication runs. The measured value of $R_{n}$ for the 250~GHz band closely matched 
the design (see Table~\ref{tab:Design_Params}). 
For the other two frequency bands, one mode of the distribution closely matched while the other mode was higher 
(lower) than design for the 150 (410)~GHz band. 
The 150~GHz detectors with a measured $R_{n}$ of 1.9~$\Omega$, instead of the nominal 1.5~$\Omega$, were calculated 
to have increased electrical cross-talk from a value of 0.5\% to 0.8\% and decreased 
loop-gain by 30\%. 
The 410~GHz detectors with a measured $R_{n}$ of 1.4~$\Omega$ were calculated 
to have increased Johnson noise by 4\% relative to the nominal expected value of 4.0 pA~$/\sqrt{\mathrm{Hz}}$.

\subsubsection{Transition Temperature}

\begin{figure}[ht!]
\centering
\includegraphics[width=0.31\textwidth]{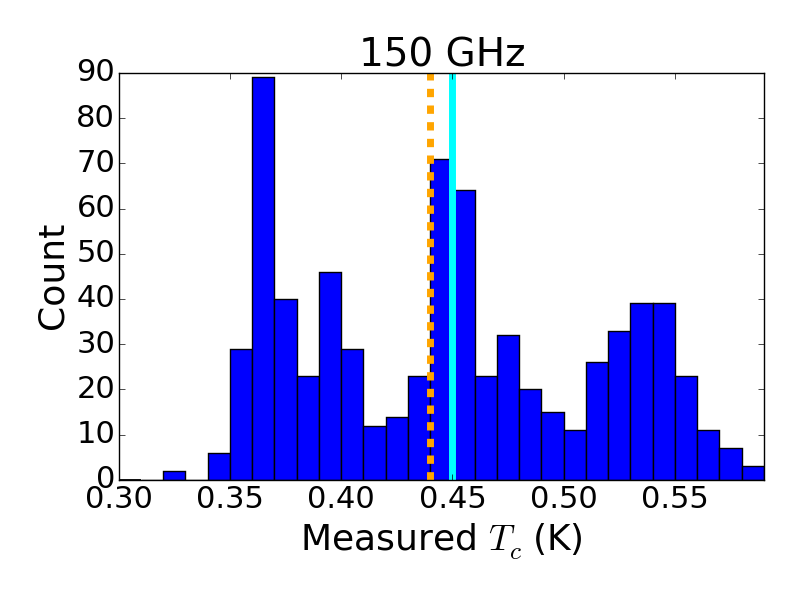}
\includegraphics[width=0.31\textwidth]{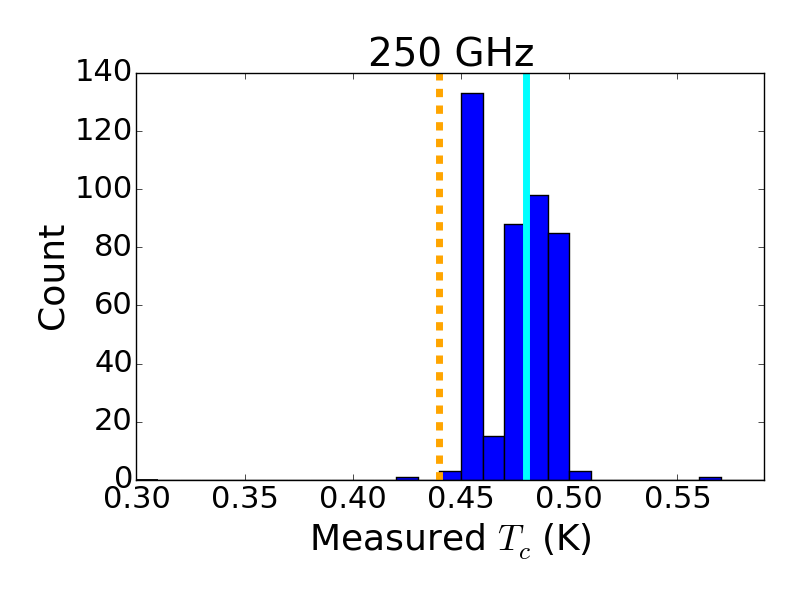}
\includegraphics[width=0.31\textwidth]{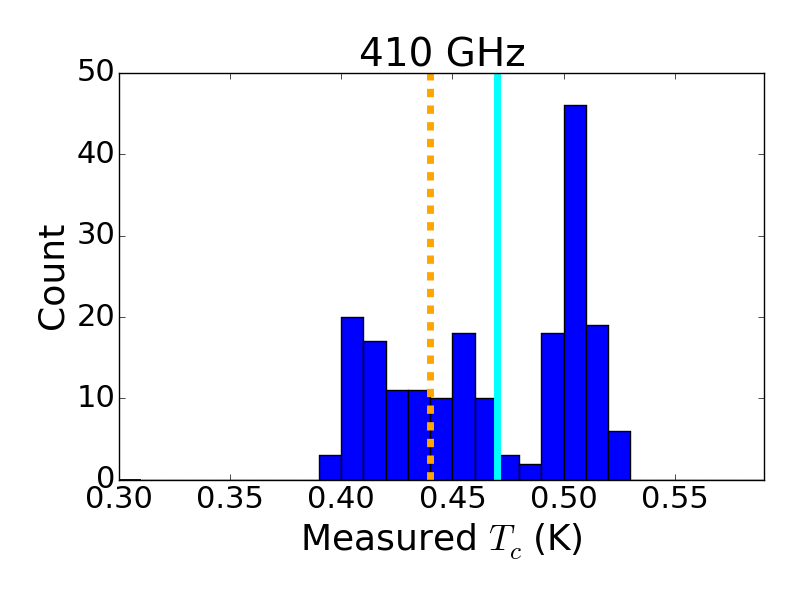}
\caption{Histogram of measured critical temperature values for the detectors in each frequency band including 
the median (vertical cyan) and design (vertical gold dashed) values. 
\label{fig:tc_histograms} }
\end{figure}

For the measurement of the critical temperature $T_{c}$ we biased the detectors with 5~nV such that the Joule heating of 1.5~fW was 
small and the bath temperature was a good proxy for the bolometer's temperature. For the same reason, the measurement was 
done in dark conditions. We slowly changed the detectors' temperature while monitoring the current through the TES. 
At the critical temperature the current showed a steep transition.
Figure~\ref{fig:tc_histograms} shows histograms summarizing the measured critical temperatures. There 
was a $\sim$20\% wafer-to-wafer spread in the measured $T_{c}$ with the medians for the three bands within 
11\% of the target value of 0.44~K. 
The 150~GHz detectors have the widest spread of measured $T_{c}$. The effect of this spread 
is twofold: (1) it increases Johnson (phonon) 
noise when $T_{c}$ is above (below) the design value, and (2) it increases (decreases) the detector 
saturation power when the $T_{c}$ is above (below) the design value.
At the high (low) edge of the distribution with $T_{c}$~=~0.59~K (0.34~K), there was an 16\% (14\%) increase (decrease) 
in the calculated Johnson noise relative to the nominal expected value of 4.0~pA/$\sqrt{\mathrm{Hz}}$ and an 80\% (68\%) 
increase (decrease) in the calculated phonon noise relative to the nominal expected value of 13~aW/$\sqrt{\mathrm{Hz}}$.
On the low edge of the distribution, the critical temperature causes the 150~GHz detectors to saturate, leaves the responsivity unchanged for the 250~GHz bolometers, and increases the responsivity by 66\% for the 410~GHz bolometers.
The responsivity is decreased by 110, 56 and 60\% for the 150, 250 and 410~GHz bolometers, respectively, on the 
high edge of the distribution, increasing the readout and Johnson noise relative to phonon and photon noise.

\subsubsection{Average Thermal Conductance}
\label{sec:thermalconductance}

We determine the average thermal conductance of the bolometers using the relation
\begin{equation}
 \overline{G}(T_{0}) = P_{sat}(T_0)/ (T_{c} - T_{0}),
\label{eqn:gbar}
\end{equation}
The `saturation power' $P_{sat}$ is the power necessary to operate the TES in the regime 
of strong electrothermal feedback in which the total power absorbed is constant.
This power depends on the bath temperature
\begin{equation}
P_{sat}(T_{0}) = P_{e} (T_{0}) + P_{abs}. 
\label{eqn:boloPowerFlow}
\end{equation}
Here $P_{e}$ is the electrical power absorbed in Joule heating and $P_{abs}$ is the radiative power absorbed.
In dark conditions we assume that $P_{abs}$~=~0, and so $P_{sat}(T_{0})$~=~$P_{e,d}(T_{0})$ is 
therefore a measurable quantity; we added the subscript $_{d}$ to $P_{e,d}$ to highlight that this is the electrical 
power measured in dark conditions. 

\begin{figure}[ht!]
\centering
\includegraphics[width=0.31\columnwidth]{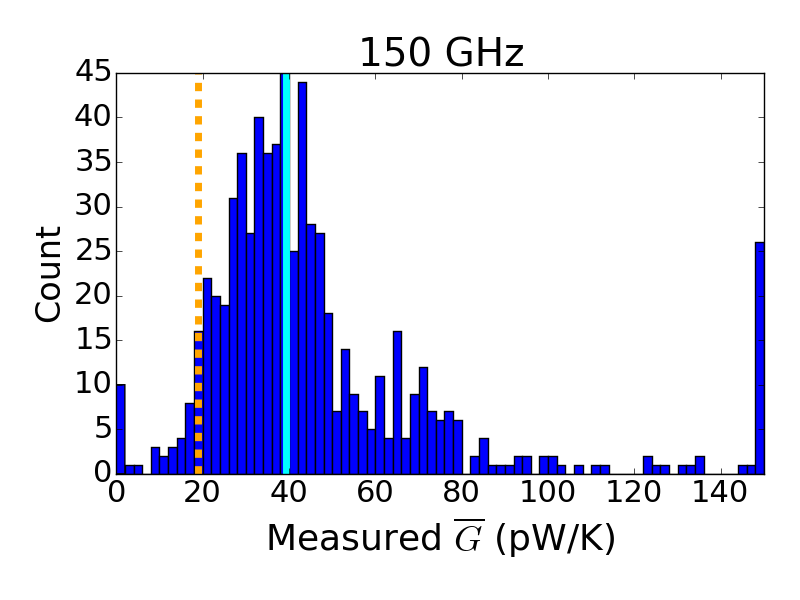}
\includegraphics[width=0.31\columnwidth]{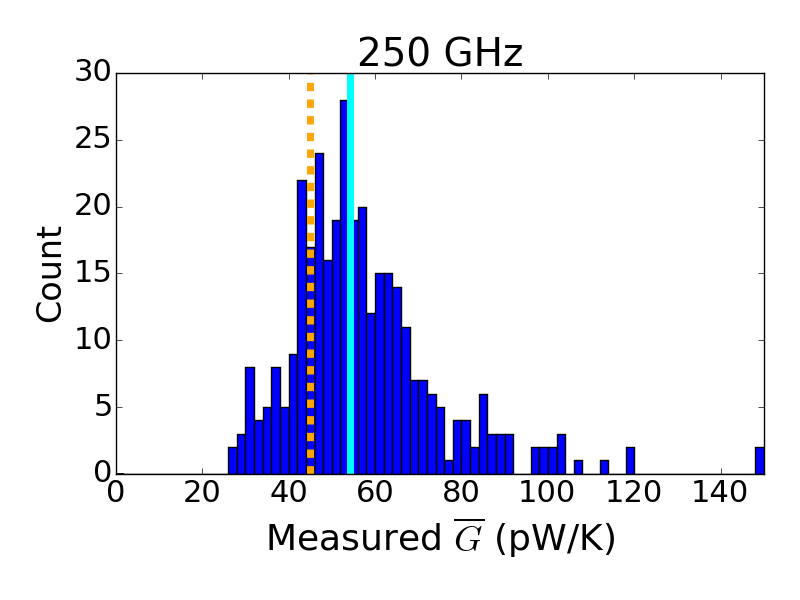}
\includegraphics[width=0.31\columnwidth]{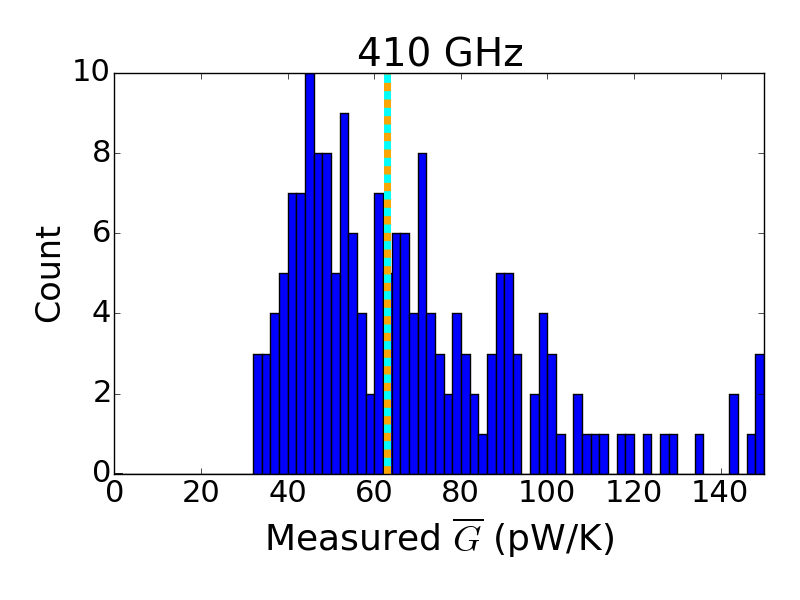}
\caption{Histograms of the measured average thermal conductance values for the three frequency bands including the 
median (vertical cyan) and design (vertical gold dashed) values. 
We piled measurements of  $\overline{G}$ exceeding 150~pW/K into the last histogram bin.
}
\label{fig:G_Histograms} 
\end{figure}

Histograms summarizing the measured average thermal conductance values for the three \ac{EBEX} frequency bands are shown in 
Figure~\ref{fig:G_Histograms}. The values of  $\overline{G}$ are given for the \ac{EBEX} bath temperature $T_{0}$~=~0.25~K.
The measurements were conducted in three different cryostats, each operating at 
a different bath temperature $T_{0'}$. We corrected the measured values $\overline{G}(T_{0'})$ to $\overline{G}(T_{0})$ using 
the scaling
\begin{equation}
P_{sat}(T_0) = \left( \frac{T^{n+1}-T_0^{n+1}}{T^{n+1}-T_{0'}^{n+1}} \right) P_{sat}(T_{0'}),
\label{eqn:ScalePsat}
\end{equation}
which assumes that the differential thermal conductance follows a power law $ G \propto T^{n}$; see Section~\ref{sec:optical_time_constant}. 
The design and median of measured values for the average thermal conductances are given in Table~\ref{tab:Design_Params}. 
There is appreciable spread in the measured values. This spread is 
a consequence of variance between wafers and is also apparent in the measurements of $T_{c}$. 
Higher thermal conductance increases phonon noise. For example, for a median 250~GHz detector, 
phonon noise increased by 
10\% relative to the nominal value of 20~aW/$\sqrt{\mathrm{Hz}}$.
Higher thermal conductance also increases the required electrical power, decreasing responsivity, which scales as 1/$V_{bias}$. 
With a lower responsivity, noise sources that produce fixed current such as readout noise, are referred to 
higher noise equivalent power values. In this manner, higher thermal conductance effectively amplifies readout noise. 

%% file: time_constants.tex
\subsubsection{Detector Time Constants}
\label{sec:time_constants}

The effective time constant of the bolometer is given by 
\begin{equation}
 \tau_{eff} = \frac{\tau_0}{1+\mathcal{L}}, 
\label{eqn:tau_eff}
\end{equation}
where $\mathcal{L}$ is the loop-gain of the bolometer~\citep{lee_appliedoptics_1998}.

During the course of the EBEX project we measured time constants in various configurations including a flashing \ac{LED} as part of `dark' measurements, chopping between two optical 
loads~\citep{hubmayr_thesis, Dan_thesis}, using the phase and amplitude methods during polarization 
calibration, as described in \ac{EP1} and in~\citet{Klein_thesis}, and assessing the shapes of cosmic-ray events during flight. 
Here we report the results from measurements conducted post-flight on three wafers. In contrast with previous measurements, 
the post-flight ones were all done in identical conditions, using the same method, and have good detector statistics. 
Each of the three wafers had detectors fabricated for a different frequency band. For the 150, 250, and 410~GHz band 
wafers there were 88, 97, and 88 active bolometers, respectively. 
\begin{figure}[ht]
\center
\includegraphics[width=0.48\textwidth]{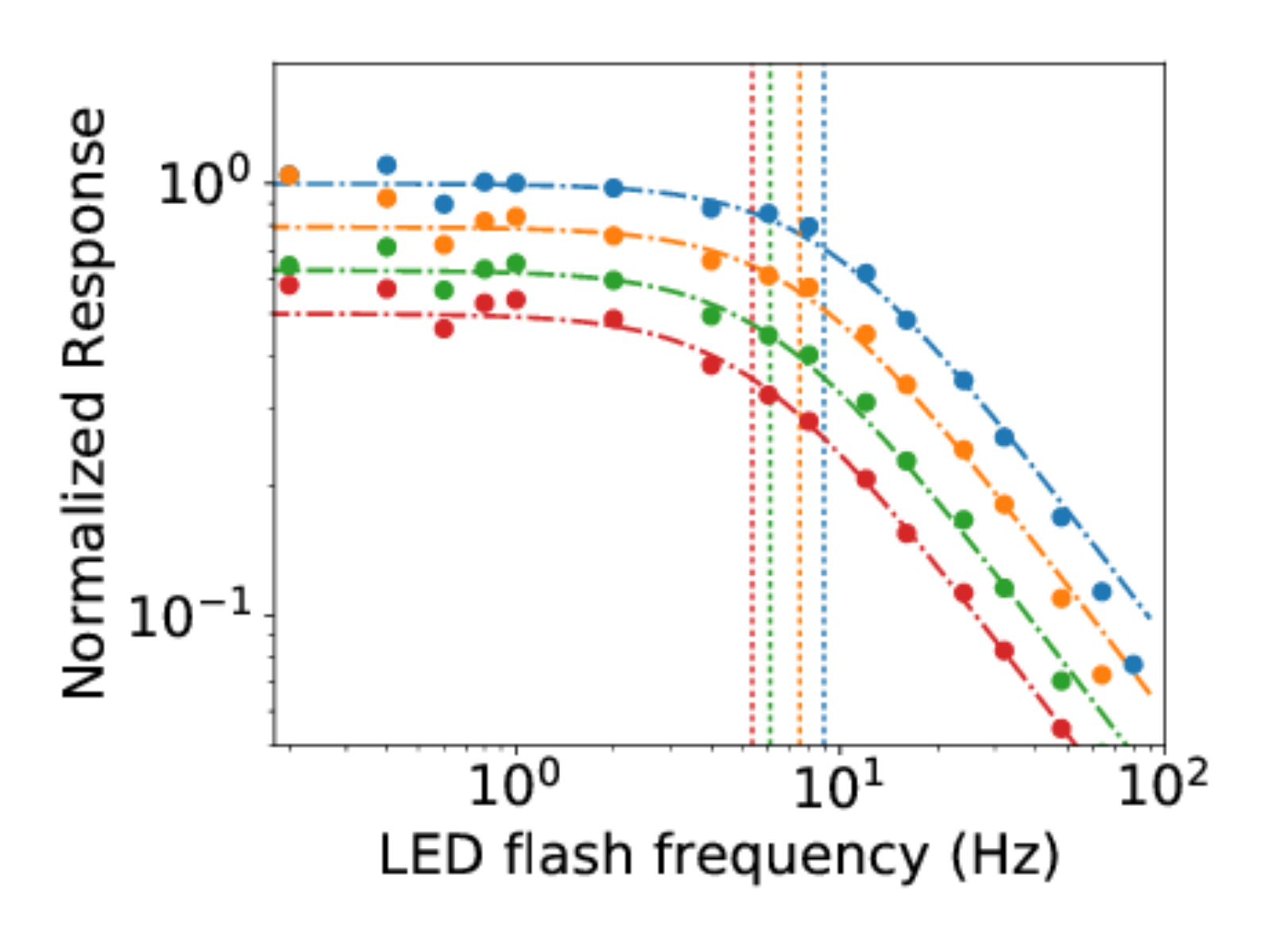}
\includegraphics[width=0.48\textwidth]{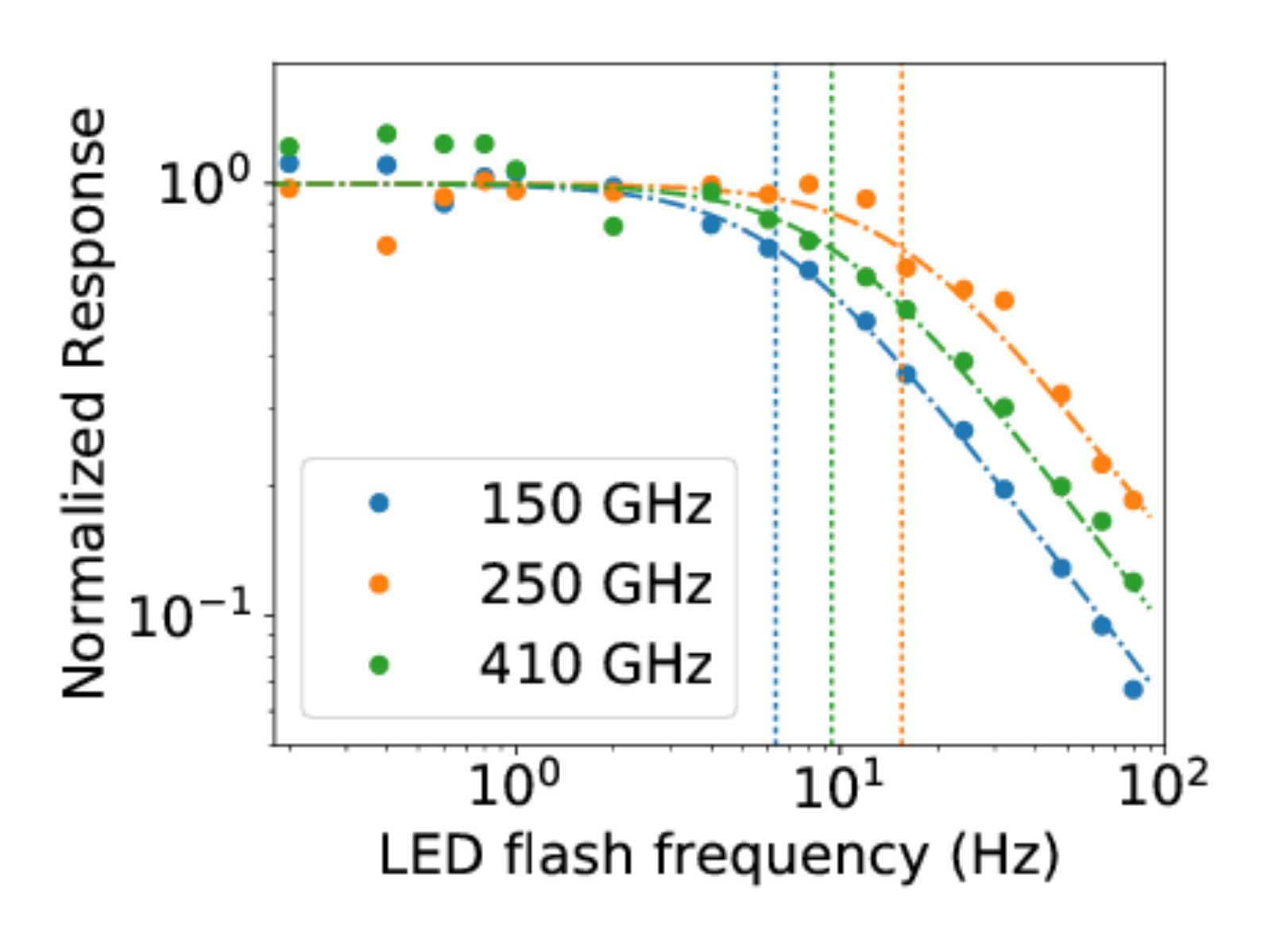}
\caption{Examples of the measured, normalized temporal response of bolometers as a function of \ac{LED} flash frequency 
(points), the single pole response fits (dot-dashed), and the 3~dB point (vertical dots). 
Left: the response of four 150~GHz bolometers biased
at 0.8~$R_{n}$. The normalized response is offset vertically for clarity. Right: the response of one
bolometer per frequency.  All three were tuned to 0.8~$R_{n}$. 
\label{fig:LED_response} }
\end{figure}

We used a dedicated cryostat to operate one wafer at a time in dark conditions and at a bath temperature of 0.32~K.
We measured $\tau_{eff}$ at various operating points along the superconducting transition as quantified by fractions 
of bolometer normal resistance $R_{n}$.
We used \ac{LED}s with a center wavelength of 940~nm and intrinsic time constant of 1~$\mu$s to excite the bolometers. 
The \ac{LED} was flashed on/off at 13--16 frequencies from 0.1 to 80~Hz for 2~min at each frequency. We calculated the 
power spectral density of the time stream, removed the background, fit a Fejer 
kernel~\citep{Weisse_FejerKernal2006} to the peak centered on the flash frequency, and extracted the power
under the peak from the fit.  We plot the response as a function of flash frequency, 
fit to a single pole function
\begin{equation}
R(f) = \frac{A}{\sqrt{1+(f/f_{c})^2}},
\label{eqn:single_pole}
\end{equation}
where $A$ is an arbitrary amplitude at $f=0$, 
extract the 3~dB cut-off frequency $f_{c}$, and convert it to the bolometer time constant  
\begin{equation} 
\tau_{eff} = \frac{1}{2 \pi f_{c}}.
\end{equation}
Figure~\ref{fig:LED_response} shows a selection of these measurements that are normalized to unity through division by 
their normalization values $A$. 
For each measurement we extracted the error $\delta f_{c}$ from the fit's covariance matrix and propagated it
to calculate $\delta \tau_{eff}$. 
We kept those measurements of $\tau_{eff}$ for which $\delta \tau_{eff} / \tau_{eff} < 0.5$ 
and $\delta \tau_{eff} < 10$~ms. For each 
wafer the cuts removed between 20\% and 50\% 
of the measurements at each point in the transition.
 
\begin{figure}[ht]
\center
\includegraphics[width=0.47\textwidth]{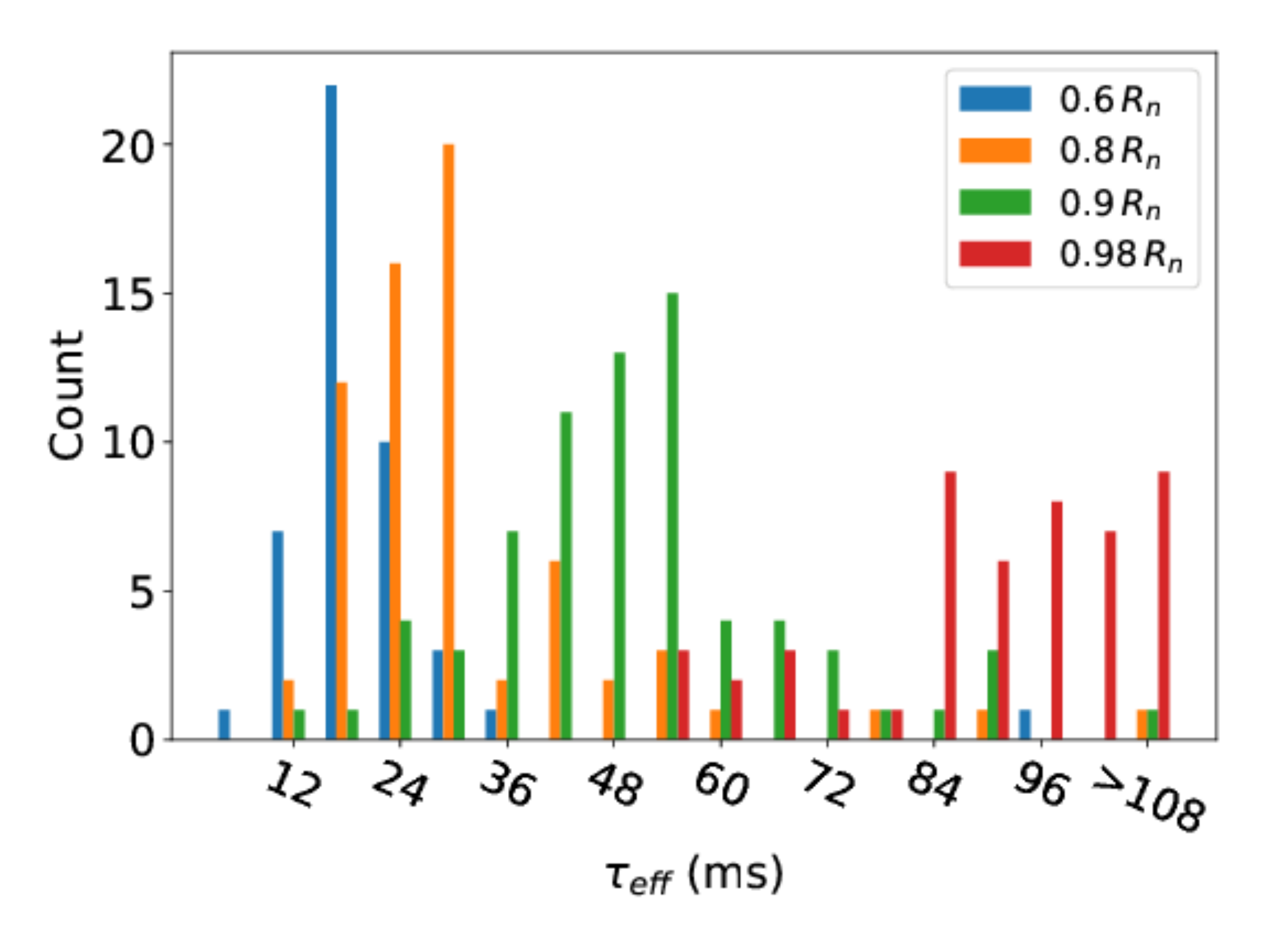}
\includegraphics[width=0.49\textwidth]{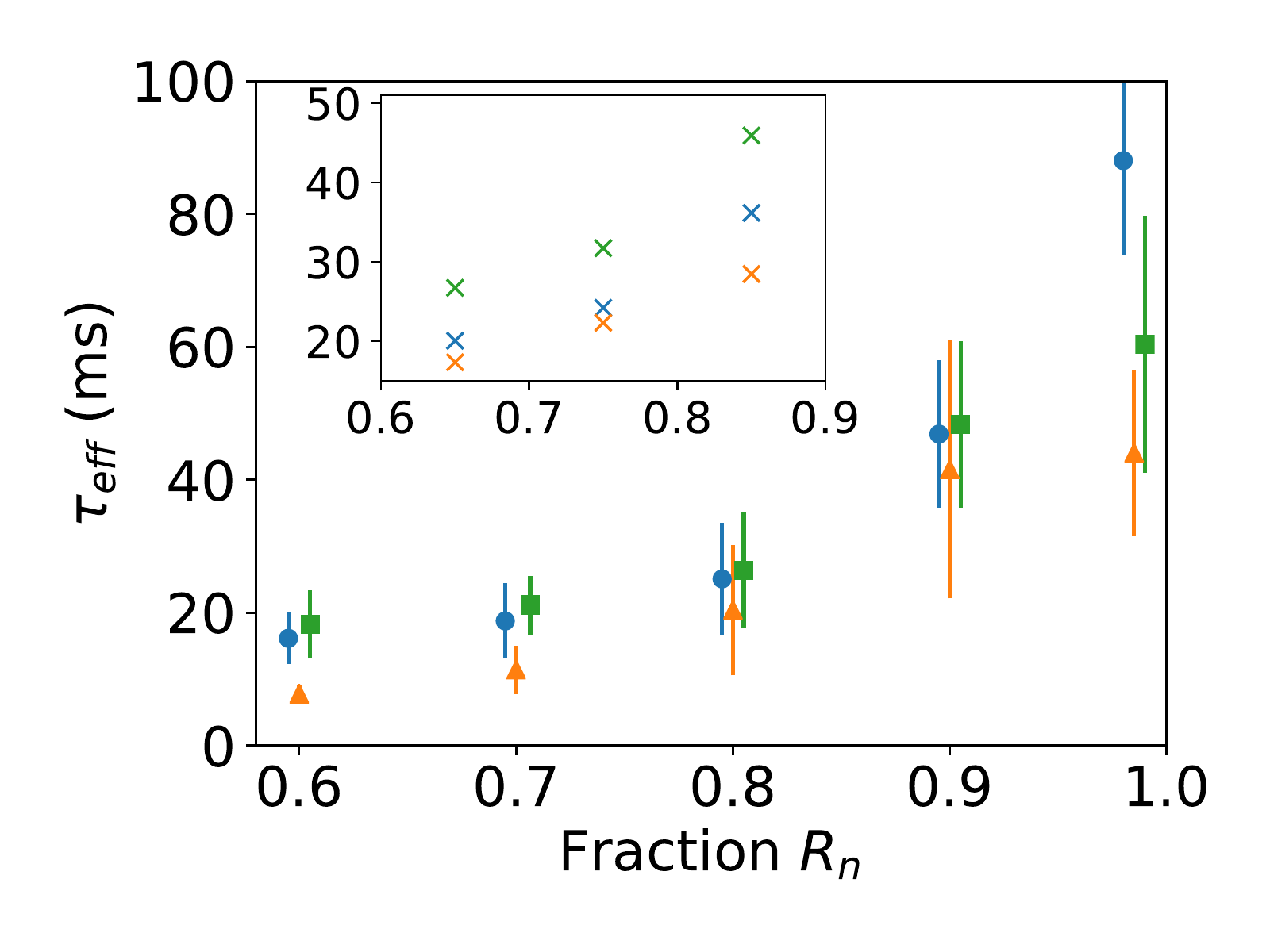}
\caption{Left: distributions of measured $\tau_{eff}$ for the 150~GHz wafer at four depths in transition.
              Right: the median $\tau_{eff}$ as a function of the fraction of $R_{n}$ for the different 
         frequency bands: 150~GHz (blue circles), 250~GHz (orange triangles), and 
         410~GHz wafer (green squares). Error bars are the standard deviation of the 
         $\tau_{eff}$ distributions at each transition depth. 
         We extrapolated laboratory measurements to flight conditions to estimate median
         time constants at the three transition depths used in flight (inset).
\label{fig:LED_taus} }
\end{figure}

The measured distributions of $\tau_{eff}$ for the 150~GHz bolometers are shown in Figure~\ref{fig:LED_taus}. 
For each wafer and transition depth we found the median of the distribution and plot the medians as a 
function of transition depth. 
On all three wafers the width of the distribution was dominated by variations between detectors rather 
than measurement errors; typical $\delta \tau_{eff}$ values were 1~ms.
For all three wafers $\tau_{eff}$ qualitatively followed the expectation of $1/(1+\mathcal{L})$, since $\mathcal{L}$
increases deeper in the transition; see Figure~\ref{fig:LED_taus}. 

We measured $\tau_0$ by assuming $\mathcal{L}$~$<<$~1 at the highest bias point, 0.98~$R_{n}$, so $\tau_{eff}$~$\approx$~$\tau_0$; 
median $\tau_0$ values are given in Table~\ref{tab:Design_Params}. They are longer than design values by factors of 4--6. 
We combined the measurements of $\tau_{0}$ with independent measurement of $G$ to infer bolometer heat 
capacities. For each bolometer we calculated $G$ according to Equation~\ref{eqn:G_dyn}
using measured $\overline{G}$, $T_c$, $n$, and a 0.32~K bath temperature.
We calculated $C$ for each bolometer and give the design and median of measured values 
in Table~\ref{tab:Design_Params}. We find that the median heat capacities are larger than design values by 
factors of 4--8. There are two possible sources for a larger than expected 
heat capacities: larger specific heats or larger volumes. We measured the thicknesses of the various layers deposited 
to make the bolometers and find that the volumes were within 15\% of design values. 
     
We provided an approximate value for the time constants in flight by extrapolating the thermal conductance to the 
in-flight 0.25~K bath temperature, and by accounting for changes in $\mathcal{L}$ due to 
the lower electrical bias of the detectors, which was a consequence
of higher in-flight optical load. We did not attempt to correct $\mathcal{L}$ for differences in $\alpha=d(\log R)/d(\log T)$
because the derivatives $\frac{dR}{dT}$ were noisy.
The calculated $\tau_{eff}$ for the three bias points used in flight are shown in Figure~\ref{fig:LED_taus}. 
During the majority of flight, bolometers were tuned to 0.85~$R_{n}$. 


%% file: optical_load.tex
\subsection{Radiative Load}
\label{sec:optical_load}

\begin{figure}[htbp]
\begin{center}
\epsscale{1.0}
\includegraphics[width=0.5\textwidth]{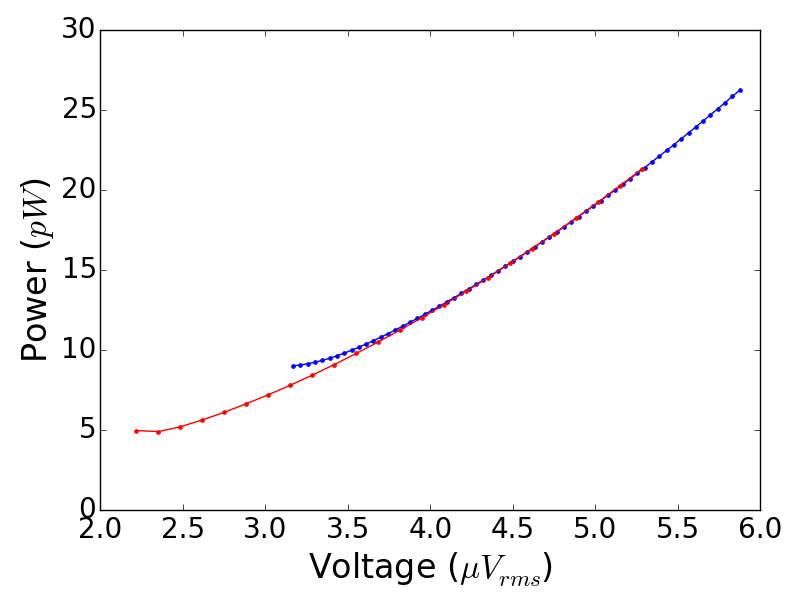}
\caption{Measurements of load curves of a 250~GHz detector in dark conditions (blue)
and during flight (red). The detector is operated with voltage bias such 
that the total power is constant, which is approximately 9~pW in dark conditions. 
Only 5~pW of electrical power are necessary in flight because absorbed radiative power 
makes up the difference.
\label{fig:dark_light_pv} }
\end{center}
\end{figure}

Equation~\ref{eqn:boloPowerFlow} quantifies the balance of 
power between a bolometer and its thermal bath when the total power absorbed $P_{abs}$ is constant. 
$P_{sat}$, which depends on the thermal conductance, the transition temperature, and bath temperature, 
has a characteristic value for each detector and was measured for 
the majority of the detectors in dark conditions. Although these measurements were done at different 
bath temperatures $T_{0'}$, they were corrected to the operating temperature in flight $T_{0}$ using 
Equation~\ref{eqn:ScalePsat}. 
We use Equation~\ref{eqn:boloPowerFlow} to  infer the radiative power absorbed in flight
\begin{equation}
P_{abs, f} = P_{sat}(T_{0}) - P_{e,f} (T_{0}) = P_{e,d}(T_{0}) - P_{e,f}(T_{0}) 
\label{eq:RadiativeLoad}
\end{equation}
where the subscripts $ _{d}$ and $_{f}$ denote dark and in-flight conditions, respectively.
Figure~\ref{fig:dark_light_pv} shows an example of a 150~GHz detector for which $P_{abs, f}$~=~4~pW. 
The electrical power dissipated is plotted as a function of the voltage across the bolometer in dark and 
flight conditions; these are commonly called `load curves'.  The difference of 4~pW is measured 
at the lowest power points of the two load curves, the points at which the bolometer 
had just entered the constant power operating regime. To ensure stability we operated the 
detectors as they just entered this regime, that is, relatively high in the superconducting 
transition. For the majority of the flight the detectors were operated at 85\% of their normal resistance. 

\begin{figure}[ht!]
\begin{center}
\includegraphics[width=0.32\columnwidth]{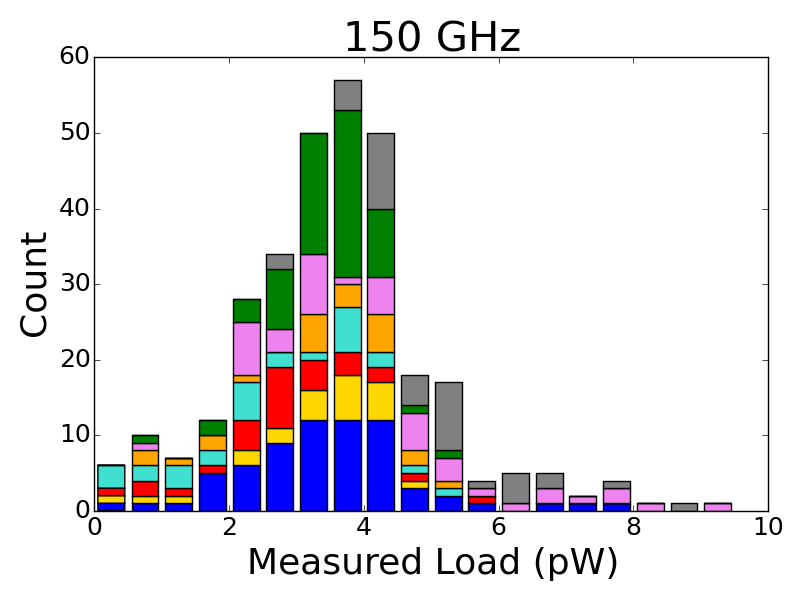}
\includegraphics[width=0.32\columnwidth]{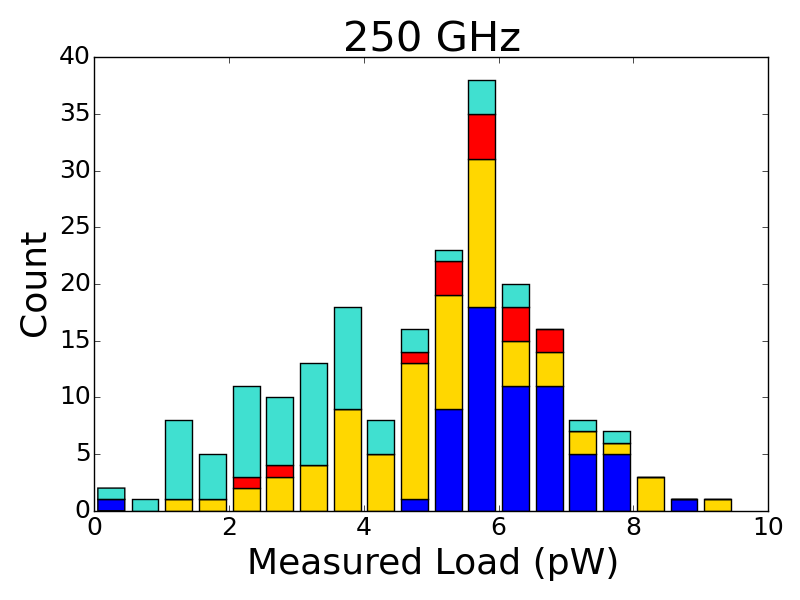}
\includegraphics[width=0.32\columnwidth]{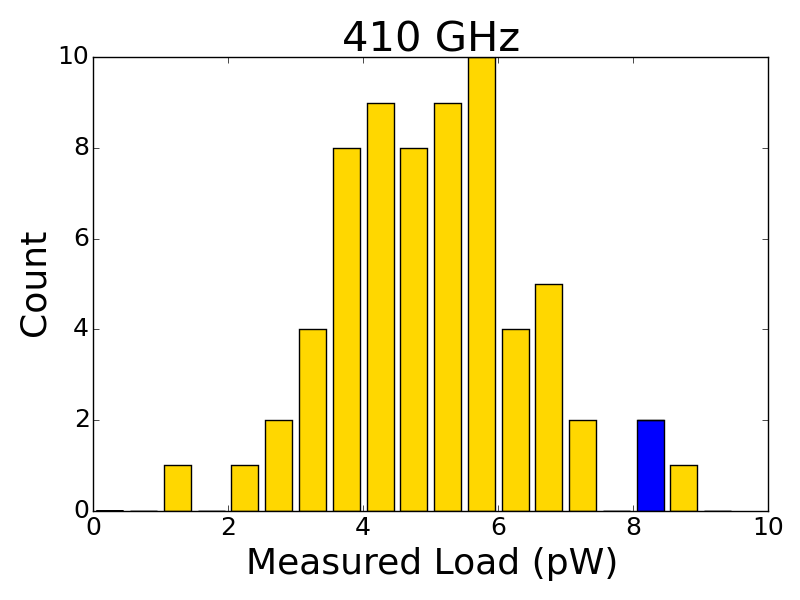}
\caption{Histograms of the measured radiative load from the first detector tuning at float. The different colors represent 
different wafers.  The medians and standard deviations of the distributions are 3.6~$\pm$~1.5,  5.3~$\pm$~1.8, and 5.0~$\pm$~1.4~pW, for the 150, 250, and 410~GHz, respectively. 
\label{fig:radiative_load_histograms} }
\end{center}
\end{figure}

Figure~\ref{fig:radiative_load_histograms} shows the distribution of the radiative load measured by all detectors.
We used  the load curves measured during the first tuning at float altitude. The elevation (zenith) angle 
of the telescope was 60$^\circ$ (30$^\circ$).
We measure an average load of 3.6, 5.3, and 5.0~pW for the 150, 250, and 410~GHz detectors, respectively.


%% file: optical_efficiency.tex
\subsection{Optical Efficiency}
\label{sec:optical_efficiency}

The end-to-end instrument efficiency for a given detector, defined as the ratio of radiative power absorbed
by the detector $P_{abs}$ to the radiative power incident on the instrument $P_{inc,\,\, inst}$, is the product 
of the telescope's transmission $\epsilon_{t}$ and the bolometer absorption efficiency $\epsilon_{b}$
\begin{equation}
\epsilon = \frac{P_{abs}}{P_{inc,\,\, inst}} = \epsilon_{t} \frac{P_{abs}}{P_{inc}} =  \epsilon_{t} \epsilon_{b}.
\label{eq:total_efficiency}
\end{equation}
Here $P_{inc}$ is the radiative power incident on a bolometer. 

The transmission of the telescope is the product of the transmissions of each of the optical elements. 
These transmission coefficients were measured or calculated from 
first principles~\citep{aubin_thesis, Zilic_thesis} 
and are summarized in Tables~\ref{tab:transmissions2013} and~\ref{tab:transmissions2010}
for the flight and ground measurement configurations, respectively.  
For absorption we used the optical path length of a representative detector in a given 
frequency band and extrapolated cryogenic loss tangent values 
to our temperatures~\citep{Jacob2002_cyrogeniclosstan}.

\begin{table}[htbp]
\begin{center}
\begin{tabular}{|l|c|c|c|c|}
\multicolumn{4}{l}{} \\ \hline
Element & 150 GHz & 250 GHz & 410 GHz\\ \hline \hline
Primary mirror & 1.00 & 1.00 & 1.00\\ \hline
Secondary mirror & 1.00 & 1.00 & 1.00\\ \hline
Cryostat window & 0.97 & 0.96 & 0.97\\ \hline
Thermal filter type 4 & 0.95 & 0.95 & 0.95\\ \hline 
Thermal filter type 4 & 0.95 & 0.95 & 0.95\\ \hline 
Thermal filter type 3 & 0.98 & 0.98 & 0.98\\ \hline
Teflon filter & 0.96 & 0.95 & 0.93\\ \hline
Thermal filter type 3 & 0.98 & 0.98 & 0.98\\ \hline
LPE1 filter & 0.98 & 0.94 & 0.98\\ \hline
Thermal filter type 4 & 0.95 & 0.95 & 0.95\\ \hline 
LPE2 filter & 0.96 & 0.96 & 0.98\\ \hline
Field lens & 0.97 & 0.97 & 0.95\\ \hline
LPE2b filter & 0.97 & 0.99 & 0.91\\ \hline
\ac{HWP} & 0.94 & 0.93 & 0.91\\ \hline
LPE2b filter & 0.97 & 0.99 & 0.91\\ \hline
Pupil lens \#1 & 0.97 & 0.96 & 0.97\\ \hline
Pupil lens \#2 & 0.97 & 0.96 & 0.96\\ \hline
Polarizing grid & 0.50 & 0.50 & 0.50\\ \hline
Camera lens & 0.97 & 0.96 & 0.97\\ \hline
Low-pass-blocker & 0.96 & 1.00 & 0.96\\ \hline
Low-pass-edge & 0.86 & 0.95 & 0.96\\ \hline
Feedhorns & 1.00 & 1.00 & 1.00\\ \hline
Waveguides & 1.00 & 1.00 & 1.00\\ \hline \hline
Total & 0.23 & 0.24 & 0.21\\ \hline  
\end{tabular}
\end{center}
\caption{Optical elements used during flight and their transmission for each of the frequency bands. These 
transmissions are used to determine $\epsilon_{t}$ in-flight. 
\label{tab:transmissions2013}}
\end{table}

\begin{table}[htbp]
\begin{center}
\begin{tabular}{|l|c|c|c|c|}
\multicolumn{4}{l}{} \\ \hline
Element & 150~GHz & 250~GHz & 410~GHz\\ \hline \hline
Thermal filter type 1 & 1.00 & 0.97 & 0.93\\ \hline 
Thermal filter type 2 & 0.97 & 0.96 & 0.90\\ \hline 
Teflon filter & 0.96 & 0.95 & 0.93\\ \hline
Thermal filter type 3 & 0.98 & 0.98 & 0.98\\ \hline 
LPE1 filter & 0.98 & 0.94 & 0.98\\ \hline
Thermal filter type 4 & 0.95 & 0.95 & 0.95\\ \hline 
LPE2 filter & 0.96 & 0.96 & 0.98\\ \hline
Field lens & 0.96 & 0.97 & 0.95\\ \hline
LPE2b filter & 0.97 & 0.99 & 0.91\\ \hline
HWP$^\dagger$& 0.54 & 0.53 & 0.51\\ \hline
Pupil lens \#1 & 0.97 & 0.95 & 0.96\\ \hline
Pupil lens \#2    & 0.97 & 0.95 & 0.96\\ \hline
Grid & 0.50 & 0.50 & 0.50\\ \hline
Camera lens & 0.97 & 0.96 & 0.97\\ \hline
Low-pass-blocker & 0.96 & 1.00 & 0.96\\ \hline
Low-pass-edge & 0.86 & 0.95 & 0.96\\ \hline
Feedhorns & 1.00 & 1.00 & 1.00\\ \hline
Waveguides & 1.00 & 1.00 & 1.00\\ \hline \hline
Total & 0.16 & 0.15 & 0.12\\ \hline  
\multicolumn{4}{l}{{\footnotesize $^\dagger$ At the time of this test the \ac{HWP} had no antireflection coating. } }
\end{tabular}
\end{center}
\caption{Optical elements used during the ground measurement of absorption efficiency 
and their transmission for each of the frequency bands. 
\label{tab:transmissions2010} }
\end{table}

The calculation of efficiencies required knowledge of the response of the instrument as a function 
of frequency. These measurements were reported in \ac{EP1}. The bolometer efficiencies we report here 
are average efficiencies
\begin{equation}
\langle \epsilon_{b} \rangle = \frac{1}{\Delta \nu} \int_{\nu_{1}}^{\nu_{2}} \epsilon_{b} (\nu) d\nu, \,\,\,\, \Delta \nu = \nu_{2} - \nu_{1},
\label{eq:average_efficiency}
\end{equation}
where $\nu_1$ and $\nu_2$ define the bandwidth of the instrument $\Delta \nu$.
An absorption efficiency per band is given 
Table~\ref{tab:efficiencysummary} and the results are discussed 
in Section~\ref{sec:efficiency_discussion}.

In this section we give 
constraints on $<\epsilon_{b}>$ using three types of measurements: (a) Galactic plane scans; (b) laboratory cold load, and 
(c) comparison of laboratory and in-flight optical loads. 
Only the laboratory cold load measurement is purely laboratory-based.

\subsubsection{Galactic Plane Scans}
\label{sec:efficiencyfromcal}

\begin{figure}[htbp]
\centering
\includegraphics[width=0.32\textwidth]{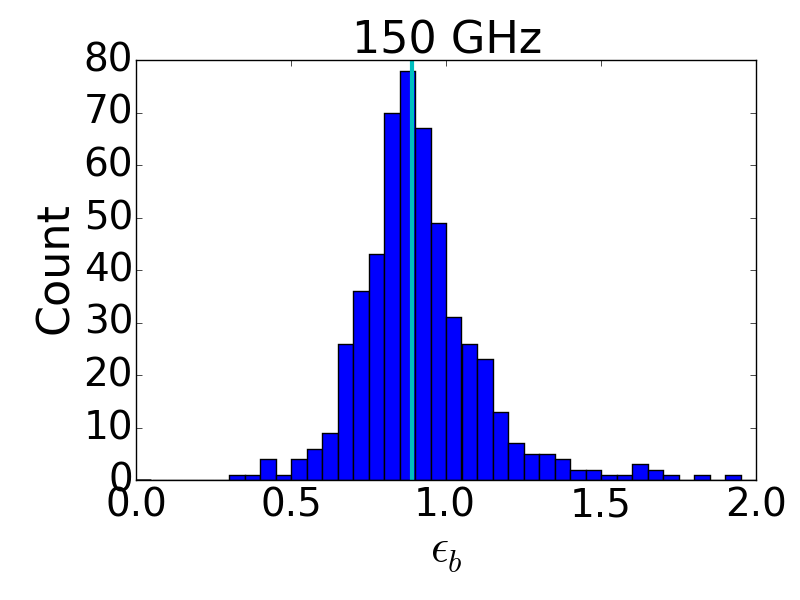}%
\includegraphics[width=0.32\textwidth]{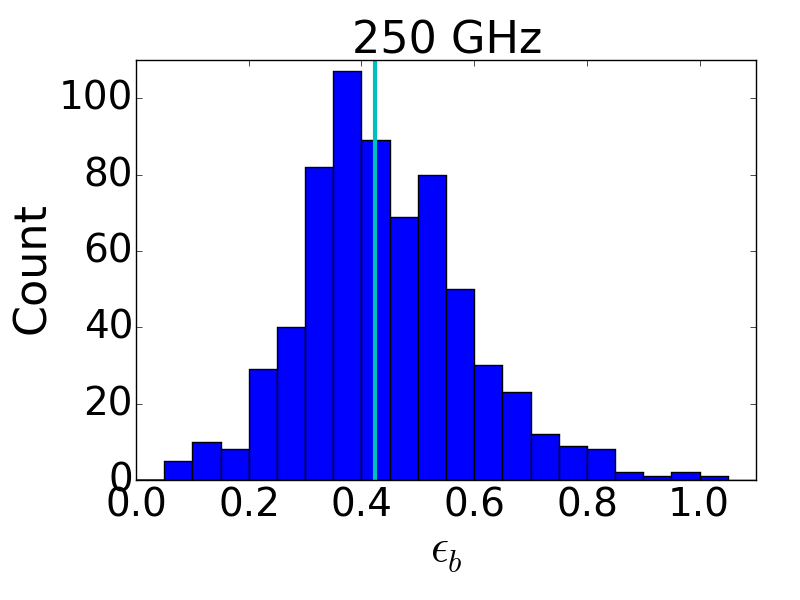}%
\includegraphics[width=0.32\textwidth]{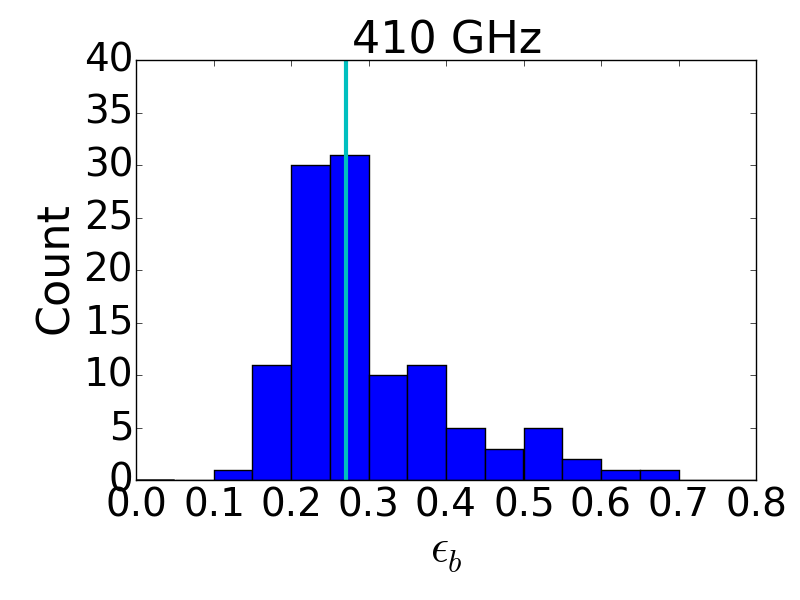}\\%
\caption{Average bolometer absorption efficiency distributions inferred from 
the calibration factors measured for all valid detectors and sections of the flight. The median efficiencies (vertical cyan) 
are given in Table~\ref{tab:efficiencysummary}. 
\label{fig:opt_eff_from_calib_dist} }
\end{figure}

We used passes across the Galactic plane to calibrate the end-to-end response for each
detector~\citep{Aubin_MGrossman2015}. We assumed the 
Galactic signal is a sum of \planck\ component maps, which have 
been scaled to and integrated over each of the measured bands,
see Appendix~\ref{sec:refMaps} for more details. 
We determined the calibration factor $CAL=d P_{inc,\, inst} / d I^{c}$~(in units of W/count) that gave the conversion from 
power incident on the instrument to current flowing through a bolometer in readout system counts. 
We related the current in each bolometer in counts to the power absorbed using the current responsivity  
$S_{I}$~(A/W) and the conversion from current measured in readout system counts to physical units 
$dI_{b}^{A}/dI_{b}^{c}$ (A/count), which was measured in the lab and assumed to be the same for all detectors
\begin{equation}
dP_{abs} = dI^{c} \frac{dI_{b}^{A}}{dI_{b}^{c}} \frac{1}{S_{I}}. 
\label{eq:calPabs}
\end{equation}
Using Equations~\ref{eq:total_efficiency} and~\ref{eq:calPabs}, the calibration factor is now
\begin{equation}
CAL = \frac{d P_{inc,\, inst}}{d I^{c}} = \frac{1}{\epsilon} \frac{dI_{b}^{A}}{dI_{b}^{c}} \frac{1}{S_{I}} =  \frac{1}{\epsilon_{t} \epsilon_{b}} \frac{dI_{b}^{A}}{dI_{b}^{c}} \frac{1}{S_{I}},
\label{eq:calPabs2}
\end{equation}
from which we find
\begin{equation}
\epsilon_{b} = \frac{1}{CAL} \frac{1}{\epsilon_{t}} \frac{dI_{b}^{A}}{dI_{b}^{c}} \frac{1}{S_{I}}.
\label{eq:calepsilon}
\end{equation}
For the responsivity we assumed the slow signal limit and that the detector had large loop-gain such that 
\begin{equation}
S_{I} \simeq \frac{\sqrt{2}}{V_{b}}
\label{eq:respon}
\end{equation}
where $V_{b}$ was the bolometer's known voltage bias corrected for the effects of stray impedance~\citep{lee_appliedoptics_1998}. 
Figure~\ref{fig:opt_eff_from_calib_dist} shows the distributions of inferred average bolometer absorption 
efficiencies using the calibration factors for all sections of flight and for all detectors.
(The calibration factors varied between different sections.) In Table~\ref{tab:efficiencysummary} we report the median 
of each distribution as the absorption efficiency for the respective band. Additional discussion is given 
in Section~\ref{sec:efficiency_discussion}.

\begin{table}[htbp]
\begin{center}
\begin{tabular}{|c|c|c|c|}\hline
Method & 150~GHz & 250~GHz & 410~GHz \\
\hline
Scans of Galactic plane & 0.9 & 0.4 & 0.3 \\ 
\hline
Cold load  & 1.0 & 0.3 & 0.2 \\ 
\hline
\end{tabular}
\end{center}
\caption{Summary of bolometer absorption efficiencies per band extracted using two different methods.
\label{tab:efficiencysummary} }
\end{table}


\subsubsection{Laboratory Cold Load}
\label{sec:MaximaColdLoad}

We coupled the receiver to a throughput filling, temperature-adjustable cold load. The cold load was constructed to 
have an emissivity close to 1 in our frequency bands. 
We set the temperature of the 
cold load $T_{cl}$ to several values between 25 and 50~K  and at each temperature measured 
the electrical power necessary to operate the bolometer in transition. Similar to Equation~\ref{eq:RadiativeLoad},
the radiative power absorbed by the bolometer in this configuration is given by 
\begin{equation}
P_{abs, cl}(T_{cl}) = 
P_{e,d} - P_{e,cl}(T_{cl}) = \epsilon_{b} \epsilon_{t} P_{inc, cl}(T_{cl}), 
\label{eqn:RadiativeColdLoad}
\end{equation}
where the subscript $_{cl}$ denotes `cold load'. 
Note that in contrast to Equations~\ref{eqn:boloPowerFlow} and~\ref{eq:RadiativeLoad} the parametrization 
of power in Equation~\ref{eqn:RadiativeColdLoad} is in terms of the load temperature. 
Following the left side of the equation, we measured the 
absorbed radiative $P_{abs, cl}$ of each bolometer 
by differencing the measured saturation power (from the measurement in dark conditions) and the  
measured electrical power in each of the cold load temperatures. This radiative load is linearly proportional 
to the radiative load incident on the instrument $P_{inc, cl}$, which we calculated 
using the spectrum of a black-body load of known temperature, the instruments' frequency bands, and 
the transmission efficiencies given in Table~\ref{tab:transmissions2010}. 
The cold load measurement was done for one wafer 
in each frequency band. 

\begin{figure}[htbp]
\centering
\includegraphics[width=0.24\textwidth]{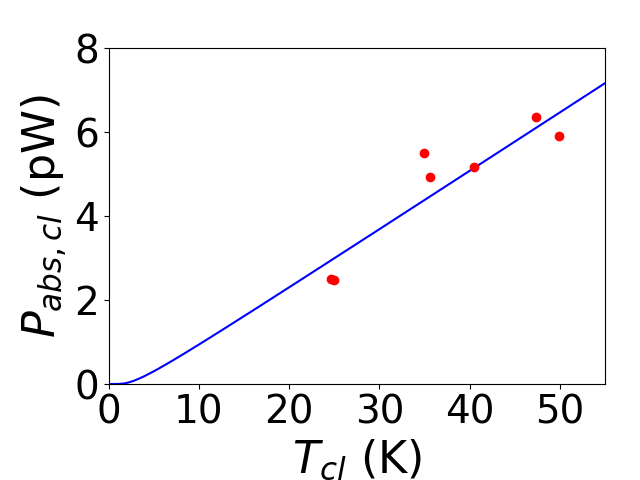}
\includegraphics[width=0.24\textwidth]{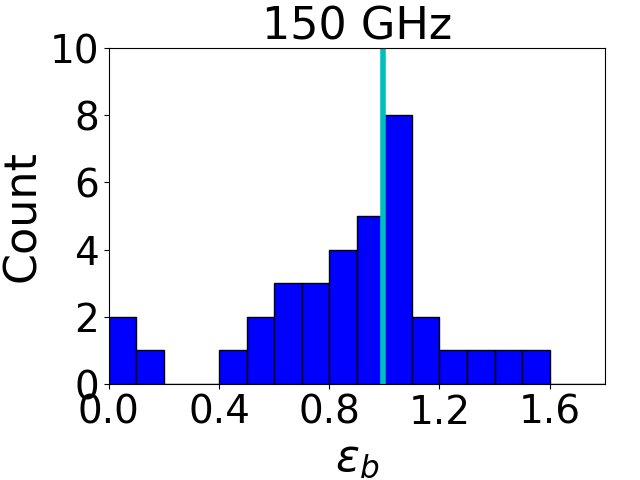}
\includegraphics[width=0.24\textwidth]{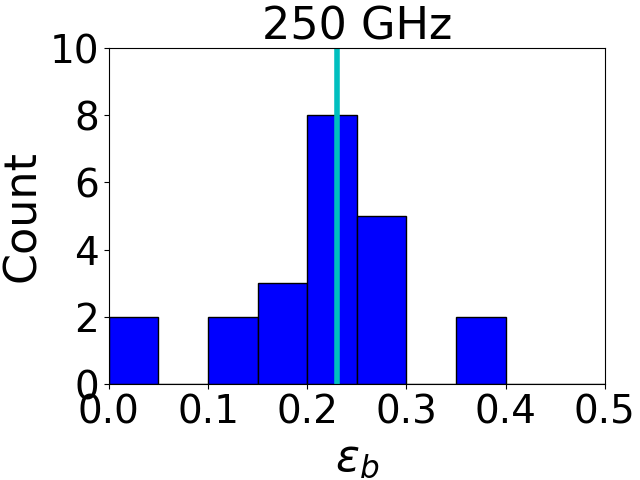}
\includegraphics[width=0.24\textwidth]{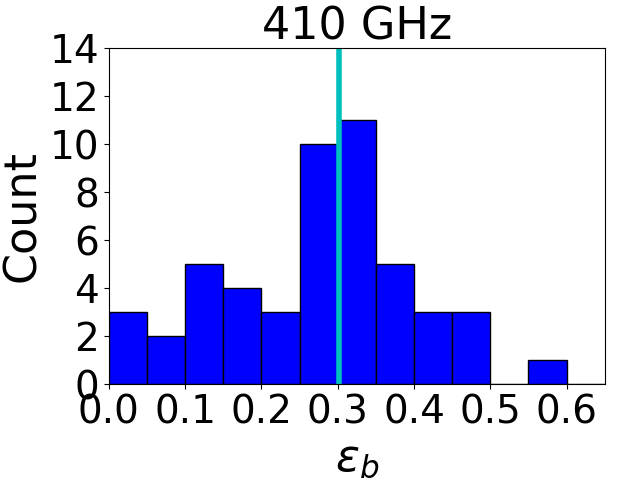}
\caption{Left: measured (red circles) and fitted (solid blue) power absorbed by a 150~GHz bolometer 
as a function of the cold load temperature.
Right: The distribution of measured bolometer absorption efficiency and their median (vertical cyan) for each 
of the frequency bands. The median values are given in Table~\ref{tab:efficiencysummary}. 
\label{fig:efficiency} }
\end{figure}

Figure~\ref{fig:efficiency} shows an example of $P_{abs, cl}$ as a function of the cold load 
temperature for one 150~GHz detector, as well as a best-fit model from which we extract its $\epsilon_{b}$.
The Figure also shows the distribution of derived bolometer absorption efficiencies for each frequency band. 
The medians of the distributions are reported in Table~\ref{tab:efficiencysummary}. Additional discussion is given 
in Section~\ref{sec:efficiency_discussion}.

\subsubsection{Comparison of Laboratory and in-Flight Optical Loads}
\label{sec:efficiencyfromloadcurve}

A combination of Equations~\ref{eq:RadiativeLoad} and~\ref{eq:total_efficiency} gives
\begin{equation}
P_{abs, f} = \epsilon_{t} \epsilon_{b} P_{inc, \, inst, f} =  P_{e,d}(T_{0}) - P_{e,f}(T_{0}), 
\label{eq:loadcurve1}
\end{equation}
where $P_{inc, \, inst, f}$ is the radiative power incident on the instrument in flight.
We find
\begin{equation}
\epsilon_{b} =  \frac{ P_{e,d}(T_{0}) - P_{e,f}(T_{0}) }{\epsilon_{t} P_{inc, inst, f}} =  
\frac{ P_{e,d}(T_{0}) - P_{e,f}(T_{0}) }{P_{inc,f }}, 
\label{eqn:loadcurve2}
\end{equation}
where $P_{inc,f}$ is the radiative power incident on the bolometer in flight. 

We calculated the power incident on the detectors by including contributions from the sky (\ac{CMB} and atmospheric emission~\citep{bao_thesis}), 
the mirrors, other optical elements along the optical path, such as the vacuum window and 
absorptive filters, and the transmission coefficients listed in 
Table~\ref{tab:transmissions2013}~\citep{aubin_thesis}; see Table~\ref{tab:loadpredictions}. 
The electrical powers in Equation~\ref{eqn:loadcurve2} were 
known from laboratory and in-flight load curve measurements (see Section~\ref{sec:optical_load}). 
We found median bolometer absorption efficiencies of 3.4, 1.1, and 0.7 for the 150, 250, and 410~GHz
bands, respectively. Bolometer absorption efficiencies larger than 1 are unphysical 
and we discuss this result in the next Section.

\begin{table}[htbp]
\begin{center}
\begin{tabular}{|c|c|c|c|}\hline
Element & 150 GHz & 250 GHz & 410 GHz\\
 & (pW) & (pW) & (pW) \\ \hline \hline
\ac{CMB}              & 0.063 & 0.032 & 0.0047\\ \hline
Atmosphere       & 0.0054 & 0.064 & 0.37\\ \hline
Mirrors      & 0.19 & 0.45 & 0.76\\ \hline
Instrument & 0.36 & 1.3 & 3.5\\ \hline \hline
Total & 0.62 & 1.8 & 4.7\\ \hline
\end{tabular}
\end{center}
\caption{Predicted optical power incident on the bolometer during flight $P_{inc, f}$.
For each element the power incident on the 
instrument is higher and is reduced by reflection and absorption by optical elements along the light path. Emission 
by the instrument is dominated by a 1/2 inch thick teflon filter that had a temperature of 110~K 
and by the \ac{HWP} that reached $\sim$50~K.
\label{tab:loadpredictions} }
\end{table}

\subsubsection{Discussion of Absorption Efficiency Results}
\label{sec:efficiency_discussion}

The calibration and cold load measurements give approximately consistent results; a relatively high absorption 
efficiency at the 150~GHz band and medium absorption at the 250 and 410~GHz bands. The calibration measurement
is based on the response to changes in the signal level as the instrument passes across the Galactic plane. The scatter
in the histograms of Figure~\ref{fig:opt_eff_from_calib_dist}, which gives
some values that are larger than 1, is due primarily to larger uncertainty in the Galactic calibration for 
some of the detectors. The other two measurements are sensitive to the total power absorbed in the detector. 
The cold load measurement also has some scatter due to noise or systematic errors in the calculation of the absorbed power. 
For example, it is sometimes challenging to accurately estimate $P_{e}$ from the load curves such as shown 
in Figure~\ref{fig:dark_light_pv}.

The results giving absorption efficiencies larger than 1 for the load curve-based measurement indicate that the denominator 
in Equation~\ref{eqn:loadcurve2}  had been underestimated and that there was excess load that was not accounted 
for by the sources listed in Table~\ref{tab:loadpredictions}. To find the magnitude of the excess absorbed load we 
difference the measured and expected values. The measured loads are 3.6, 5.3, and 5.0~pW for the 
150, 250, and 410~GHz bands, respectively; see Section~\ref{sec:optical_load}. The expected absorbed load is 
inferred from the product of the expected load incident on the bolometer, as given in Table~\ref{tab:loadpredictions}, and 
the bolometer absorption efficiencies coming from the Galaxy calibration; see Table~\ref{tab:efficiencysummary}. 
We find excess absorbed load of 3.1, 4.6, and 3.6~pW for the 150, 250, and 410 bands, respectively. These 
excess loads were not observed with the laboratory cold load. We note that the design values for the thermal conductances were 
based on power saturation values of 4, 9, and 12~pW, respectively; see Section~\ref{sec:thermal_isolation}. Relative
to our pre-instrument construction estimates, we find a reduction in sky signal and increase in instrumentally induced load 
that give a total load consistent with our margin of safety. The 
reduction in sky signal was a result of having bands with transmission that was approximately half that expected relative 
to top-hat bands with the same edges.


We hypothesize that beam spill-over past the vacuum window and onto warm surfaces outside 
of the receiver -- here we view the 
beam as emanating from the focal plane toward the sky -- gave rise to the excess load. This spill-over was not detected 
with the cold load because the load was mounted onto the vacuum window flange and therefore intercepted 
the entire throughput. 



%% file: noise.tex
\subsection{Noise Performance}
\label{sec:noise_performance}

We used flight data to quantify individual-detector and overall instrument noise and compare it to expectations. 
The \ac{NEP} $N$, which for brevity we also refer to as `noise', consists of several terms 
\begin{eqnarray}
N \,\, (\mbox{W} / \sqrt{\mbox{Hz}}) & = & 
          \left[ N_{\mathrm{photon}}^2 + N_{\mathrm{phonon}}^2 + N_{\mathrm{Johnson}}^2 + N_{\mathrm{readout}}^2 \right]^{1/2} \nonumber \\
          & = & 
          \left[ \left( 2h\nu P_{abs} + \xi \frac{P_{abs}^2}{\Delta \nu} \right)
            + \gamma 4k_{B} T^2 G + \frac{1}{S_I^2} \cdot \frac{4k_BT}{R}
            + \frac{1}{S_I^2} \cdot N_{\mathrm{readout}}^2  \right]^{1/2}
\label{eqn:noisebudget}
\end{eqnarray}
where $h$ is Planck's constant, 
$\nu$ is the center of the observation frequency band, 
$\xi$ is a unitless number between zero and one quantifying the photon correlation,
$\gamma$ is a unitless number between zero and one accounting for the temperature gradient along the link from the \ac{TES} to the bath,
$k_{B}$ is Boltzmann's constant,
$T$ is the \ac{TES} temperature,
and $R$ is the \ac{TES} resistance~\citep{mather_appliedoptics_1982}. 

Time ordered data from flight is used to directly assess $N$, that is, the left side of Equation~\ref{eqn:noisebudget}. 
We used several configurations to assess the contributions of specific 
terms on the right side of Equation~\ref{eqn:noisebudget}. For each configuration we compared the measurement to expectations. 
We then combined the contributions in quadrature and compared to the total measured noise $N$. 
We determined the combination of $N_{\mathrm{Johnson}}$ and $N_{\mathrm{readout}}$ using measurements when 
the detectors were biased above their superconducting transition, $N_{\mathrm{phonon}}$ using the measured 
thermal conductances and critical temperatures, and $N_{\mathrm{photon}}$ using the measured radiative load.   

To assess any noise contribution $X$ using flight data,
we divided the data into segments of 172~s length and made use of all possible segments.
Momentary glitches such as cosmic ray hits were replaced with white noise realizations. Sections
of data that had more than 10\% of their samples flagged as glitches were excluded from the analysis.   
For each section we removed an offset and a gradient, applied a Hann window, and calculated the spectral density. 
For the assessment of the readout and Johnson noise, the data was converted from raw digital units (counts) 
to current referred to the \ac{SQUID} input (A) using Equation~\ref{eqn:counttocurrent}. 
For the assessment of the total \ac{NEP}, these data were converted from current through the \ac{SQUID} to 
power absorbed at the detector by applying the current responsivity $S_{I}$
\begin{equation}
X~[ \mathrm{W} / \sqrt{ \mathrm{Hz} }]  =  \frac{
        X~[ \mathrm{counts} / \sqrt{\mathrm{Hz} } ] \cdot dI_{b}^{A}/dI_{b}^{c} \left[ \mathrm{A} / \mathrm{count} \right] }{
        S_{I} \left[ \mathrm{A} / \mathrm{W} \right] }.
\label{eqn:counttopower}
\end{equation}
For each section of data we find the mean level of the spectral density between 
3.9 and 4.4~Hz, which is one of the polarization frequency sidebands.

\begin{figure}[ht!]
\begin{center}
\includegraphics[height=2.5in]{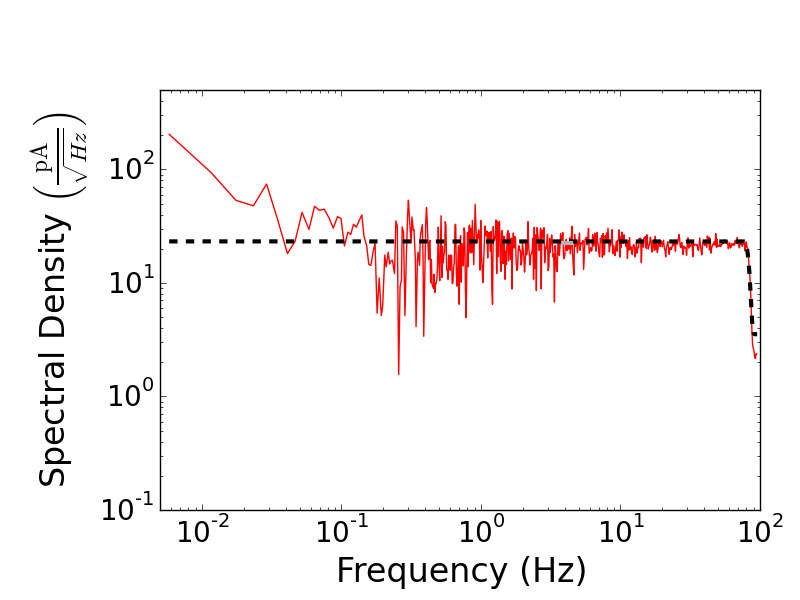}
\includegraphics[height=2.5in]{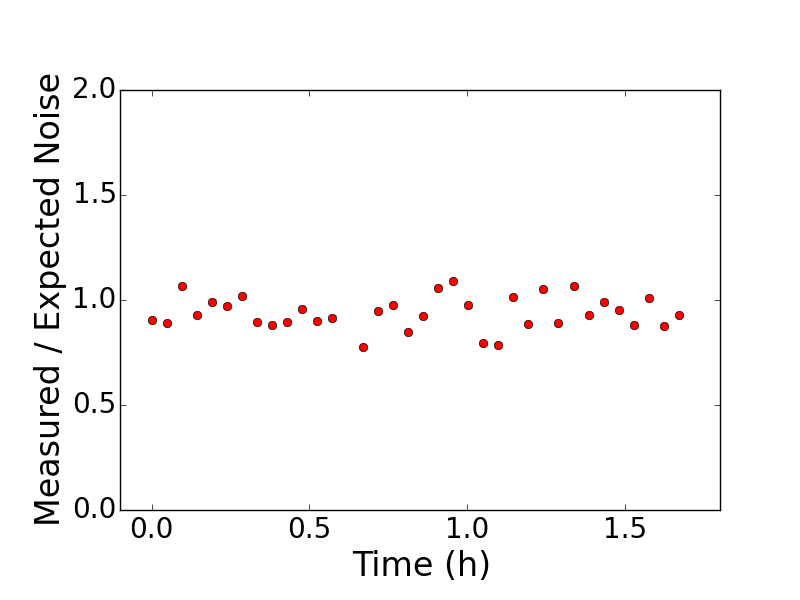}
\caption{Left: the current spectral density of one 150~GHz detector during one 172~s section of the flight when 
it was overbiased (solid red) and the expected quadrature sum of the Johnson 
and readout noise (dashed black). 
We quote noise levels averaged over a narrow band from 3.9 to 4.4~Hz (thick solid gray overlaping with the expectation). 
Right: the measured to expected readout and Johnson noise ratio as a function of time for the same detector. 
\label{fig:one_bolo_overbias_noise} }
\end{center}
\end{figure}

\subsubsection{Johnson and Readout Noise}
\label{sec:JRN}

When the detectors were biased above their superconducting transition -- we will use the shorthand `overbiased' -- 
the current responsivity was low and the last two terms in Equation~\ref{eqn:noisebudget} should 
dominate the overall noise. 
The spectral density for one section of such data for one detector 
is shown in the left panel of Figure~\ref{fig:one_bolo_overbias_noise}.
The measurement is shown in units of A/$\sqrt{\mathrm{Hz}}$ following Equation~\ref{eqn:counttocurrent} 
since both $N_{\mathrm{readout}}$ and $N_{\mathrm{Johnson}}$ are intrinsically current noise sources.
To assess noise level over time we calculated such spectral density for all valid sections 
over 1.7~hours of data, about 35 sections per detector.  
We find that within $\sim$20\% the noise levels recorded are stable; see the right panel of 
Figure~\ref{fig:one_bolo_overbias_noise}.
Higher variations are expected as compared to the resistor case
in Section~\ref{sec:readout_performance} since overbiased detectors have residual sensitivity to stage temperature variations and incident power.
  
\begin{figure}[ht!]
\begin{center}
\includegraphics[height=2.5in]{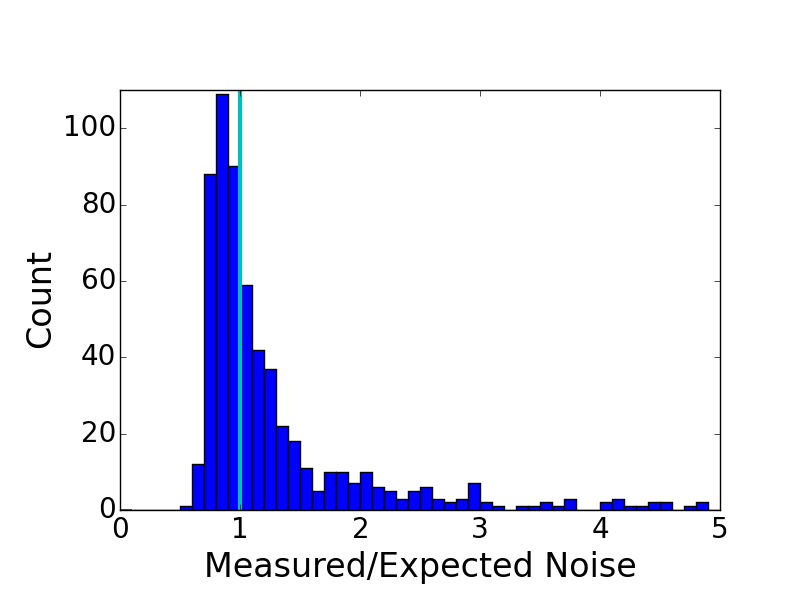}
\caption{Histogram of the median ratio of measured to expected readout and Johnson noise for all bolometers. 
There are 68 bolometers (not shown) with a ratio greater than 5.
The median value (vertical cyan) is 1.0
and the mode is centered at 0.85.
\label{fig:overbias_noise_hist} }
\end{center}
\end{figure}

\begin{table}[ht!]
\begin{center}
\begin{tabular}{| c | c | c |}
\hline  
Wafer & Expectation & Measured/expected ratio  \\
& (pA/$\sqrt{\mathrm{Hz}}$) &  \\ \hline
\hline 150-09 & 20 & 1.1  \\
\hline 150-14 & 18 & 1.3  \\
\hline 150-15 & 18 & 0.9  \\
\hline 150-39 & 22 & 0.8  \\
\hline 150-43 & 24 & 1.0  \\
\hline 150-47 & 24 & 1.0  \\
\hline 250-23 & 22 & 1.0  \\
\hline 250-24 & 26 & 0.9  \\
\hline 250-29 & 21 & 1.2  \\
\hline 410-28 & 32 & 1.0  \\ \hline
\hline All detectors &  & 1.0  \\
\hline
\end{tabular}
\end{center}
\caption{The expected combination of Johnson and readout noise and the ratio of measured to expected 
levels as measured in-flight for overbiased detectors. 
For each wafer the ratio is for the median of the distribution for the detectors on that wafer. 
The first three digits for the wafer number gives its operating frequency band. 
\label{tab:overbias_noise_table} }
\end{table}


For each detector we calculated the predicted readout noise using the electronic specifications of 
each element of the readout chain~\citep{aubin_thesis}. 
To estimate the Johnson noise we assumed a detector resistance $R_n$ as inferred from the network analysis, 
and a detector temperature $T =  T_{c} + \Delta P / \overline{G}$, where $\Delta P$ is 
the difference between the electrical power dissipated when the detector is overbiased
and $P_{sat}$ measured in dark conditions. 
We added the predicted contributions from readout and Johnson in quadrature
to find a combined noise term (in A/$\sqrt{\mathrm{Hz}}$). We multiplied this term by a factor 
1.7. 
This is the factor which accounts for the excess noise that was 
observed with the resistor measurements described in Section~\ref{sec:readoutnoise}. 
The expected levels shown in 
Figures~\ref{fig:one_bolo_overbias_noise} and ~\ref{fig:overbias_noise_hist} include this extra factor. 
For each detector and for all sections over the given 1.7 hours of data we formed the ratio of measured to expected 
noise and found the median. A histogram of the medians for all detectors is shown in Figure~\ref{fig:overbias_noise_hist}. 
Table~\ref{tab:overbias_noise_table} gives the median ratios for each wafer.  The 
median ratio for all detectors is 1.0. 

There was excess noise identified with resistor channels in Section\ref{sec:readout_performance}
amounting to a factor of 1.7 
higher than predictions.
When this excess is added to predictions of bolometer noise in the overbias state 
we find consistency with measurements.
We noted in Section~\ref{sec:readoutnoise} that noise measurements in the overbias state after the 
North American engineering flight did not exhibit this excess noise. The ratio of measured to predicted noise 
was 1.0$\pm0.1$~\citep{Aubin_TESReadout2010}. These measurements were done in the flight receiver, but 
the receiver was in the laboratory, without other electromagnetic interference (EMI) sources such as attitude control motors
and telemetry transmitters. We therefore hypothesize that EMI contributed additional noise that is most apparent in 
configurations in which the Johnson and readout noise terms dominate. 


\subsubsection{Noise Equivalent Power in Transition}
\label{sec:noiseintransition}

The total noise $N$ when the detector operates in its superconducting transition is given by Equation~\ref{eqn:noisebudget}. 
We report on measured $N$ using flight data after removal of the \ac{HWPSS}
(see Section~\ref{sec:timedomaindata}).  
We used all the data that passed quality cuts to be used for astrophysical analysis but also included 
additional data that would otherwise not be used because of, for example, large pointing errors or  
scan speeds that were too low. 

\begin{figure}[ht!]
\begin{center}
\includegraphics[height=2.5in]{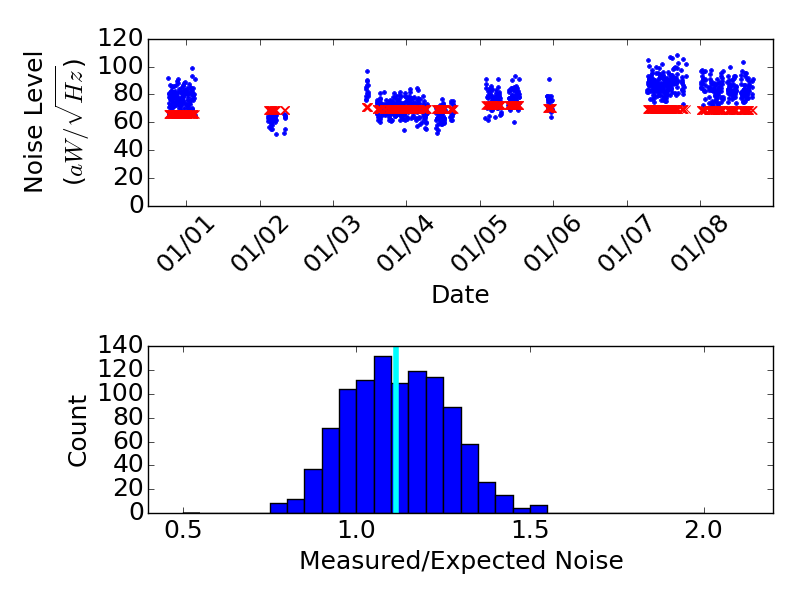}
\includegraphics[height=2.5in]{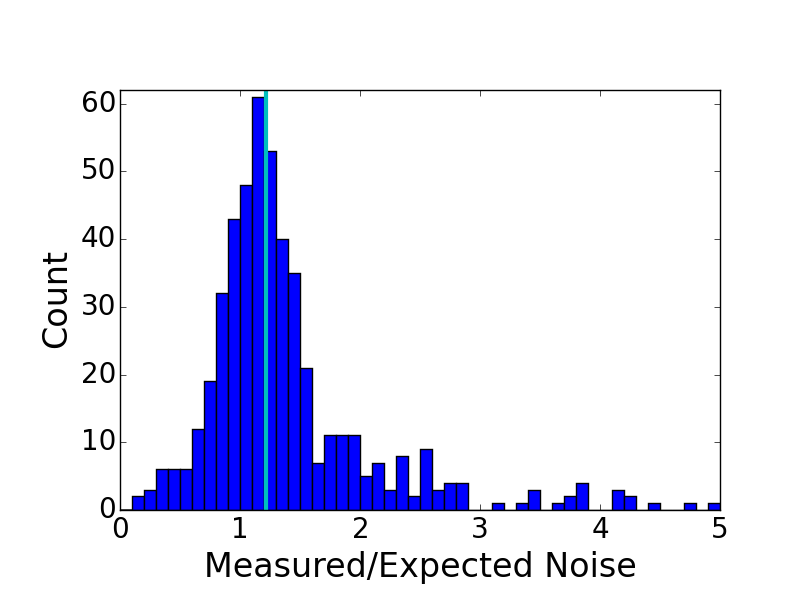}
\caption{Left top: in-transition measured (blue dots) and expected (red crosses) \ac{NEP} for one 150~GHz detector 
throughout flight. Gaps indicate times that data is absent or rejected. 
Left bottom: distribution of measured-to-expected ratio for this detector throughout flight, 
and the median (vertical cyan) ratio of 1.1.
Right: the distribution of the median ratio of measured to expected in-transition noise throughout flight for all bolometers.
The median ratio (vertical cyan) is 1.2. 
\label{fig:in_transition_noise} }
\end{center}
\end{figure}

The \ac{NEP} as a function of time throughout flight for one 150~GHz detector is shown in Figure~\ref{fig:in_transition_noise} 
along with the expected value. We describe the calculation of the expected value in the next paragraph. 
As before, we found a representative value for the ratio of measured-to-expected \ac{NEP} throughout flight for a 
given detector by finding the median 
over all sections. We combined the medians for each detector to find a median per wafer, which 
is reported in Table~\ref{tab:in_transition_noise_table}.  A median measured \ac{NEP} per frequency band is given 
in Table~\ref{tab:in_transition_noise_split}. Figure~\ref{fig:in_transition_noise} 
gives a histogram for the medians of measured-to-expected ratio for all detectors. 

\begin{table}[ht!]
\begin{center}
\begin{tabular}{|c|c|c|c|}
\hline  Wafer & Expected NEP [aW/$\sqrt{\mathrm{Hz}}$] & Measured/expected ratio \\ \hline
\hline 150-09 & 69 & 1.1   \\
\hline 150-14 & 48 & 0.7  \\
\hline 150-15 & 57 & 1.1  \\
\hline 150-39 & 77 & 1.1  \\
\hline 150-43 & 56 & 1.2  \\
\hline 150-47 & 64 & 1.3  \\
\hline 250-23 & 92 & 1.1  \\
\hline 250-24 & 78 & 1.4  \\
\hline 250-29 & 98 & 1.4  \\
\hline 
410-28 & 103 & 1.6  \\ \hline
\hline
All detectors &  & 1.2  \\
\hline
\end{tabular}
\end{center}
\caption{The median expected \ac{NEP} and measured-to-expected 
ratio for each wafer and combined for all detectors.
\label{tab:in_transition_noise_table} }
\end{table}


We calculated an expected \ac{NEP}
by finding the theoretical Johnson and readout current noise terms and multiplying 
them by (1) a factor of 1.7 
coming from the excess noise in the resistor 
measurements (see Section~\ref{sec:readoutnoise}), and (2) by an additional factor per wafer as 
listed in Table~\ref{tab:overbias_noise_table}. 
The second factor has been identified through the noise measurement with overbiased detectors (see Section~\ref{sec:JRN}). 
To convert from current noise to \ac{NEP} we divided the Johnson and readout noise terms by the 
current responsivity and assumed the high loop-gain limit.
We assumed $\gamma =0.5$ to predict the phonon noise~\citep{mather_appliedoptics_1982}. 
We calculated photon noise using the measured radiative load per detector, as described in Section~\ref{sec:optical_load}. 
We quote the predicted level for a photon correlation factor of 1. 
The median predictions decreases by 20\% (7\%) when using a correlation factor of 0 for the 
150 and 250 (410)~GHz bands. 


%
For the 150 band the measured noise is broadly consistent with calculations after accounting for the 
excess noise identified through measurements with resistors. No further excess noise is 
identified when operating the bolometers in transition. 
The average noise increase for the 3 (1) wafers at 250 (410)~GHz is about 30 (60)\% 
after accounting for the 
excess noise. This level is at the border or beyond the 20-30\% uncertainties in our estimates. 
This additional source of noise is being investigated, although the small number of available wafers 
may give a biased impression of trends. 


\begin{table}[ht!]
\begin{center}
\begin{tabular}{|c|c|c|c|}\hline
Noise source & \multicolumn{3}{c|}{ {Value ($aW/\sqrt{\mathrm{Hz}}$)} } \\
 & 150 GHz & 250 GHz & 410 GHz \\ \hline \hline
Readout & $\sqrt{2} \cdot 15$ & $\sqrt{2} \cdot15$ & $\sqrt{2} \cdot 23$ \\ \hline
Johnson & $\sqrt{2} \cdot 7.6$ & $\sqrt{2} \cdot 8.2$ & $\sqrt{2} \cdot 13$ \\ \hline
Phonon & 22 & 25 & 32 \\ \hline
Photon & 32 & 49 & 63 \\ \hline
Excess & 43 & 64 & 133 \\ \hline \hline
Total & 62 & 88 & 160 \\ \hline
\end{tabular}
\end{center}
\caption{The median \ac{NEP} contributions to noise measured in-transition for each frequency band.
\label{tab:in_transition_noise_split} }
\end{table}





%% file: net_and_maps.tex
\section{Noise Equivalent Temperature and Map Depth}
\label{sec:temperaturenoise}

We estimated the \ac{NET} in two ways: (1) using the \ac{TOD} in counts and the calibration factors that convert count
to temperature; (2) converting the measured \ac{NEP} to \ac{NET}. The term `temperature' refers to equivalent sky 
\ac{CMB} temperature fluctuations.  
In this section we describe the two methods and show that they give consistent results. We then use the measured \ac{NET} and 
the instantaneous attitude information to produce depth maps for each of the frequency bands.  

\subsection{NET using the TOD and Calibration}
\label{sec:net_from_tod}

\begin{figure}[ht!]
\begin{center}
\includegraphics[width=0.5\columnwidth]{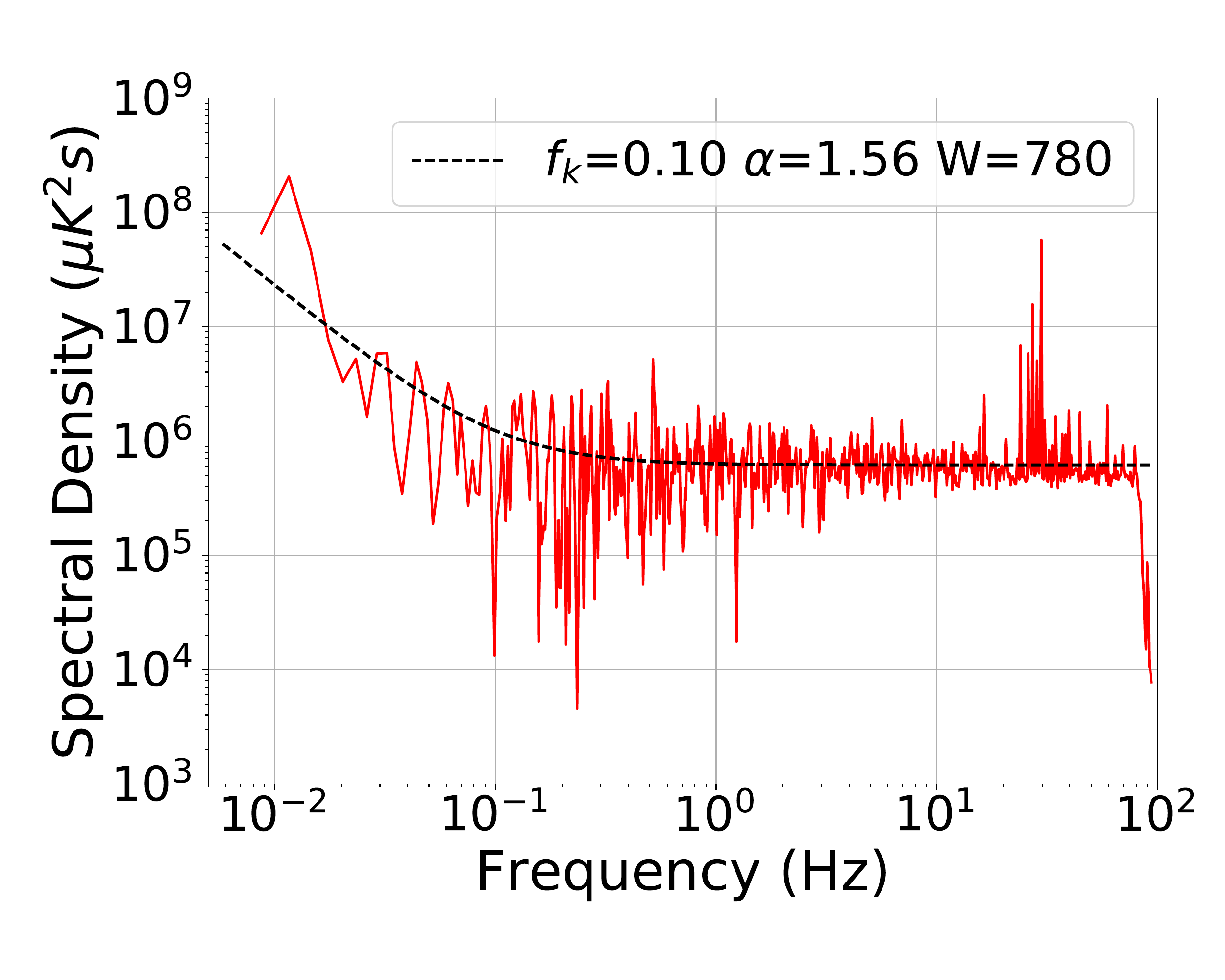}
\caption{Example PSD (solid red) for one section of a 250~GHz bolometer data with fit (dashed black) 
to the noise model of Equation~\ref{eqn:psdmodel}.
\label{fig:max_psfit} }
\end{center}
\end{figure}

For each detector we calculated power spectral densities of 5.7~min sections of calibrated, 
deglitched, \ac{HWPSS}-subtracted \ac{TOD}, 
and fit them with a three-parameter noise model $M(f)$ consisting of red and white noise terms as a function of 
frequency
\begin{equation}
M(f) = W^{2} \left[ 1+\left( \frac{f_k}{f} \right)^{\alpha} \right]. 
\label{eqn:psdmodel}
\end{equation}
The parameters are $W^{2}$, the white noise level (in K$^{2}$/Hz); $f_{k}$, the frequency cutoff of the red noise power law 
(also referred to as $f_{knee}$, in Hz); and $\alpha$, the red noise spectral index.
We used all calibrated data throughout 
flight that passed quality cuts for astrophysical analysis, see~\citet{joy_thesis}, and added other data with valid noise information 
but that would not pass cuts for making maps. Such data included sections during times at which 
the attitude reconstruction was poor or the scan speeds were below or above specific thresholds.
A polynomial with up to cubic terms in time
was subtracted from the \ac{TOD} before applying a Hann window and calculating the \ac{PSD}. 
Figure~\ref{fig:max_psfit} shows an example of a \ac{PSD} and its model fit for one section of the data of a 250~GHz detector.  

\begin{figure}[ht!]
\begin{center}
\includegraphics[width=0.99\columnwidth]{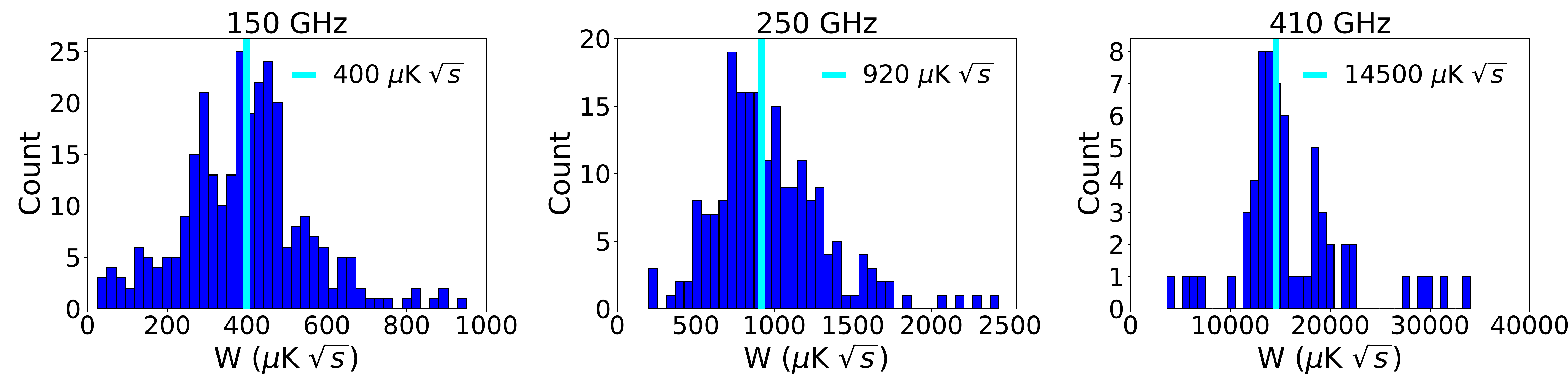}
\caption{Distribution of median \ac{NET} for all available detectors in a given frequency band and the median of
the distribution (vertical cyan).
\label{fig:net_all} }
\end{center}
\end{figure}

\begin{figure}[ht!]
\begin{center}
\includegraphics[width=0.32\columnwidth]{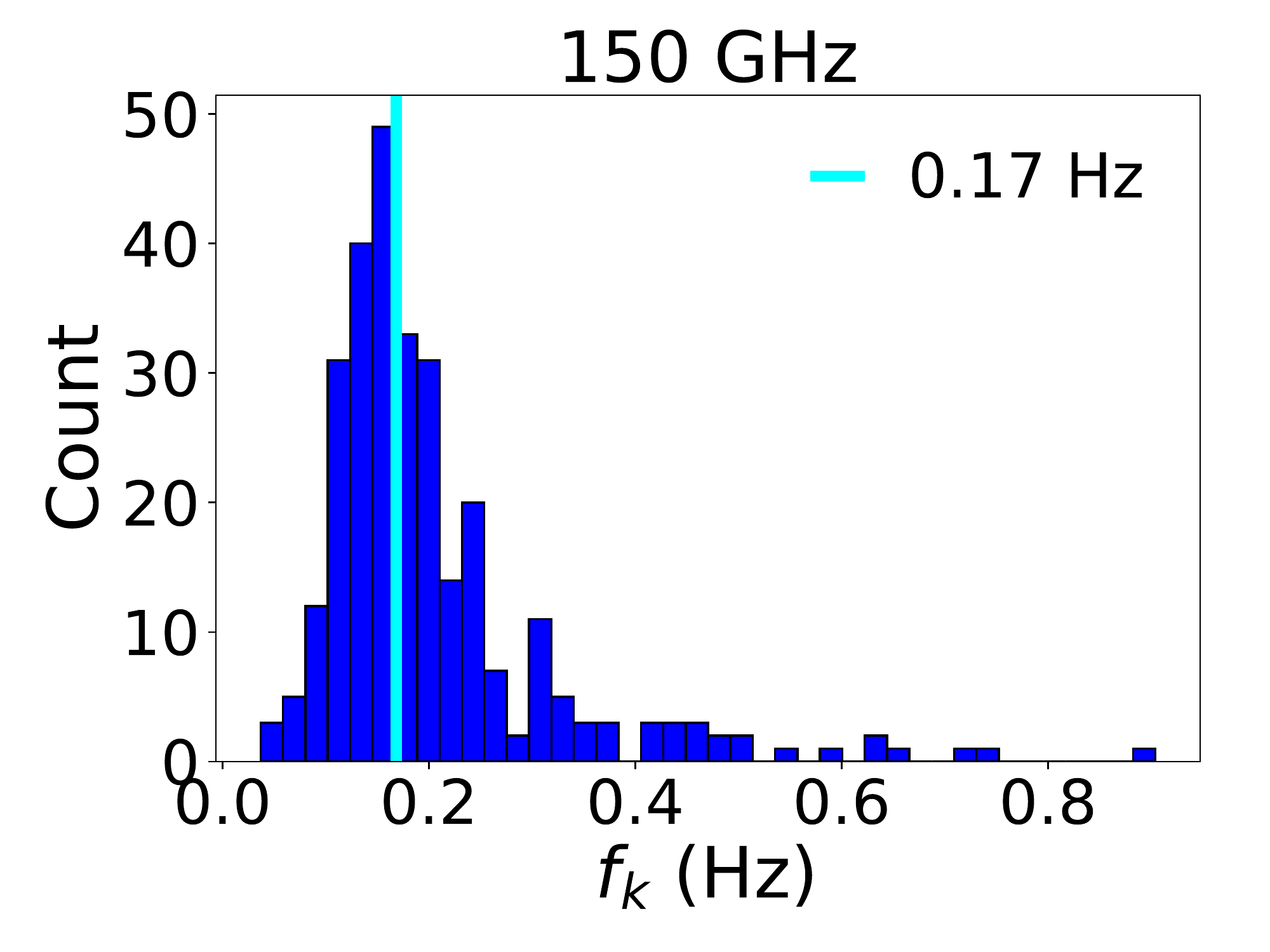}
\includegraphics[width=0.32\columnwidth]{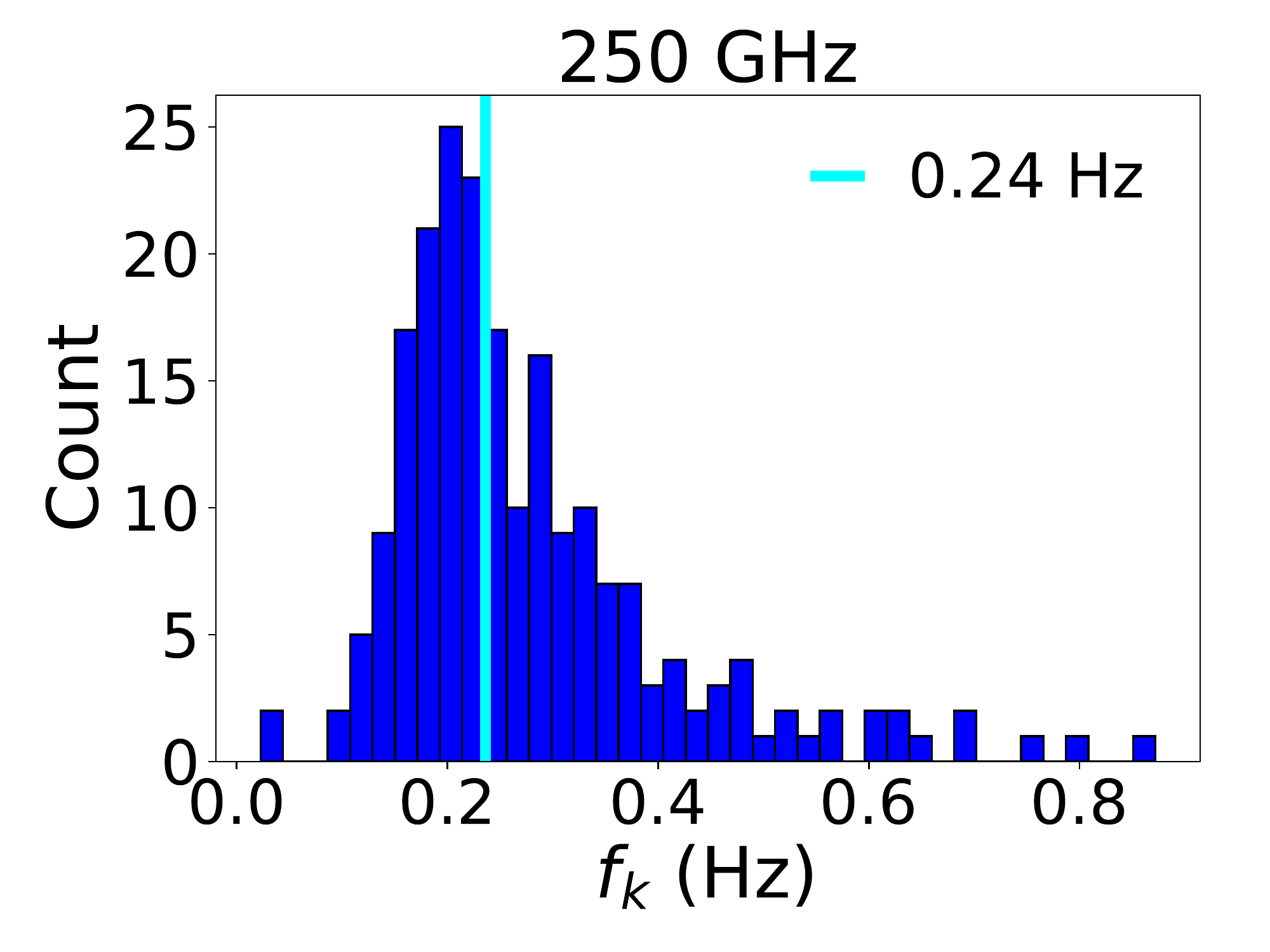}
\includegraphics[width=0.32\columnwidth]{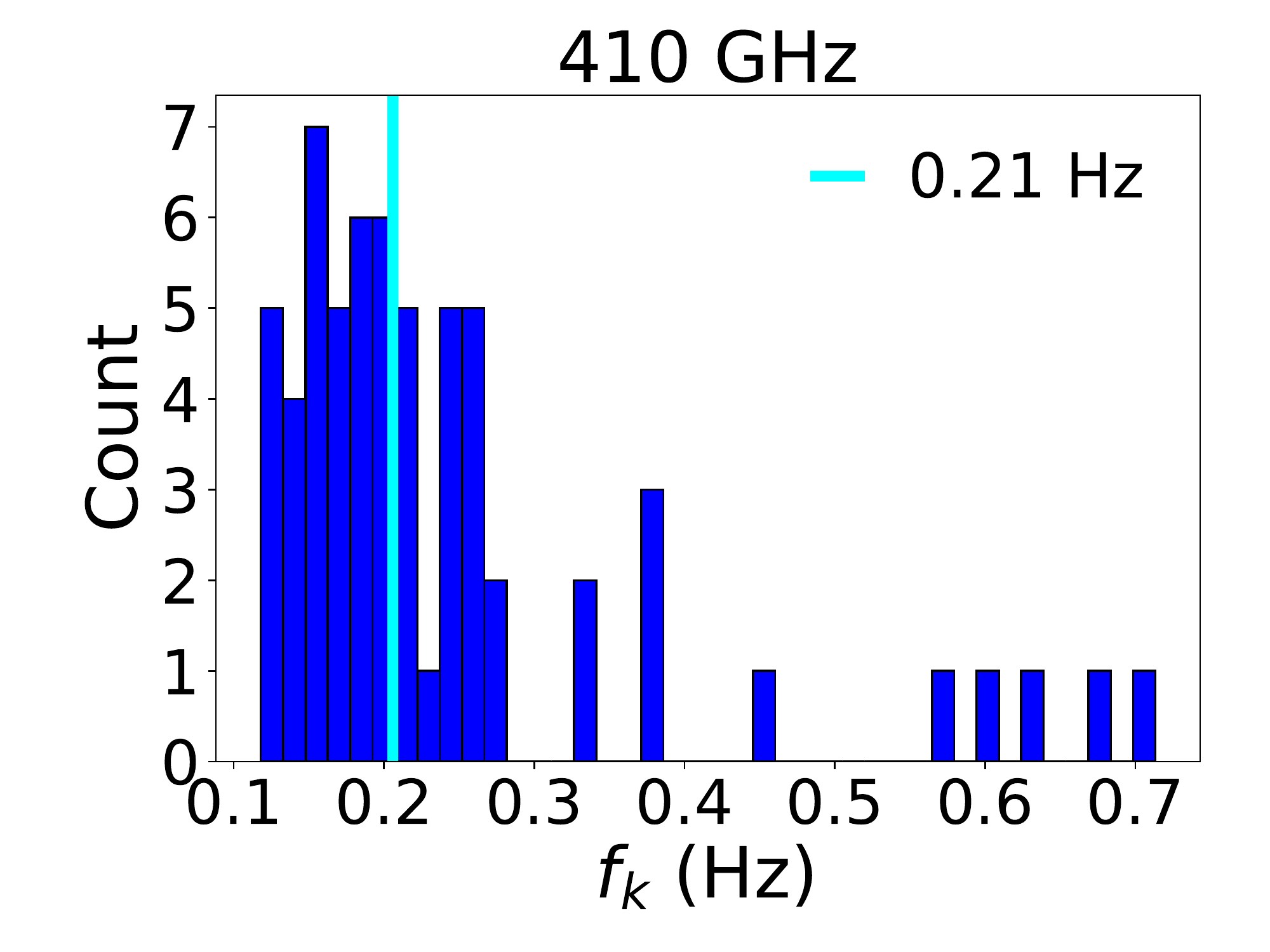}
\caption{Distribution of median knee frequencies for all available detectors in a given frequency band and the median of
the distribution (vertical cyan).
\label{fig:fknee_all} }
\end{center}
\end{figure}

\begin{figure}[ht!]
\begin{center}
\includegraphics[width=0.32\columnwidth]{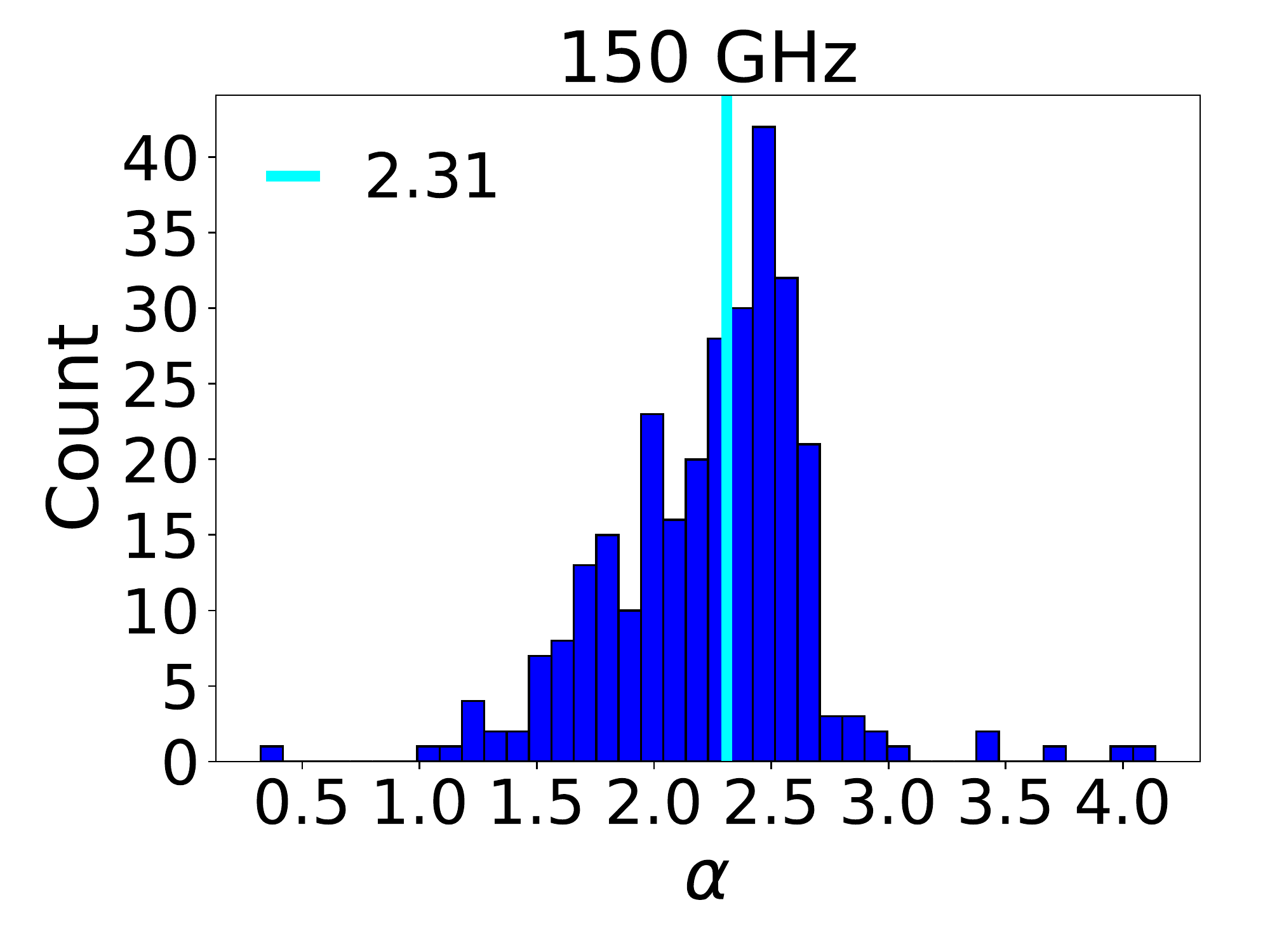}
\includegraphics[width=0.32\columnwidth]{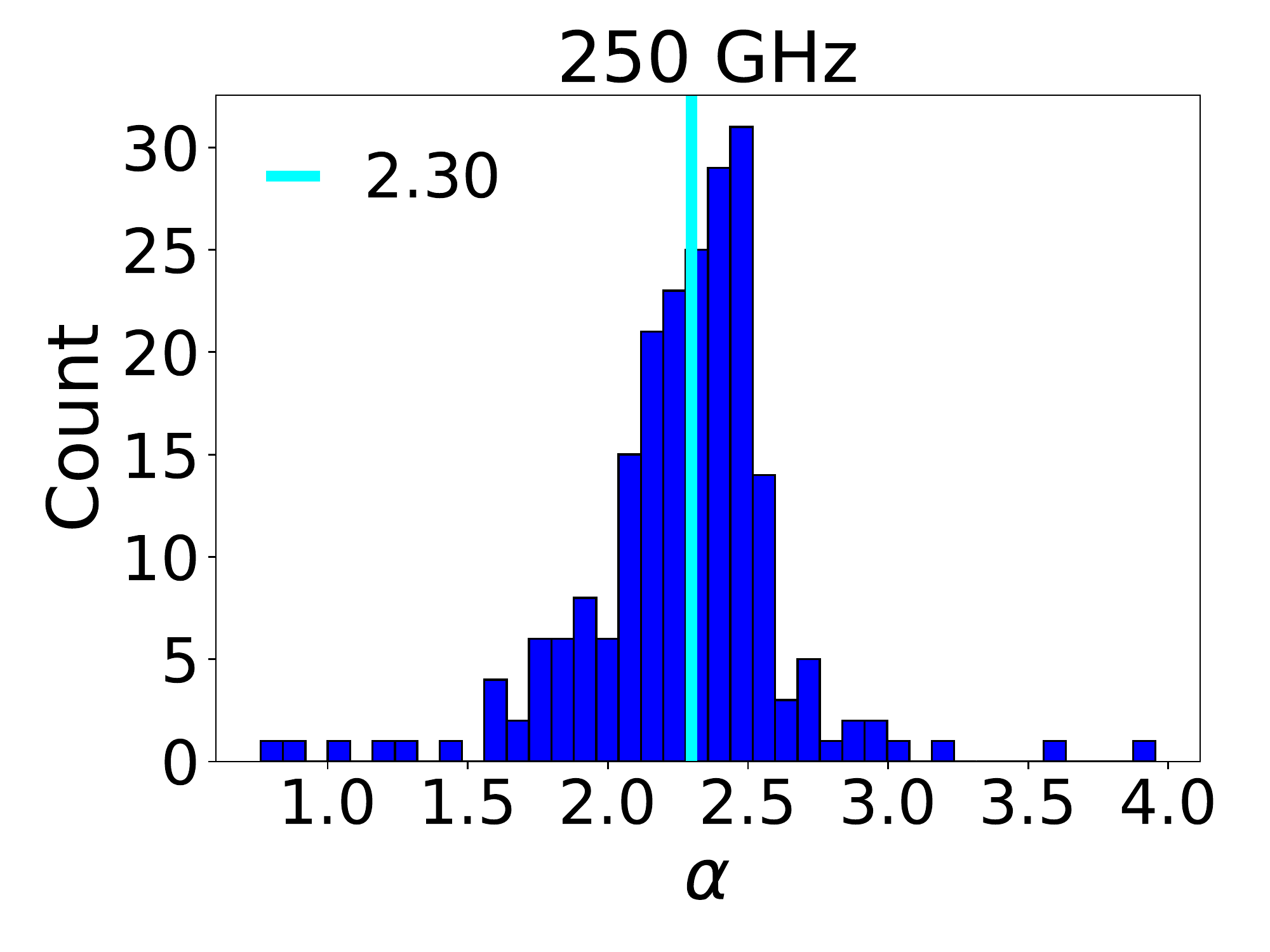}
\includegraphics[width=0.32\columnwidth]{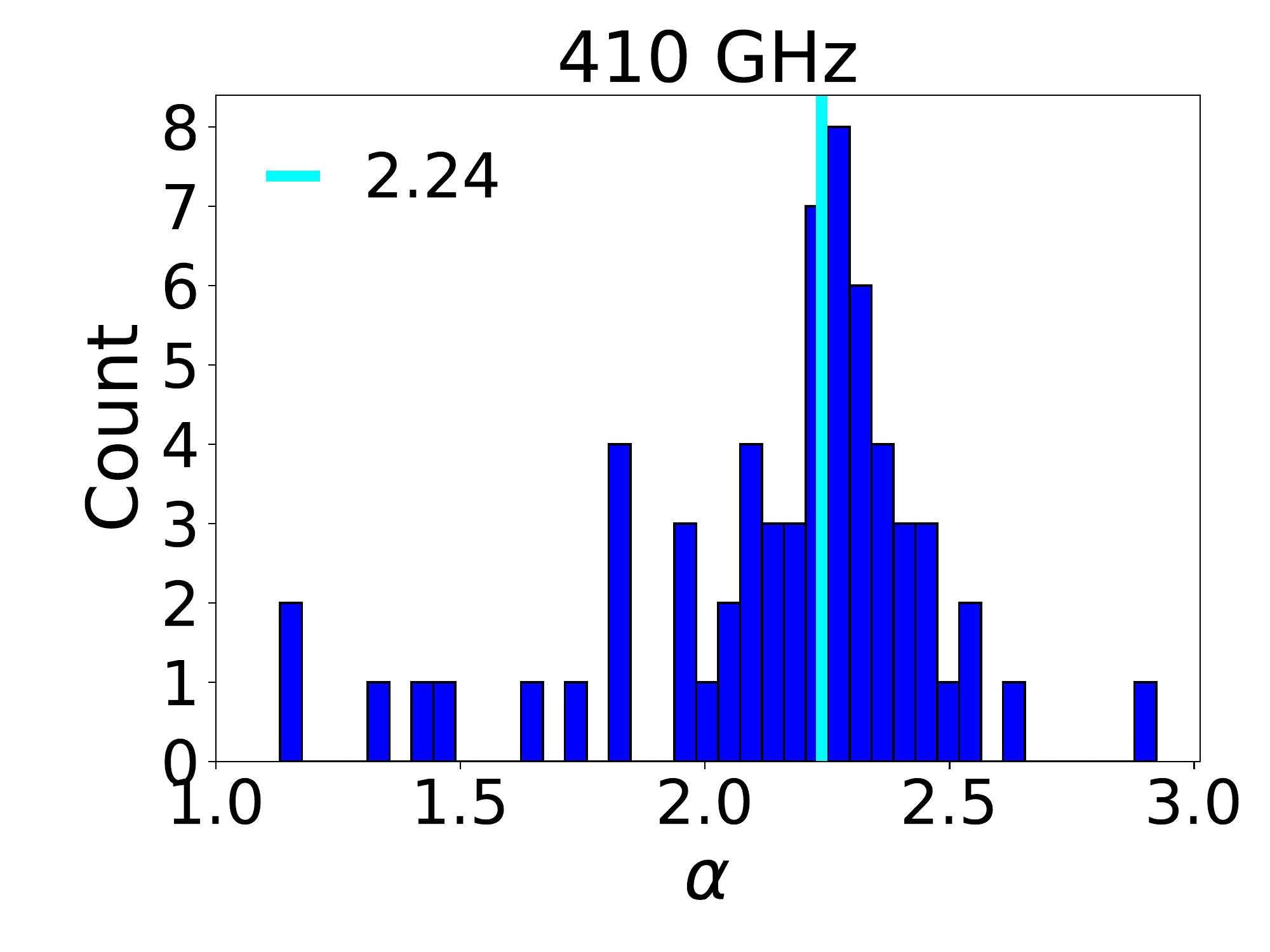}
\caption{Distribution of median red noise power indices for all available detectors in a given frequency 
band and the median of the distribution (vertical cyan).
\label{fig:alpha_all} }
\end{center}
\end{figure}

The fitting -- conducted on at least 10 sections of data for each detector, but typically over many hours
of data -- generated a set of parameters $W$, $f_k$, and $\alpha$, with which we characterized the noise 
performance. For each 
detector we find the three median parameter values over all flight. We histogram these median 
values for all available detectors per frequency band in Figures~\ref{fig:net_all},~\ref{fig:fknee_all}, and~\ref{fig:alpha_all}. 
The Figures also give the medians of the distributions. 

The data show knee frequencies near 200~mHz and red noise power law index near 2.3. These relatively 
high values are not intrinsic to the detectors or readout, but rather are a 
consequence of the azimuth motor malfunction, described in \ac{EP3}. In the absence of 
controlled azimuth pointing the payload executed full rotations causing strong Sun-synchronous drifts in the \ac{TOD}. 
Removal of a cubic order polynomial from the \ac{TOD} removes some of the drifts, but remnants are manifest in 
the statistics of $f_{k}$ and $\alpha$. A correlation analysis between the noise properties of different detectors showed 
strong coherency at low frequencies supporting the model of drifts due to Sun-synchronous signals.

\subsection{NET using the NEP}

We converted the measured \ac{NEP} to \ac{NET} using
\begin{equation} 
\mathrm{NET}(\mu \mathrm{K} \sqrt{\mathrm{s}}) = \mathrm{NEP}(\mathrm{W}/\sqrt{\mathrm{Hz}}) 
\left( \frac{dT}{dP}\right) \Bigr|_{T_{\mathrm{CMB}}} \frac{1}{\sqrt{2}} ,
\label{eqn:netfromnep}
\end{equation}
where $\left( \frac{dT}{dP}\right) \Bigr|_{T_{\mathrm{CMB}}}$ is a conversion factor from absorbed power to temperature 
calculated using the throughput, measured bands, and efficiency. For NEP
we used the median calculated for all detectors in a given frequency, as listed in Table~\ref{tab:comparenet}.

\subsection{Discussion of NET}

The two approaches for finding the NET are not independent. Finding NET using the measured calibration factors 
makes no assumption about the instrument's transmission efficiency and specific detector absorption efficiency. Finding NET using 
the NEP does not explicitly depend on the measured calibration factors. However, 
the conversion from {\it absorbed} power to temperature {\it on the sky}, as given in Equation~\ref{eqn:netfromnep}, depends 
on the transmission and absorption efficiencies, which were derived -- at least in one way of finding efficiencies -- 
using the calibration factors. Therefore the comparison we are providing is a check on end-to-end consistency, 
rather than a comparison of results using two entirely independent methods.

The measured \ac{NET}s are consistent within 10\% for the 150 and 250~GHz bands. There is a larger difference 
for the 410~GHz. The estimate from the NEP depends on the product of the instrument's throughput and efficiency 
through the multiplicative factor $dT/dP$; see Equation~\ref{eqn:netfromnep}. For the throughput we assumed a single 
mode ($A \Omega = \lambda^{2}$) for the propagating electromagnetic wave. We hypothesize that at the highest frequency band we are underestimating the throughput by assuming a complete cut-off of the 2nd mode. A 25\% higher 
throughput brings the NET from NEP and NET from calibration to values that agree within our overall uncertainty.  

\begin{table}[ht!]
\begin{center}
\begin{tabular}{|c|c|c|c|}
\hline   
                                                                      & 150~GHz  & 250~GHz & 410~GHz  \\
\hline 
NEP (aW/$\sqrt{\mathrm{Hz}}$) & 62 & 88 & 160 \\
\hline 
NET from NEP ($\mu$K~$\sqrt{\mathrm{s}})$                  & 380  & 1000  & 23000  \\
\hline 
NET from Calibration  ($\mu$K~$\sqrt{\mathrm{s}})$ & 400 & 920 & 14500 \\
\hline
\end{tabular}
\end{center}
\caption{The median measured \ac{NEP} per frequency band and comparison of \ac{NET} calculated 
from the \ac{NEP} using the measured efficiency, frequency bands and assumed throughput, and 
the \ac{NET} measured using the calibration.
\label{tab:comparenet} }
\end{table}

\subsection{Map Depth}

The high fidelity attitude reconstruction from the EBEX2013 flight was discussed in \ac{EP3}. 
We binned all the data that passed quality cuts for making maps using HEALPix~\citep{healpix2005} with $N_{side}=64$. 
With each sample we associated an equivalent temperature noise $N_{s}$ (in $\mu$K) equal to the 
product of the \ac{NET} calculated during that time section and the square root of the sampling rate. 
We calculated depth per pixel $D_{p}$ following
\begin{equation}
D_{p} = \left(  \sum \frac{1}{N_{s,p}^{2}}  \right)^{-1/2}, 
\label{eqn:mapdepth}
\end{equation}
where the sum is over all samples that had pointing associated with a given pixel $p$. 

\begin{figure}[ht!]
\begin{center}
\includegraphics[width=0.32\columnwidth]{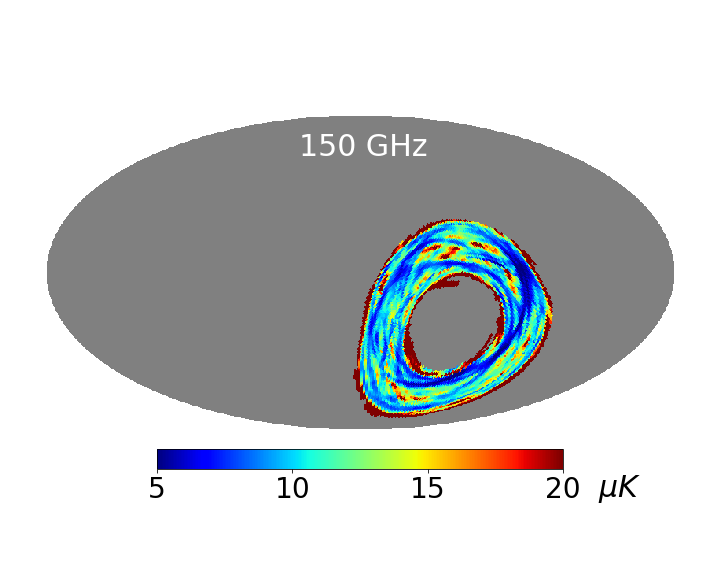}
\includegraphics[width=0.32\columnwidth]{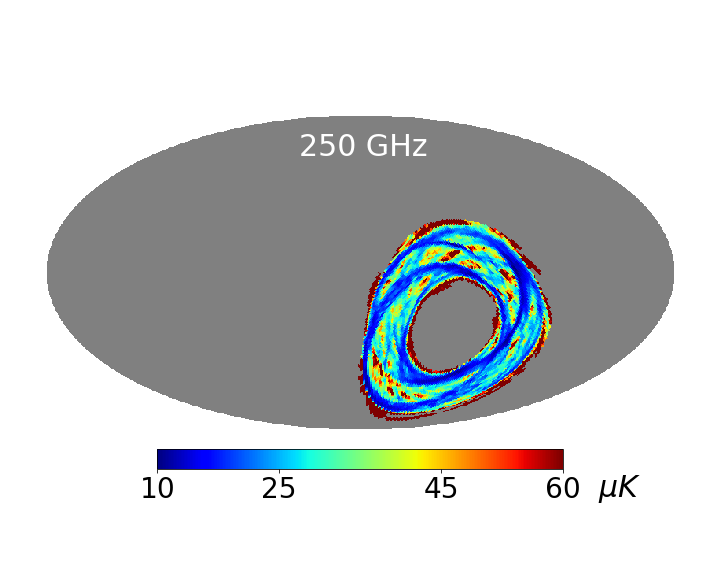}
\includegraphics[width=0.32\columnwidth]{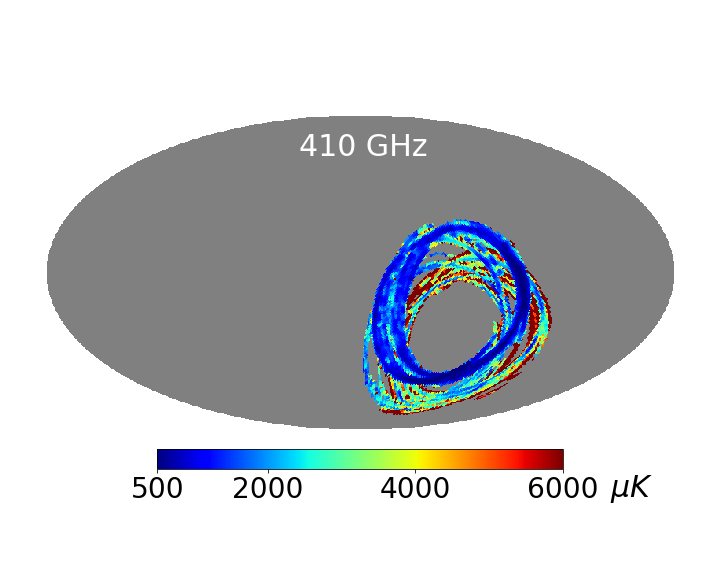}
\caption{Depth maps in Galactic coordinates using the \ac{NET} and attitude reconstruction from the EBEX2013 flight. 
The maps use HEALPix $N_{side}=64$ pixels and the color scale is linear. At this pixelization the median pixel noise is 
11, 28 and 1982 $\mu$K  for the 150, 250, and 410~GHz bands, respectively. 
\label{fig:depth} }
\end{center}
\end{figure}

The depth maps with pixels of $\sim$1~deg$^2$
for the three frequency bands are shown in Figure~\ref{fig:depth}. The noise distribution 
is strongly inhomogeneous, a consequence of an azimuth motor malfunction, as described in \ac{EP3}.  
The median depth values per pixel for the 150 and 250~GHz bands -- 11 and 28~$\mu$K, respectively --
are several factors larger than those reported by {\it Planck} and the expected CMB {\it E}-mode signal.
A signal-to-noise ratio above 1 is only achieved for a fraction of the pixels on the Galaxy at 410~GHz.
The depth maps do not encode $1/f$ noise and filtering that further degrades the signal-to-noise ratio. 
We therefore decided not to publish maps. 

%% file: summary2.tex

\section{Summary}
\label{sec:summary}


EBEX was a \ac{CMB} polarimeter that had two focal planes consisting of 14 detector wafers. Each focal plane had 4 
wafers operating at 150~GHz, two at 250~GHz, and a central wafer at 410~GHz. We have presented the design of bolometers
optimized for the low optical load in the stratosphere and for operation at 410~GHz, a frequency that is unique
for balloon payloads. To handle the thin wafers suitable for this frequency band we have developed a technique to bond
them to thicker wafers. We have shown distributions of measured thermal conductances, normal resistances, transition 
temperatures, and time constants. For the 250 and 410~GHz bands the median thermal conductances were 
within 20\% of design values. For the 150~GHz band, for which our target was 2.4 times lower than for the 250~GHz band, 
the median measured value was 2 times higher than  design. This result reflects the additional challenge in making very low thermal 
conductance spider-web bolometers operating at 0.25~K.  The lower thermal conductance is a consequence of the lower optical load 
expected at 150~GHz; see Table~\ref{tab:loadpredictions}. The measured 
transition temperatures were within 10\% of design values and median measured (design) saturation powers 
were 7.8 (4), 12 (9) and 14 (12)~pW, for the 150, 250, 
and 410~GHz, respectively. 
The measured time constants were 3.5 to 5.7 times longer than design values. Either published values
for specific heats of the materials that made up the bolometer are incorrect, or the devices had been contaminated 
by unknown materials during fabrication. 

We have measured in-flight median optical loads of 3.6, 5.3, and 5.0~pW, at the three bands, respectively. Given 
the final configuration of all optical elements inside the instrument and the measured frequency bands, 
we have calculated the expected contributions from the sky, warm telescope, and instrument emission. The warm 
mirrors, a teflon filter, and the \ac{HWP} dominated the expected load. We have also identified 
excess optical load in flight, but no excess load was measured when we coupled a 4~K load onto the vacuum 
window of the receiver in the laboratory. We therefore attribute the additional load to spillover of throughput onto 
warm surfaces outside of the receiver. 

Two measurement methods for finding the bolometer absorption efficiencies 
gave consistent results within 10\% for all frequency bands. They indicate high absorption 
efficiency ($\sim$ 0.9) for the 150~GHz and medium (medians of $\sim$0.35 and $\sim$0.25) for the 250 and 410~GHz bands. 
The measured distributions are broad occasionally giving apparently unphysical absorption values larger than 1. 
However, noisy measurements that give wide distributions can give values larger than 1. 

\ac{EBEX} was the first experiment to implement the digital version of a frequency domain multiplexing system
aboard a balloon payload. For several years it had the highest multiplexing factor of 16 until SPT3G 
began implementing a multiplexing factor of 64 in 2016. We have developed novel, superconducting, 
low inductance microstrips. We have measured gain stability of better than 0.25\% over 13 hours, and better than 2\% 
over the duration of the flight, even as the temperature of the readout boards experienced temperature excursions
of up to 20$^\circ$C. 

The rotation of the \ac{HWP} introduced a \ac{HWP} synchronous 
signal in the \ac{TOD}. To assess the noise performance we have fit
and removed the synchronous signal. We have given a detailed accounting of the measured noise. The readout noise
was measured to match predictions in the laboratory and for dark SQUIDs in flight.
We have found excess noise in-flight when analyzing resistor data 
and when the detectors were in an over-biased state, for which the readout and Johnson noise are expected 
to be dominant.  The median excess noise was a factor 1.7 larger than the expected total of Johnson and readout noise terms. 
When we included an extra noise term to account for the excess noise observed in the over-bias state
the median measured noise in-transition was 20\% higher than expectation. 
We suspect that electromagnetic pickup from telemetry electronics coupling to the long microstrips was the cause for the excess noise.



We have calculated the median \ac{NET} in two ways and found consistency for the 150, and 250~GHz bands. 
The pre-flight expectation for 150~GHz, taking into account the high thermal conductivity detectors, 
but without excess noise or load, was 210~$\mu$K$\sqrt{s}$. For the target thermal conductivity of 
19~pW/K, the expectation was 180~$\mu$K$\sqrt{s}$, a factor of 2.2 lower that the achieved value of 
400~$\mu$K$\sqrt{s}$. Our expectation of 180~$\mu$K$\sqrt{s}$ was higher by a factor of 1.3 compared 
to the best reported by BOOMERANG~\citep{crill2003} (which also had bolometers operated at 
a bath temperature near 0.25~K) because EBEX had two warm mirrors, not one, higher internal emission 
from a teflon filter due to a much larger vacuum window, higher temperature for its rotating \ac{HWP},
and higher detector thermal conductivity than required at 150~GHz.
We chose high detector thermal conductance because it 
mitigated the risk of un-anticipated optical load at the expense of higher noise.
This choice proved prudent; we did experience excess load but continued to operate the majority of the bolometers.  
A next generation balloon-borne instrument that has only one ambient temperature mirror, 
detectors with appropriately low thermal conductivity,
and that applies the lessons learned from EBEX2013 (including having lower 
internal emission from filters and \ac{HWP}, and mitigating excess load and noise) 
can achieve 140~$\mu$K$\sqrt{s}$ at 150~GHz. 

EBEX was the first balloon experiment to implement a kilo-pixel array of bolometers in the focal plane. It was 
among the first to face the challenges imposed by a significantly larger focal plane, optical elements, and vacuum 
window. The EBEX experience suggests the following path for a future experiment with kilo-pixel TES arrays 
to make optimal use of the balloon environment: (1) reduce instrument emissions through the development 
and implementation of high thermal conductance, low mm-wave emissivity
and reflectivity IR filters and lenses, and by maintaining optical elements at cold temperatures if practical. However, for an experiment
that targets a resolution finer than $\sim$25 arcmin at 150~GHz, at least one reflector will almost certainly still need
to be maintained at ambient temperatures; (2) implement optical elements that are appreciably larger than the nominal optical 
bundle to mitigate diffraction near apertures 
and thus reduce spill-over outside the designated throughput, even at the expense of a larger cryostat, 
larger window, and potentially heavier
payload;  (3) pay attention to EMI mitigation, as wiring that couples the SQUIDs to the detector focal plane 
are susceptible to the transmitter-heavy environment on-board the payload. (Since the 
EBEX2013 flight there have been additional reports of excess noise due to telemetry electronics
in other millimeter and sub-millimeter experiments~\citep{galitzki2014, gambrel_thesis}.) 
The EMI integrity of the wiring cavity of the cryostat should be excellent or experiments should 
impose radio-silence after initial check-up and calibration. Significant time should be allocated to  
testing in full flight configuration, including telemetry electronics. 

Because of the malfunction of the azimuth motor, a result of thermal design error that is discussed in \ac{EP3}, 
the instrument scanned $\sim$6000 deg$^2$, giving shallow depth with highly inhomogeneous noise. 
We have presented depth maps that gave a median noise of 11 and 28~$\mu$K per HEALPix $(N_{side} = 64)$ pixel at 150 and 250~GHz.
The map noise is higher compared to {\it Planck}'s noise as well as to that of other contemporaneous experiments
and we did not find it compelling to publish our own maps or power spectra. 

\ac{EBEX} pioneered the use of \ac{TES} bolometers on a balloon-borne platform. It was the first 
experiment to fly a small array of these detectors in a test-flight in 2009 and 
a kilo-pixel array during its EBEX2013 flight. Nearly 1000 bolometers 
were operating shortly after the payload reached float altitude. \ac{EBEX} pioneered the use of
the digital frequency domain multiplexing readout system.  This system is now implemented 
on several operating ground-based instruments~\citep{Benson2014, Stebor2016, Inoue2016} and is baselined for a 
proposed space mission~\citep{Suzuki2018}. 
\ac{EBEX} was also the first astrophysics experiment to implement a superconducting magnetic bearing. 
This system too is baselined for a ground-based and a proposed space mission~\citep{Hill2018,lida2017}. \ac{EBEX} was 
a successful technology pathfinder for future \ac{CMB} space missions. 


%% file: acronyms2.tex
\begin{acronym}
    \acro{ACS}{attitude control system}
    \acro{ADC}{analog-to-digital converters}
    \acro{ADS}{attitude determination software}
    \acro{AHWP}{achromatic half-wave plate}
    \acro{AMC}{Advanced Motion Controls}
    \acro{ARC}{anti-reflection coating}
    \acro{ASD}{amplitude spectral density}
    \acro{ATA}{advanced technology attachment}
    \acro{BRC}{bolometer readout crates}
    \acro{BLAST}{Balloon-borne Large-Aperture Submillimeter Telescope}
    \acro{CANbus}{controller area network bus}
    \acro{CMB}{cosmic microwave background}
    \acro{CMM}{coordinate measurement machine}
    \acro{CSBF}{Columbia Scientific Balloon Facility}
    \acro{CCD}{charge coupled device}
    \acro{DAC}{digital-to-analog converters}
    \acro{DASI}{Degree~Angular~Scale~Interferometer}
    \acro{dGPS}{differential global positioning system}
    \acro{DfMUX}{digital~frequency~domain~multiplexer}
    \acro{DLFOV}{diffraction limited field of view}
    \acro{DSP}{digital signal processing}
    \acro{EBEX}{E~and~B~Experiment}
    \acro{EBEX2013}{EBEX2013}
    \acro{ELIS}{EBEX low inductance striplines}
    \acro{EP1}{EBEX Paper 1}
    \acro{EP2}{EBEX Paper 2}
    \acro{EP3}{EBEX Paper 3}
    \acro{EP4}{EBEX Paper 4}
    \acro{ETC}{EBEX test cryostat}
    \acro{FDM}{frequency domain multiplexing}
    \acro{FPGA}{field programmable gate array}
    \acro{FCP}{flight control program}
    \acro{FOV}{field of view}
    \acro{FWHM}{full width half maximum}
    \acro{GPS}{global positioning system}
    \acro{HPE}{high-pass edge}
    \acro{HWP}{half-wave plate}
    \acro{HWPSS}{half-wave plate synchronous signal}
    \acro{IA}{integrated attitude}
    \acro{IP}{instrumental polarization} 
    \acro{JSON}{JavaScript Object Notation}
    \acro{LDB}{long duration balloon}
    \acro{LED}{light emitting diode}
    \acro{LCS}{liquid cooling system}
    \acro{LC}{inductor and capacitor}
    \acro{LPE}{low-pass edge}
    \acro{MLR}{multilayer reflective}
    \acro{MAXIMA}{Millimeter~Anisotropy~eXperiment~IMaging~Array}
    \acro{NASA}{National Aeronautics and Space Administration}
    \acro{NDF}{neutral density filter}
    \acro{NEP}{noise equivalent power}
    \acro{NET}{noise equivalent temperature}
    \acro{PCB}{printed circuit board}
    \acro{PE}{polyethylene}
    \acro{PME}{polarization modulation efficiency}
    \acro{PSD}{power spectral density}
    \acro{PSF}{point spread function}
    \acro{PTFE}{polytetrafluoroethylene}
    \acro{PV}{pressure vessel}
    \acro{PWM}{pulse width modulation}
    \acro{RMS}{root mean square}
    \acro{RSS}{rotation synchronous signal}
    \acro{SLR}{single layer reflective}
    \acro{SMB}{superconducting magnetic bearing}
    \acro{SQUID}{superconducting quantum interference device}
    \acro{SQL}{structured query language}
    \acro{STARS}{star tracking attitude reconstruction software}
    \acro{TCPIP}[TCP/IP]{Transmission Control Protocol/Internet Protocol}
    \acro{TDRSS}{tracking and data relay satellites}
    \acro{TES}{transition edge sensor}
    \acro{TM}{transformation matrix}
    \acro{TOD}{time ordered data}
    \acro{VME}{Versa Module European}

\end{acronym}

%% file: reference_maps.tex

\section{Calibration}
\label{sec:refMaps}

The calibration produces two equivalent conversions, one from measured counts to signal power incident on the instrument, and the other 
from measured counts to CMB temperature fluctuation of magnitude $\mathrm{K}_{\mathrm{CMB}}$. 
In Section~\ref{sec:efficiencyfromcal} we 
use $CAL$, which is the conversion from counts to power; $CAL=d P_{inc,\, inst} / d I^{c}$~(W/count). 
For completeness, for the purpose of this appendix, we denote as $CAL_{T}$ the quantity $CAL_{T}=d T_{inc,\, inst} / d I^{c}$~(K/count), 
where $T_{inc}$ refers to CMB temperature fluctuation of magnitude $T_{inc}$~(K).  
To find these calibration factors using passes across the Galactic plane we build reference signal maps at our frequency 
bands from the \planck~component maps described in~\citet{planck_x}.
We use the R2 \planck\ component maps processed with the Commander framework from https://pla.esac.esa.int/\#maps.
The reference signal maps are a combination of thermal dust, free-free, CO, and the \ac{CMB}. Within the \ac{EBEX} 
bands and near the Galactic plane the maps are entirely dominated by the dust component. 

We now describe the making of the dust intensity map as a function of frequency. Other signal maps, 
corresponding to other components, are made in the same way using the extrapolation factors 
provided in~\citet{planck_x}.
The thermal dust amplitude map $A_{D}$ referred to a baseline frequency $\nu_{0D} = 545$~GHz is first converted from 
its native $K_{RJ}$ units to intensity (W/sr m$^2$ Hz). 
The resulting map is then extrapolated to frequency $\nu$ using the frequency scaling factor 
\begin{equation}
\left( \frac{e^{\frac{h \nu_{0D}}{k T_D} - 1}}{e^{\frac{h \nu}{k T_D} - 1}} \right) \left( \frac{\nu}{\nu_0} \right)^{\beta_D + 3}, 
\end{equation} 
and using the \planck -provided maps of dust temperature $T_{D}$ and spectral index $\beta_{D}$. 

The total incident signal map (at each frequency) is a sum of the individual component signal maps.
To find the total incident signal power we integrate the total incident reference signal map over the 
EBEX bands (as measured and reported in EP1) and multiply by the throughput of the instrument. 
A linear regression of the power in the reference map vs counts  measured by a given bolometer when 
passing the galactic plane gives $CAL$ for that bolometer. 

The total incident signal map (at each frequency) is alternatively converted to 
$\mathrm{K}_{\mathrm{CMB}}$ using the spectral radiance of a blackbody  $B_{\nu} (\nu, T)$.
It is then integrated as a function of frequency and weighted by the normalized transmission of the instrument. 
A linear regression of the temperatures in the reference map to the counts measured by a given bolometer when passing the galactic 
plan give $CAL_{T}$ for that bolometer.


